\documentclass[journal]{vgtc}                     %

\usepackage{times}                     %

\usepackage{amsmath,amsfonts}
\usepackage{color, colortbl}
\usepackage{soul}
\usepackage{setspace}
\usepackage[normalem]{ulem}

\newcommand\blfootnote[1]{%
  \begingroup
  \renewcommand\thefootnote{}\footnote{#1}%
  \addtocounter{footnote}{-1}%
  \endgroup
}
\onlineid{1008}

\vgtccategory{Research}

\vgtcpapertype{Systems \& Rendering}

\title{\texorpdfstring{GASP: Gradient-Aware Shortest Path Algorithm for \\ Boundary-Confined 2-Manifold Reeb Graph Visualization}{GASP: Gradient-Aware Shortest Path Algorithm for Boundary-Confined 2-Manifold Reeb Graph Visualization}}

\author{%
    \authororcid{Sefat E Rahman}{0009-0000-3263-7100},
    \authororcid{Tushar M. Athawale}{0000-0003-3163-6274}, and
  \authororcid{Paul Rosen}{0000-0002-0873-9518}
}

\authorfooter{
  \item
  	Sefat E Rahman and Paul Rosen are with the University of Utah.
  	E-mail: \{~sefat.rahman~|~paul.rosen~\}@utah.edu.
  \item
  	Tushar M. Athawale is with Oak Ridge National Laboratory.
  	E-mail: athawaletm@ornl.gov.
}

\abstract{%
  Reeb graphs are an important tool for abstracting and representing the topological structure of a function defined on a manifold.
  We have identified three properties for faithfully representing Reeb graphs in a visualization: they should be constrained to the boundary, compact, and aligned with the function gradient.
  Existing algorithms for drawing Reeb graphs are agnostic to or violate these properties.
  In this paper, we introduce an algorithm to generate Reeb graph visualizations, called \textit{GASP}, that is cognizant of these properties, thereby producing visualizations that are more representative of the underlying data.
  To demonstrate the improvements, the resulting Reeb graphs are evaluated both qualitatively and quantitatively against the geometric barycenter algorithm, using its implementation available in the Topology ToolKit (TTK), a widely adopted tool for calculating and visualizing Reeb graphs.

}


\teaser{
 \centering
 \rotatebox{90}{\small \hspace{15pt}\textsf{GASP Algorithm}\hspace{18pt}\textsf{Geometric Barycenter}} 
 \includegraphics[width=0.805\linewidth]{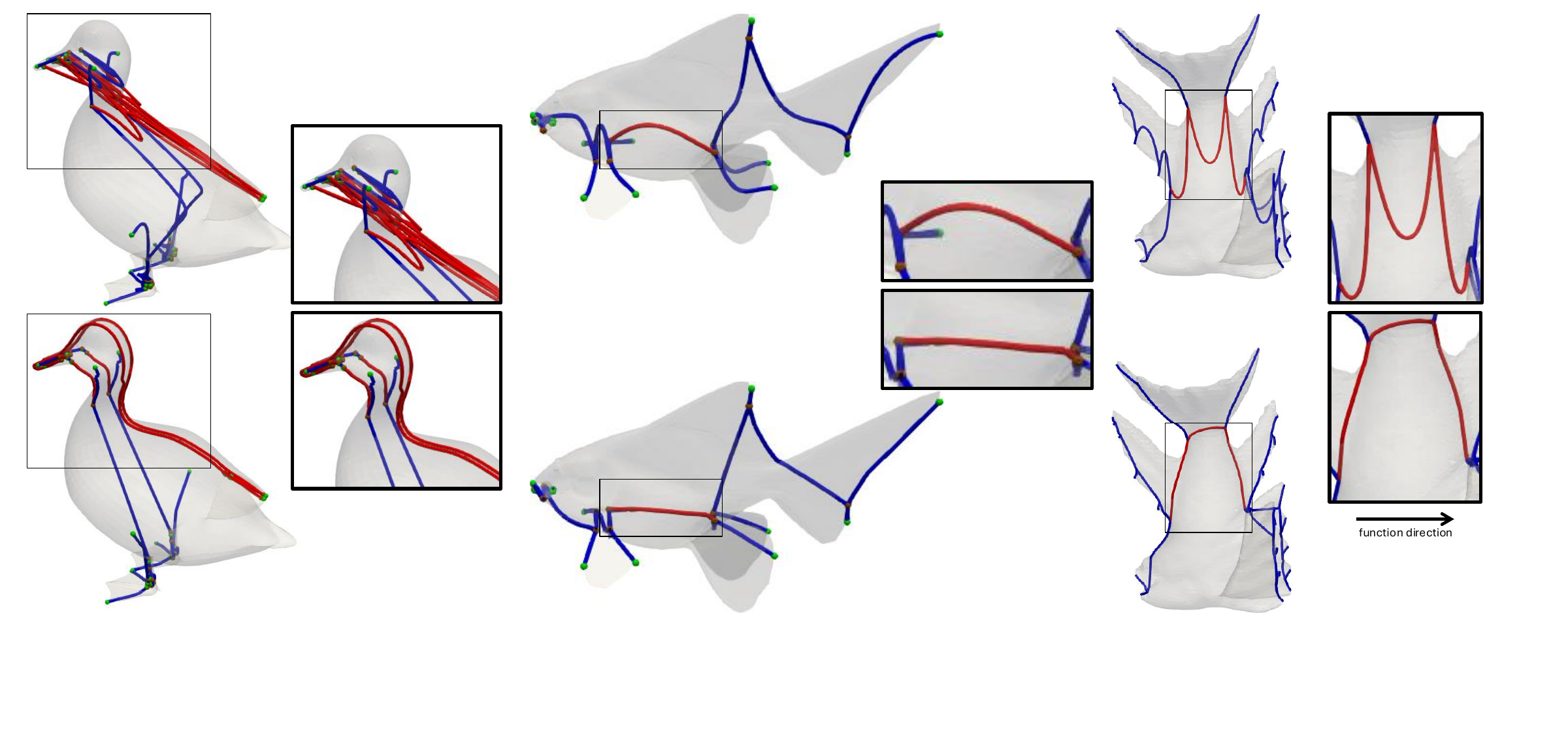}

 \vspace{-10pt}
 \subfloat[\texttt{duck} ($x$ function)\label{fig:teaser:outside}]{\hspace{0.35\linewidth}}
 \subfloat[\texttt{fish} ($z$ function)\label{fig:teaser:length}]{\hspace{0.3\linewidth}}
 \subfloat[\texttt{windfish} ($x$ function)\label{fig:teaser:direction}]{\hspace{0.225\linewidth}}
 \hspace{10pt}
 
 \caption{
 Examples showing that the GASP algorithm (bottom) better accommodates the three desired properties for visualizing Reeb graphs, as compared to the geometric barycenter (GB) method, implemented in the Topology ToolKit (top). 
 The complete Reeb graphs include both \textit{\textcolor{red}{red}} and \textit{\textcolor{blue}{blue}} arcs, but \textit{\textcolor{red}{red}} arcs highlight improvements over GB.
 (a)~GASP constrains the path of the Reeb graph arcs such that they remain either inside or on the surface of the object, whereas many arcs generated by GB do not conform to the boundary of the model.  
 (b)~GASP chooses the shortest path between two critical points (while constrained by the boundary), but some GB arcs are unnecessarily long due to the desire for smoothness. 
 (c)~GASP maintains consistency with the gradient of the function, while in part due to smoothing, GB arcs travel nearly perpendicular to the direction of the function (left-to-right in this case).}
 \label{fig:teaser}

}

\definecolor{MyGray}{gray}{0.9}

\newcommand{\grad}[1]{{\nabla {#1}}}

\newlength\mylen

\graphicspath{{figs/}{figures/}{pictures/}{images/}{./}} %

\newcommand{\Cspace}{\mm{{\mathbb C}}}
\newcommand{\Rspace}        {{\mathbb R}}

\newcommand {\mm}[1] {\ifmmode{#1}\else{\mbox{\(#1\)}}\fi}

\newcommand{\Mgroup}        {{\mathcal M}}

\begin{document}

\crefname{figure}{Fig.}{Fig.}
\Crefname{figure}{Fig.}{Fig.}
\crefname{section}{Sec.}{Sec.}
\Crefname{section}{Sec.}{Sec.}

\maketitle

\setstretch{0.965}
\section{Introduction}
\label{sec:intro}

The
Reeb graph is a powerful tool for understanding the shape and structure of objects and data. 
It provides a compact structural abstraction of a function defined on a manifold (i.e., a skeleton of an object)~\cite{kitazawa2023notes}.
A wide range of applications have utilized Reeb graphs~\cite{9677901,agarwal_et_al:LIPIcs.FSTTCS.2017.8,Gelbukh_2023}, 
from analyzing 3D printing models~\cite{9926485} to assisting in disease studies, such as Alzheimer's~\cite{Shailja2022.03.11.482601}.
However, faithfully visualizing the Reeb graph structure is of the utmost importance for interpretation~\cite{TA:2024:topoSensitivity}.

Since the Reeb graph is drawn based on a scalar function defined over a manifold, it should closely reflect those objects. To ensure this, we identified three essential properties and one optional property that a Reeb graph visualization should possess. 
First, because the scalar field is defined on the manifold, the arcs of the Reeb graph should remain confined within the manifold. Any arc extending beyond it would imply the Reeb graph exists outside the manifold (see \Cref{fig:teaser:outside} top).
Second, the Reeb graph arcs should be as short as possible to minimize visual clutter and to facilitate easier interpretation (see \Cref{fig:teaser:length} top). Third, the arc orientations should closely align with the direction of the associated scalar field gradient. A mismatch between the gradient direction and the Reeb graph arcs can be confusing and make it harder to understand the scalar field (see \Cref{fig:teaser:direction} top). Fourth, smoother arcs provide a clearer presentation of the paths, making them easier to track, but this property is optional, as it should not be at the cost of the other properties. \blfootnote{This manuscript has been authored by UT-Battelle, LLC under Contract No. DE-AC05-00OR22725 with the U.S. Department of Energy. The publisher, by accepting the article for publication, acknowledges that the U.S. Government retains a non-exclusive, paid up, irrevocable, world-wide license to publish or reproduce the published form of the manuscript, or allow others to do so, for U.S. Government purposes. The DOE will provide public access to these results in accordance with the DOE Public Access Plan (\url{http://energy.gov/downloads/doe-public-access-plan}).}

Several algorithms for drawing Reeb graphs and contour trees (loop-free Reeb graphs) have been proposed (see \Cref{sec:reeb-vis}). However, none explicitly identify or address all of these properties. Prior work on Reeb graph visualization has focused on planar drawings or included only brief mentions within the context of Reeb graph construction~\cite{pascucci2007robust, ReebGraphSimplification}. To the best of our knowledge, this paper is the first to focus exclusively on visualizing $2$-manifold Reeb graphs in 3D embeddings.

Motivated by the goal of producing Reeb graph visualizations that adhere to the three essential properties, we developed an algorithm, called \textit{GASP}, which produces Reeb graphs that are boundary-constrained and \ul{G}radient-\ul{A}ware with \ul{S}hort \ul{P}aths. GASP works by first decomposing the model into topological cylinders, each corresponding to a Reeb graph arc. These arcs are then drawn as the shortest paths between critical points over a graph constructed from successive (iso)contours.
GASP arcs generally align with the first property (keeping arcs within the model) by leveraging the contours to constrain arc placement within the boundary (see \Cref{fig:teaser:outside} bottom). GASP addresses the second property (minimizing arc length) by computing the shortest paths between critical points (see \Cref{fig:teaser:length} bottom).
GASP satisfies the third property (following the gradient) by progressing from one contour to the next, ensuring that arcs align with the direction of the function (see \Cref{fig:teaser:direction} bottom). 
Finally, although GASP arcs are generally smooth, it does not directly address the optional property of smoothness.

Given its widespread use, we compare the GASP algorithm to the geometric barycenter~\cite{pascucci2007robust} algorithm, as implemented within Topological ToolKit (TTK)~\cite{ttk}, for visualizing the Reeb graph. For each of the three essential properties, confining arcs within the model (see \Cref{fig:teaser:outside}), shorter arc length (see \Cref{fig:teaser:length}), and alignment with the functional gradient (see \Cref{fig:teaser:direction}), GASP outperforms the geometric barycenter algorithm at only a small cost to the optional property of smoothness.

\vspace{3pt}
\noindent
In summary, the key contributions of this paper are:
\begin{itemize}[noitemsep, itemsep=3pt]
  \item The identification of three essential and one optional property for faithfully visualizing Reeb graphs.  
  \item A Reeb graph visualization algorithm, \textit{GASP}, that is boundary-constrained and \ul{G}radient-\ul{A}ware with \ul{S}hort \ul{P}aths.  
  \item An open-source implementation of GASP, and a set of new quantitative benchmarks for evaluating Reeb graph visualization quality that uses 30 triangle meshes with height and geodesic functions.
\end{itemize}

\section{Background on Reeb Graphs}
\label{sec:reeb}

We briefly present the technical definitions of isocontours and critical points that are the building blocks of a Reeb graph construction. We then describe Reeb graph construction, computation, and visualization.

\subsection{Isocontours and Critical Points}  
\label{sec:levelSets}
Isocontours are a foundational tool in visualization. Isocontours represent a locus of points in a $2$-dimensional manifold that attain a fixed function value, known as an isovalue. We restrict our discussion to scalar functions sampled on a $2$-manifold, as other types of functions and manifolds are beyond the scope of this paper. Mathematically, if $f:\Mgroup\rightarrow \Rspace$ is a scalar function defined on a $2$-manifold, $\Mgroup$, then the isocontour $\mathcal{L}$ for isovalue $k$ can be represented as $\mathcal{L}(k)\equiv\{P: P\in \Mgroup \land f(P)=k\}$, where $P$ denotes domain positions with function value~$k$. The change in data values over the domain can be understood by visualizing the evolution of isocontours for isovalues spaced at regular intervals. \Cref{fig:reebGraphConcept:a} visualizes isocontours (in \textcolor{orange}{\textit{orange}}) of a height function, $f$, for various isovalues, $f_i$, mapped on a torus. 
A single isovalue may produce a single or multiple connected components. As observed in \Cref{fig:reebGraphConcept:a}, the isocontours immediately beyond split and merge saddles comprise two connected components.

Critical points, $C$, are special points of a field that have a close relation with the evolution of isocontours. Technically, critical points denote domain positions, $P$, where the field gradient vanishes. Let $f: \Mgroup \to \Rspace$ be a Morse function; $\grad{f}$ denotes its gradient. A point $C \in \Mgroup$ is considered {\em critical} if $\grad{f}(C)=0$; otherwise it is {\em regular}. A critical point is categorized into three types: local minimum, local maximum, or saddle. In particular, if $f(C)$ is smaller than all of its neighbors, then it is a local minimum. Similarly, if $f(C)$ is greater than all of its neighbors, then it is a local maximum. If $f(C)$ is smaller than one neighbor and greater than the next neighbor in alternating fashion, with the neighbors visited sequentially in a clockwise/counterclockwise manner, it is a saddle. Critical points represent the important domain positions because they result in the birth, splitting, merging, and death of connected components in isocontours. \Cref{fig:reebGraphConcept:a} visualizes the critical points as black spheres. As observed, isocontours (depicted in \textcolor{orange}{\textit{orange}}) appear, split, merge, and disappear at the local minimum, split saddle, merge saddle, and local maximum, respectively, as the isovalue increases. The evolution of isocontours with respect to critical points is captured by Reeb graphs, as described in the next section.

\subsection{The Reeb Graph} 
\label{sec:reebGraphs}

\begin{figure}[!t]
    \begin{center}
    \includegraphics[height=2.45cm]{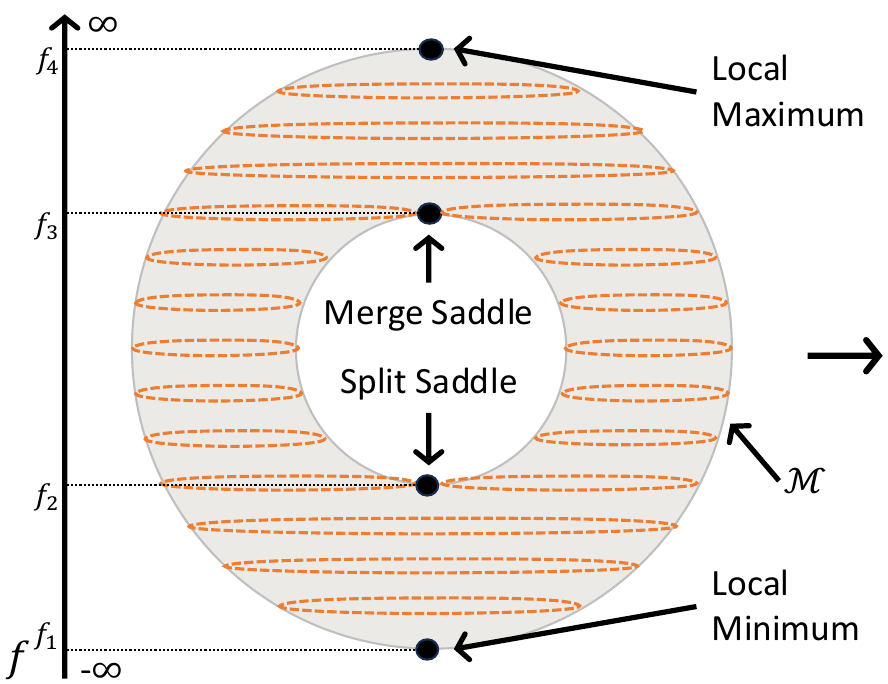}%
    \includegraphics[height=2.45cm]{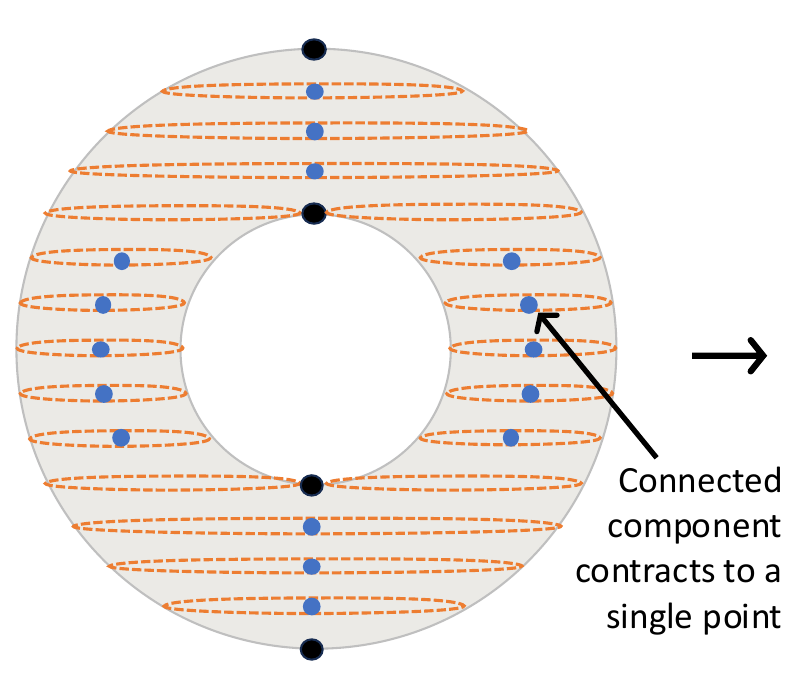}%
    \includegraphics[height=2.45cm]{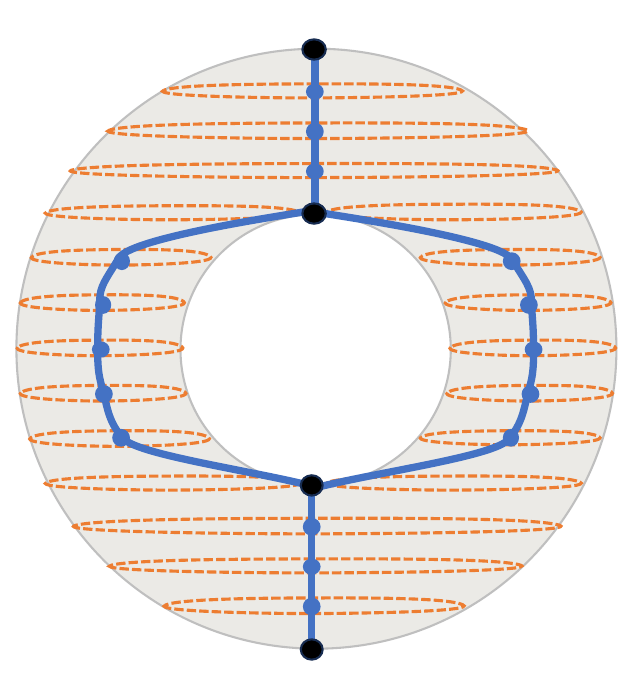}
    \end{center}

    \vspace{-20pt}
    \hspace{0.085\linewidth}
    \subfloat[\label{fig:reebGraphConcept:a}]{{\hspace{0.22\linewidth}}}
    \hspace{0.1\linewidth}
    \subfloat[\label{fig:reebGraphConcept:b}]{{\hspace{0.21\linewidth}}}
    \hspace{0.08\linewidth}
    \subfloat[\label{fig:reebGraphConcept:c}]{{\hspace{0.235\linewidth}}}
    
    \caption{Reeb graph construction. (a)~Isocontours are visualized in \textcolor{orange}{\textit{orange}}, and the \textit{black} spheres denote the positions of critical points. (b)~Connected components are contracted to a single point depicted as \textcolor{blue}{\textit{blue}} dots. (c)~Critical points and the point representation of connected components provide the Reeb graph as a structural abstraction.}
    \label{fig:reebGraphConcept}
\end{figure}

The Reeb graph is a structural abstraction that provides insight into the topological skeleton of scalar field data. Formally, the Reeb graph tracks the connected components of isocontours in the domain as $f$ is swept from $-\infty \rightarrow +\infty$~\cite{computational_topology_intro}. 
On a simply connected manifold, $\Mgroup$, the Reeb graph of a Morse function, $f$ is loop-free (a generalization of contour tree) ~\cite{carr2003computing}.
As shown in \Cref{fig:reebGraphConcept:b}, each connected component of an isocontour (\textcolor{orange}{\textit{orange}} ellipses) can be contracted to a single point (depicted as \textcolor{blue}{\textit{blue}} dots). Connecting the points representing connected components forms arcs or edges ($E$) of a Reeb graph, as portrayed in \Cref{fig:reebGraphConcept:c}. In summary, the arcs ($E$) of a Reeb graph indicate connected components in the domain with the same topology, and critical points ($C$) in the Reeb graph represent nodes that indicate topological changes. Reeb graphs, therefore, provide a compact and abstract visual representation of the data topology.

\subsection{Reeb Graph Computation}
\label{sec:reeb-compute}

Considering the high utility of Reeb graphs in the analysis of complex data, several prior works focused on methods for efficient computation and visualization of Reeb graphs. Shinagawa and Kunii~\cite{TA:Shinagawa:1991} proposed $O(n^2)$ algorithm for Reeb graph construction, where $n$ represents the number of vertices in a $2$-manifold triangular mesh. Since then, multiple new algorithms were proposed that increased the efficiency of Reeb graph constructions~\cite{TA:Kree:2004, TA:Doraiswamy:2008, DORAISWAMY2009606, TA:Tierny:2009:LoopSurgery, minimalContouringReebGraphs}. 
Harvey and Wang~\cite{harvey2010randomized} achieved an optimal expected running time of $O(m \log m)$ through a randomized dynamic-connectivity strategy, and Parsa~\cite{parsa2012deterministic} derandomized this approach to obtain the same bound in the worst case.  
Gueunet et al.~\cite{gueunet2019task} presented a task-based parallel algorithm for Reeb graph calculation and visualization, leveraging merge and split trees constructed from scalar field data. 
It uses Fibonacci heaps to efficiently handle dynamic operations like merging and splitting components. 
TTK implements this approach for Reeb graph computation. 
Although this work targets parallel performance, its sequential core follows Parsa’s framework, which itself builds on the randomized dynamic-connectivity formulation introduced by Harvey and Wang.

\subsection{Reeb Graph Visualization}
\label{sec:reeb-vis}

Several approaches have been used to visualize Reeb graphs, contour trees, and general graphs. Strodthoff and J\"{u}ttler~\cite{STRODTHOFF2015186:layeredReebGraph} devised a layered-Reeb graph construction algorithm for efficient computation and visualization of Reeb graphs for solids with boundary embedded in 3D. Shinagawa et al.~\cite{Shinagawa:Reebgraphdrawing} devised a 2D layout for the representation of Reeb graphs. Pascucci et al.~\cite{Pascucci2009} devised ``toporrery'' as a way to draw contour trees in a radial manner that avoids the self-intersection of edges in planar methods. Heine et al.~\cite{Heine:contourtreedrawing} leveraged advances in graph drawing techniques~\cite{dot:graphdrawing, Auber2004:tulip} for drawing contour trees in 2D layouts. 3D medial axis algorithms~\cite{cornea2024curve} deal with a related problem, in that they produce a skeleton of a manifold. However, they do not have the constraints introduced by the scalar function defined on the manifold.
Rehal and Sen~\cite{rehal2023generation} proposed computing all non-homotopic paths between points on a surface or in its ambient space using routing graphs derived from handle and tunnel loops, offering a topology-aware approach conceptually related to skeleton and graph generation methods.

Similar to our approach, the embedded layout of Reeb graphs, as described by Doraiswamy and Natarajan~\cite{ReebGraphSimplification}, ensures that arcs lie within their respective monotone cylinders in the input domain. This approach leverages the LS-graph to trace paths confined to the interior of the corresponding cylinders, facilitating intuitive visualization within the original dataset's spatial context. The proposed technique addresses the first desired property of a Reeb graph identified by us, but no experimental results were presented to support the claim. Further, the other two properties, i.e., generating shorter arcs and alignment with the gradient of the function, are not clearly addressed.

\begin{figure}[!t]
    \centering

    \begin{minipage}[m]{0.15\linewidth}
        \includegraphics[width=\linewidth]{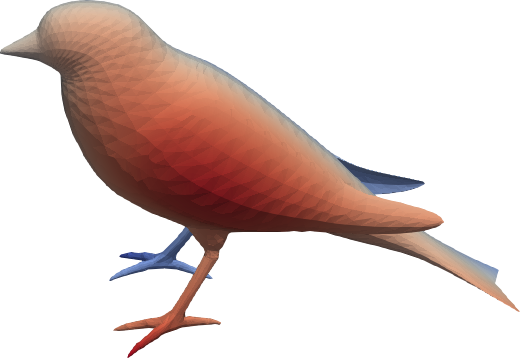}
    \end{minipage}
    \hfill
    \begin{minipage}[m]{0.825\linewidth}
        \subfloat[Straight Edges\label{fig:TTK-Bird-lines}]{\includegraphics[width=0.33\linewidth]{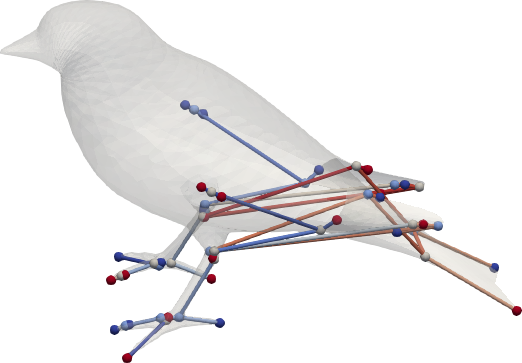}}
        \hfill 
        \subfloat[Arcs 15 Subdivisions\label{fig:TTK-Bird-arcs}]{\includegraphics[width=0.33\linewidth]{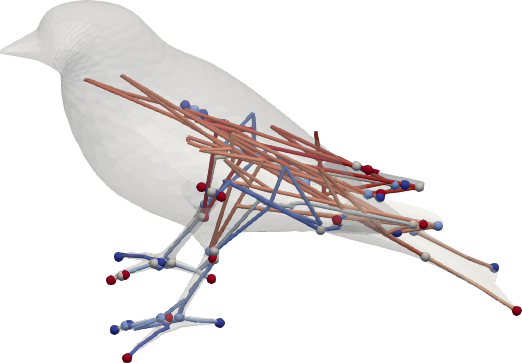}}
        \hfill
        \subfloat[Smoothing 15 Iterations\label{fig:TTK-Bird-Curves}]{\includegraphics[width=0.33\linewidth]{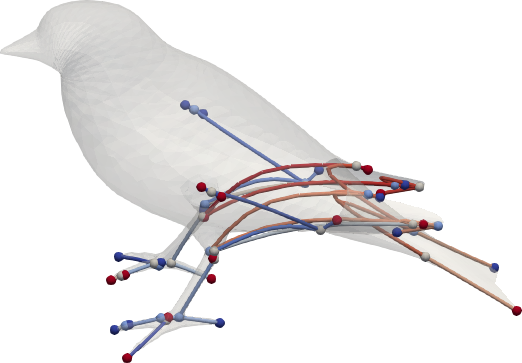}}
    \end{minipage}
    
    \caption{Illustration of TTK Reeb graph drawing for \texttt{bird} ($x$ function). (a)~TTK starts by drawing straight edges between the critical points. (b)~Next, TTK applies the arc sampling to subdivide the edges into arcs. (c)~Finally, a geometric smoother is applied to improve the arc aesthetics.}
    \label{fig:UseCase1-TTK-Bird}
\end{figure}

\paragraph{Geometric Barycenter Algorithm} The Topology ToolKit (TTK)~\cite{ttk} is a frequently used tool for visualizing Reeb graphs that uses the geometrical barycenter algorithm~\cite{pascucci2007robust}.
TTK's method for visualizing Reeb graphs involves three steps~\cite{julien2024, pascucci2007robust}. First, Reeb graph edges connect critical points directly via straight lines (see \Cref{fig:TTK-Bird-lines}). Arcs are then produced using an arc sampling parameter that evenly divides the straight edge into sub-components (see \Cref{fig:TTK-Bird-arcs}). Then, each sample is placed at the geometrical barycenter of the isocontours, and, finally, the arcs are smoothed geometrically to improve the aesthetics of the representation (see \Cref{fig:TTK-Bird-Curves}).
TTK produces smooth, aesthetically pleasing Reeb graph arcs; however, in many cases, it fails to produce Reeb graph arcs with the three essential properties we identified. First, the process used by TTK often results in arcs that extend beyond the model's boundary (see the \textcolor{red}{red} arcs in \Cref{fig:teaser:outside} top). 
Further, the sampling and smoothing processes often lead to unnecessarily long paths, which cause visual clutter (see the \textcolor{red}{red} arcs in \Cref{fig:teaser:length} top). 
Finally, the process of smoothing can introduce significant deviations from the function gradient direction (see the \textcolor{red}{red} arcs in \Cref{fig:teaser:direction} top).

To the best of our knowledge, no prior work has thoroughly addressed the challenge of drawing Reeb graphs in 3D, focusing on the three essential properties we have identified.

\section{Literature Review}

In this section, we briefly discuss prior work on Reeb graphs for data analysis and 
topology-based visualizations.

\paragraph{Reeb Graphs for Data Analysis}
Reeb graphs play a pivotal role in data analysis. Reeb graphs and contour trees (as well as other topology-based techniques) are used as a way to compare and measure the similarity of features across complex scalar fields, as surveyed in~\cite{TA:Yan:2021:ScalarFieldComparisonTopoDescriptors}. Various metrics, including interleaving~\cite{TA:2013:Morozov:interleavingDistance}, edit~\cite{TA:2020:Sridharamurthy:editDistance}, and stable~\cite{TA:2023:Bollen:stableDistance} distance, have been proposed to measure the distance between merge trees that are foundational to the construction of Reeb graphs and contour trees. Recently, functional distortion metric~\cite{TA:2014:Bauer:reebGraphDistance} and intrinsic interleaving distances between merge trees~\cite{TA:2023:Lan:labledInterleavingDistanceReebGraph} were used to compare Reeb graphs quantitatively. Chen et al.~\cite{TA:Chen:2013:fluidParticleTracking} analyzed time-dependent multi-fluid data by studying Reeb graphs of fluid density distributions. Weber et al.~\cite{Weber2011} proposed feature tracking using Reeb graphs in time-varying data for combustion applications. Reeb graphs are also a fundamental tool in shape analysis in computer graphics and other domains~\cite{BIASOTTI20085:shapeanalysis, extendedSurfaceGraphs:SurfaceRendering, landscape:ReebGraph}. Natali et al.~\cite{NATALI2011151:pointCloud} proposed utilizing Reeb graphs for extracting the topological skeleton of point cloud data. Given the widespread use of Reeb graphs in data analysis, accurate visualization methods for Reeb graphs are a critical component of that analysis.

\paragraph{Topological Visualizations}
Topological visualization is a powerful tool used to concisely convey the scale and position of important data features to facilitate analysis of complex scientific data~\cite{Bremer09tvcg, rosen2021using, NatarajanMolecularDynamics,TA:2022:Nauleau:turbulentFlowTopoAnalysis}. While isocontour~\cite{Lorensen:1987:MCA}, critical point~\cite{TA:Morse:1930:criticalPoints}, and Morse-Smale complex~\cite{TA:Edelsbrunner:2003:MorseSmaleComplexes} techniques explicitly convey the position of important data features, persistence diagram and persistence curve techniques~\cite{persplot_stability} convey the scale of data features. Reeb graphs~\cite{computational_topology_intro} and contour trees~\cite{carr2003computing} provide topological abstractions that convey feature positions through critical points and scale through arc length in a planar representation. The discrete graph-based representation of these topological visualization types, however, can make their perception and data analysis tasks difficult. Topological simplification~\cite{EdelsbrunnerLetscherZomorodian2002,TA:Gyulassy:2005:scalarFieldSimplification, ReebGraphSimplification} is, therefore, a standard method used to reduce the level of detail or noise in data to make topological visualizations more interpretable. Recent work by Athawale et al.~\cite{TA:2024:topoSensitivity} and previous work on vector field topology~\cite{TA:Laidlaw:2005:2dvectorFieldVisUserStudy, TA:Forsberg:2009:3dvectorFieldVisUserStudy} showed a limited capability of users to perceive topological features conveyed by various methods. The studies concluded a need to improve existing visualization designs to make topological visualizations more perceptible. In this paper, we propose a novel algorithm, GASP, to enhance the faithfulness of Reeb graph visualization compared to the geometric barycenter algorithm.

\begin{figure}[!b]
    \centering

    \hfill
    \includegraphics[width=0.75\linewidth]{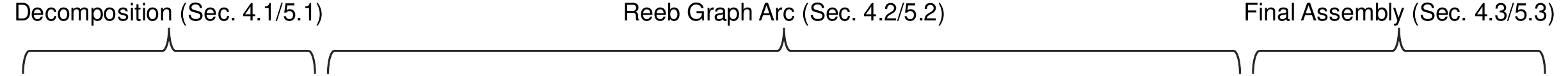}
    
    \begin{minipage}[m]{0.25\linewidth}
        \subfloat[\label{fig:conceptual:input}]{\hspace{10pt}\includegraphics[height=1.7cm]{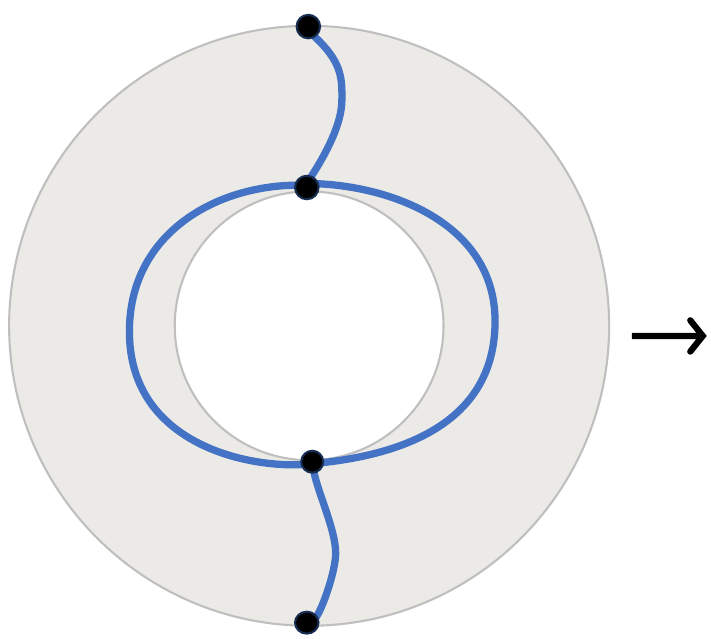}}
    \end{minipage}
    \begin{minipage}[m]{0.135\linewidth}
        \subfloat[\label{fig:conceptual:decomposition}]{\begin{minipage}[m]{\linewidth}\includegraphics[width=\linewidth]{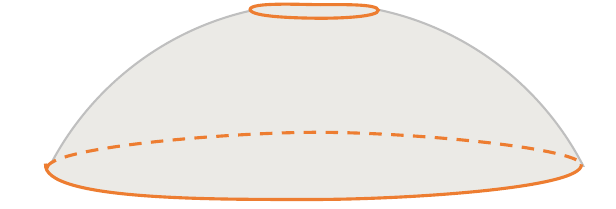}
        \includegraphics[width=\linewidth]{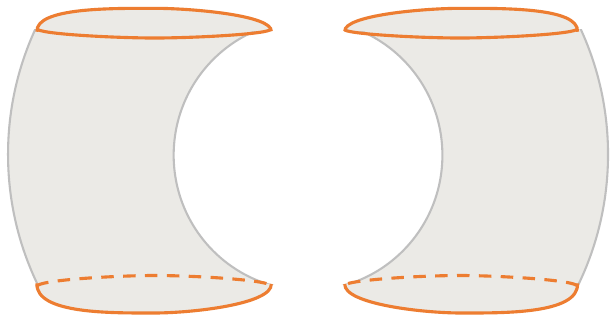}
        \includegraphics[width=\linewidth]{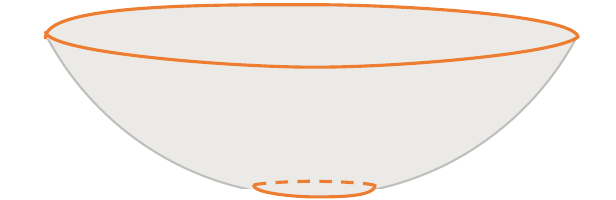}
        \end{minipage}}
    \end{minipage}
    \begin{minipage}[m]{0.14\linewidth}
        \subfloat[\label{fig:conceptual:contours}]{\includegraphics[height=1.7cm]{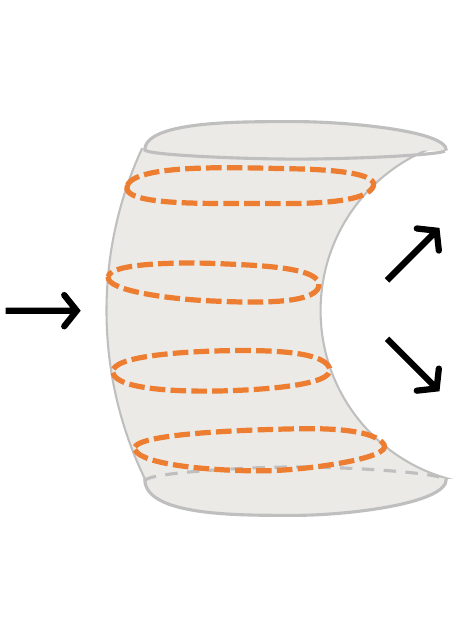}}
    \end{minipage}    
    \begin{minipage}[m]{0.07\linewidth}
        \subfloat[\label{fig:conceptual:boundary}]{\includegraphics[width=\linewidth]{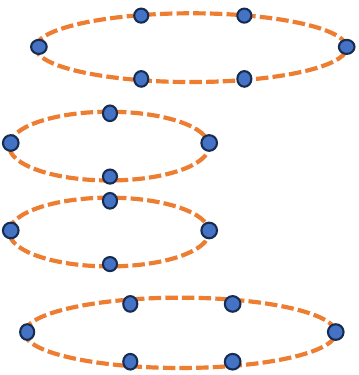}}

        \subfloat[\label{fig:conceptual:interior}]{\includegraphics[width=\linewidth]{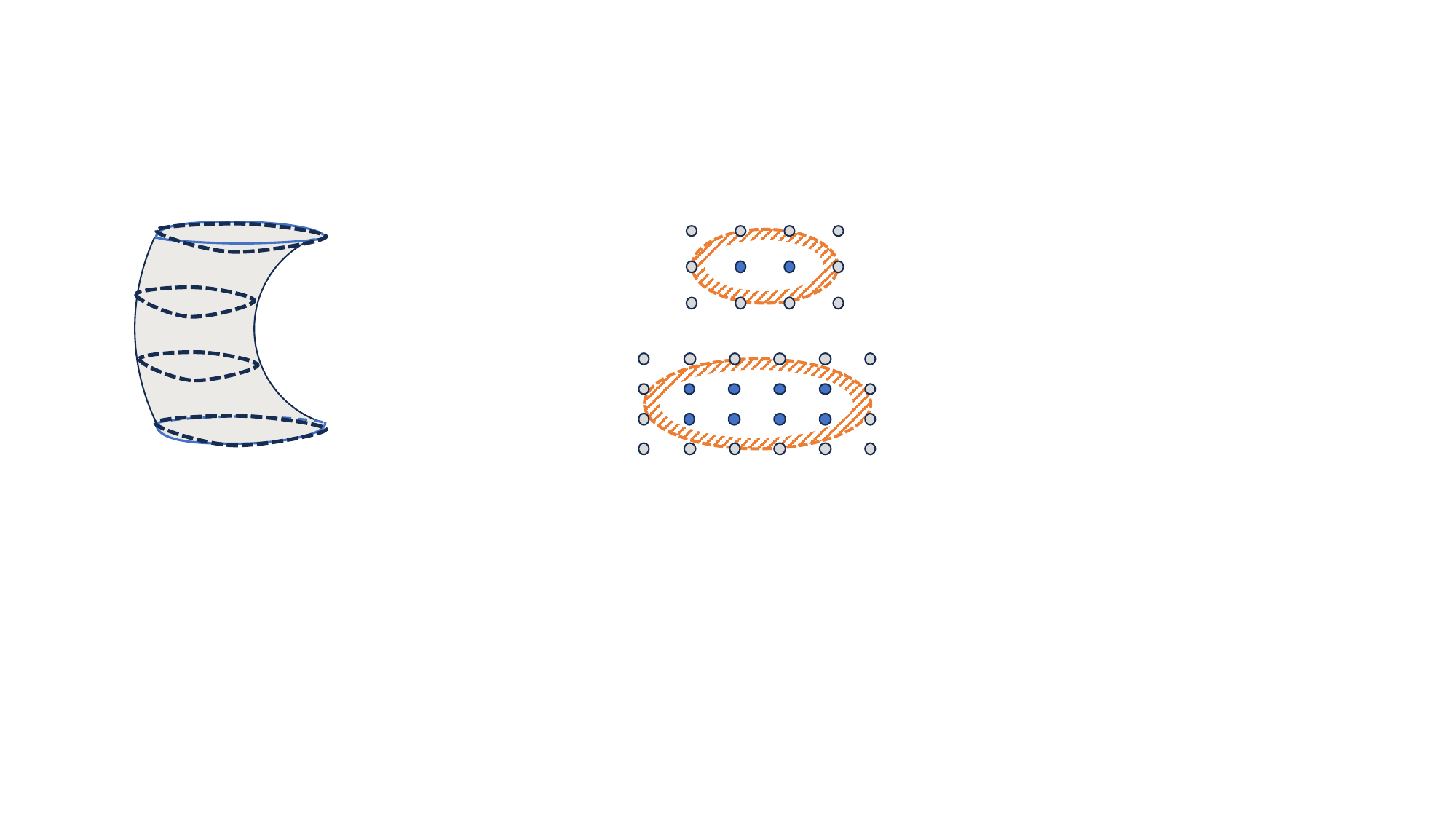}}        
    \end{minipage}
    \begin{minipage}[m]{0.17\linewidth}
        \subfloat[\label{fig:conceptual:path}]{\includegraphics[height=1.7cm]{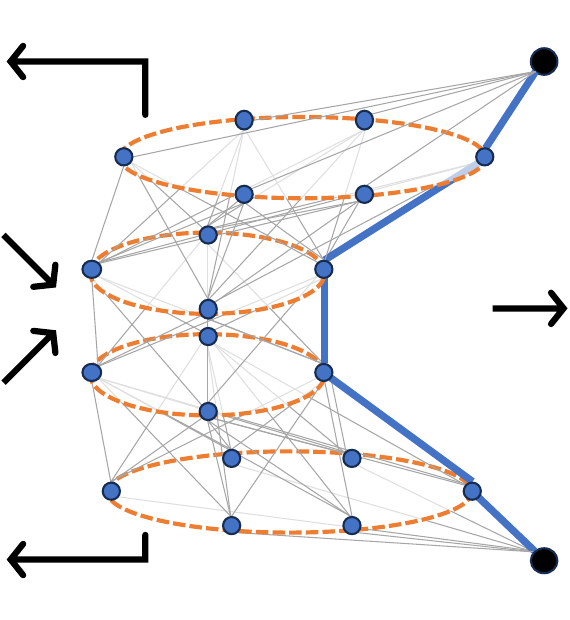}}
    \end{minipage}
    \begin{minipage}[m]{0.19\linewidth}
        \subfloat[\label{fig:conceptual:assembly}]{\includegraphics[height=1.7cm]{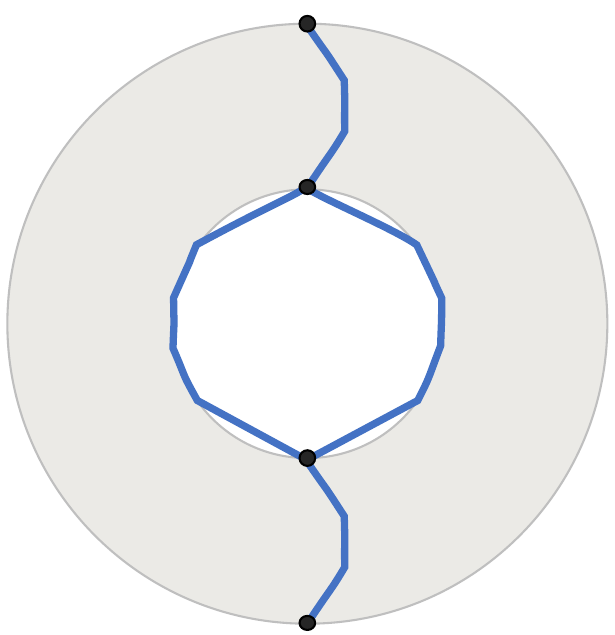}}
    \end{minipage}

    \caption{GASP algorithm description. (a)~A Reeb graph is calculated. While shown on the mesh, the calculation provides only a graph object with critical points as nodes and arcs as edges. (b)~GASP first decomposes the mesh into a set of topological cylinders, one per edge. (c)~A series of isocontours is calculated for each topological cylinder, and candidate points for the arc path are calculated by (d)~the \textit{boundary approach} using the contours themselves or (e)~the \textit{interior approach} by finding points inside the contours. (f)~Once the candidate points are found, graph edges are formed between the critical points and the adjacent contours and between adjacent contours, and the output arc is calculated as the shortest path between the critical points. (g)~Finally, the arcs from all topological cylinders are assembled into a final Reeb graph.}
    \label{fig:conceptual}
\end{figure}

\section{GASP (Gradient-Aware Shortest Path) Reeb Graphs}
\label{sec:method-conceptual}

We target generating Reeb graphs that are better representations of the model and function used to generate them. To do this, we begin with a high-level conceptual description.
The input to our approach is a 2-manifold embedded in 3D space, $\Mgroup$, a Morse function, $f$, and the Reeb graph consisting of a set of critical points, $C$, and Reeb graph edges, $E$, which connect pairs of critical points (i.e., $E_i=(C_j,C_k)$). Our process, shown in \Cref{fig:conceptual}, contains three main steps.

\subsection{Decomposition}

To begin, our approach decomposes the model, $\Mgroup$, into a subset of objects, one per Reeb graph edge, $E_i=(C_j,C_k)$ (see \Cref{fig:conceptual:decomposition}). Without a loss of generality, we assume $f(C_j)<f(C_k)$. The decomposition is done by slicing the model at $f(C_j)+\epsilon$ and $f(C_k)-\epsilon$, and selecting the portion of the model associated with the Reeb graph edge. Here, $\epsilon$ is assumed to be a very small number. One basic property of Reeb graphs is that each portion of the model associated with a Reeb graph edge, $E_i$, will be a \textit{topological cylinder}, $\Cspace_i$~\cite{parallel_reeb}.

\subsection{Reeb Graph Arc in a Topological Cylinder}
\label{sec:concept:arc}

Next, GASP focuses on drawing a Reeb graph arc within a single topological cylinder, $\Cspace_i$, associated with Reeb graph edge, $E_i=(C_j,C_k)$.

\subsubsection{Contours}
The first step of generating the Reeb graph for $\Cspace_i$ involves calculating a series of isocontours on the cylinder. The goal is to create contours with similar spacing across the entire model. The number of isocontours is selected by $n=\lceil (f(C_k)-f(C_j))/S+1 \rceil$, where $S$ is a user-set spacing parameter.
Those $n$ contours are evenly spaced from $f(C_j)+\epsilon$ to $f(C_k)-\epsilon$. \Cref{fig:conceptual:contours} shows an example.

\subsubsection{Candidate Points}
\label{sec:candidate-points}
The next step involves generating a series of candidate points, $p$, using the contours from the prior step, which will be used to identify the Reeb graph path. We present two variations. The first variation, called the \textit{boundary approach} (see \Cref{fig:conceptual:boundary}), selects a series of candidate points along the contour boundary itself. As will be noted in the implementation, we directly use the points produced by isocontouring. The second variation, called the \textit{interior approach} (see \Cref{fig:conceptual:interior}), generates a series of candidate points in a grid-like pattern inside the contour and some user-set distance from the boundary (denoted by the \textcolor{orange}{\textit{orange}} stripes). 

\subsubsection{Path Graph and Path Extraction}
The next step of the process extracts a path graph, which is used to find the arc path. This is done by forming a graph between the candidate points found in the prior step. Two types of edges are created. First, between two adjacent contours, path graph edges are formed between all pairs of candidate points in neighboring levels. Second, the critical points are connected to all candidate points in the adjacent contour. All edges are weighted by the Euclidean distance between their endpoints. The resulting graph will look similar to \Cref{fig:conceptual:path}. 
The final step of the process extracts the shortest path between the two critical points using Dijkstra's algorithm, as denoted by the \textcolor{blue}{\textit{blue}} path in \Cref{fig:conceptual:path}.

\subsection{Final Assembly}

The final Reeb graph is assembled by combining the individual arcs formed per topological cylinder, sharing common points only at the critical points (see \Cref{fig:conceptual:assembly}).

\paragraph{Summary}
This approach has several desirable properties. First, decomposing the mesh into individual cylinders simplifies the problem significantly. Second, using the isocontours to derive the path of the arc results in an arc that is generally inside or near the surface of the manifold, with only minor deviations possible in non-convex regions. Third, selecting the shortest path along adjacent isocontours results in shorter arcs that also travel in the same direction as the function. The second and third properties, in particular, directly address the three Reeb graph properties we highlighted in \cref{sec:intro}.

\begin{figure}[!t]
    \centering
    \subfloat[Original Models\label{fig:decomp:orig}]{
    \includegraphics[height=1.5cm]{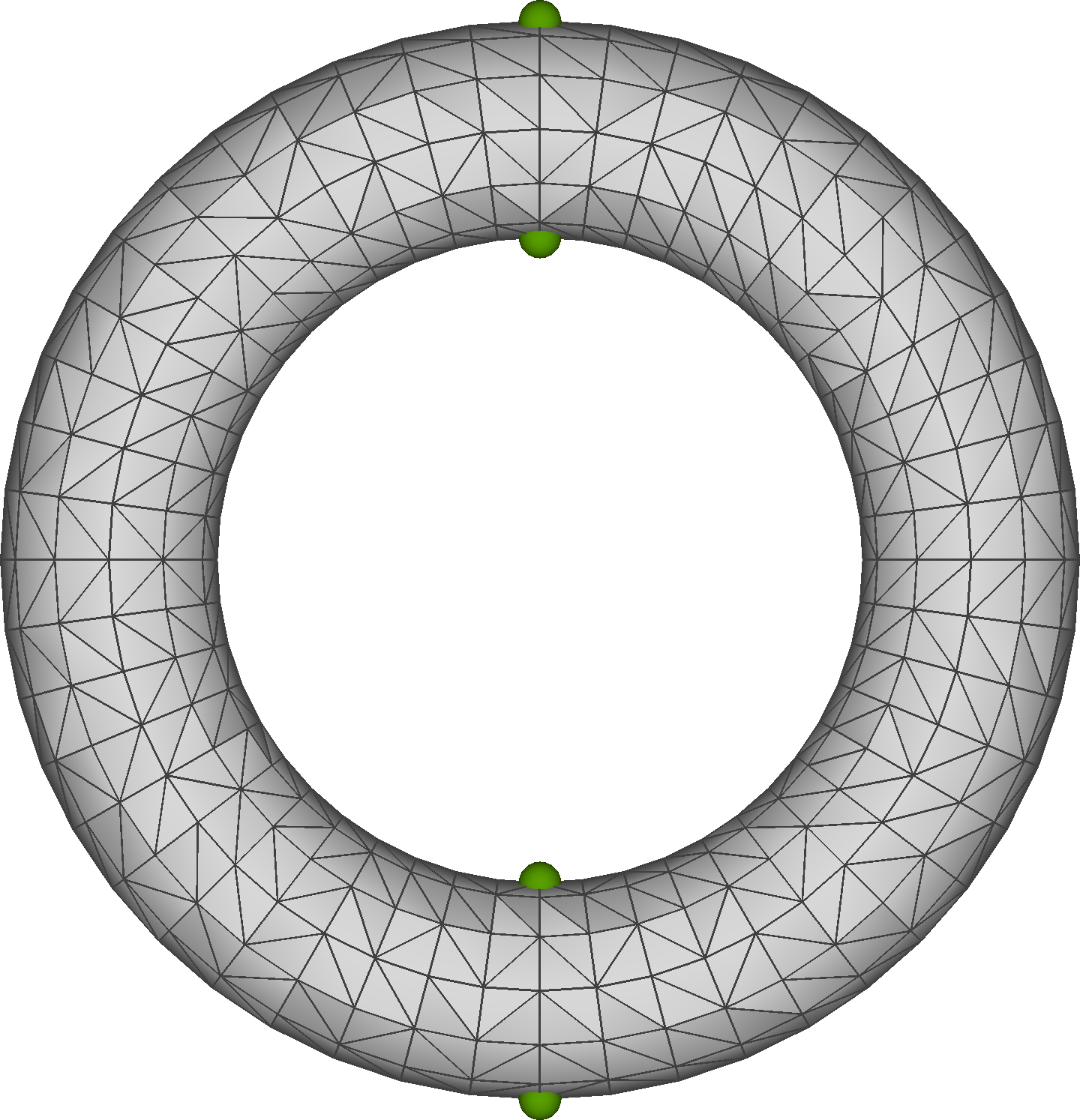} 
    \includegraphics[height=1.5cm]{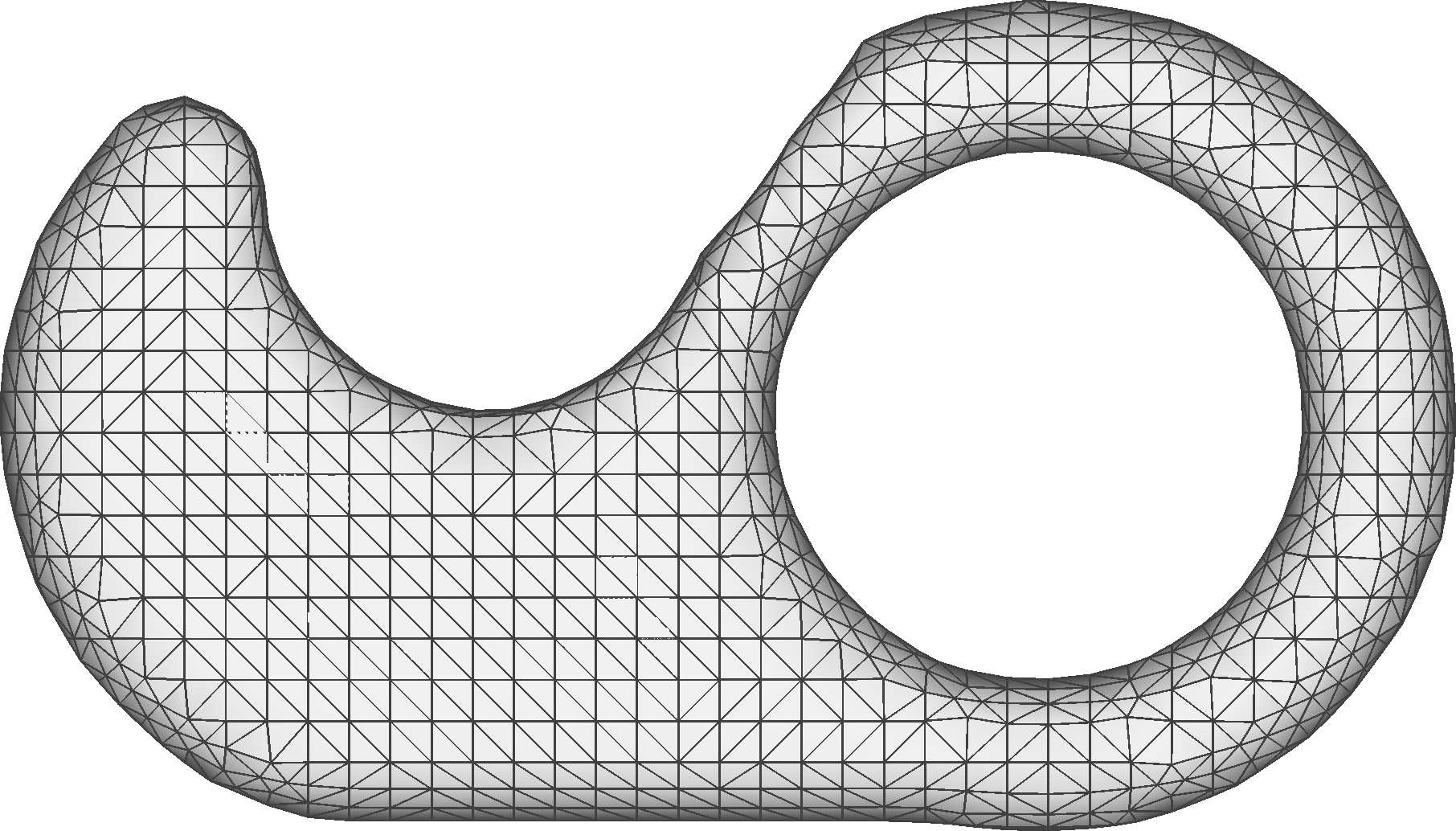}}
    \hfill
    \subfloat[Rough Cut\label{fig:decomp:rough}]{\includegraphics[height=1.5cm]{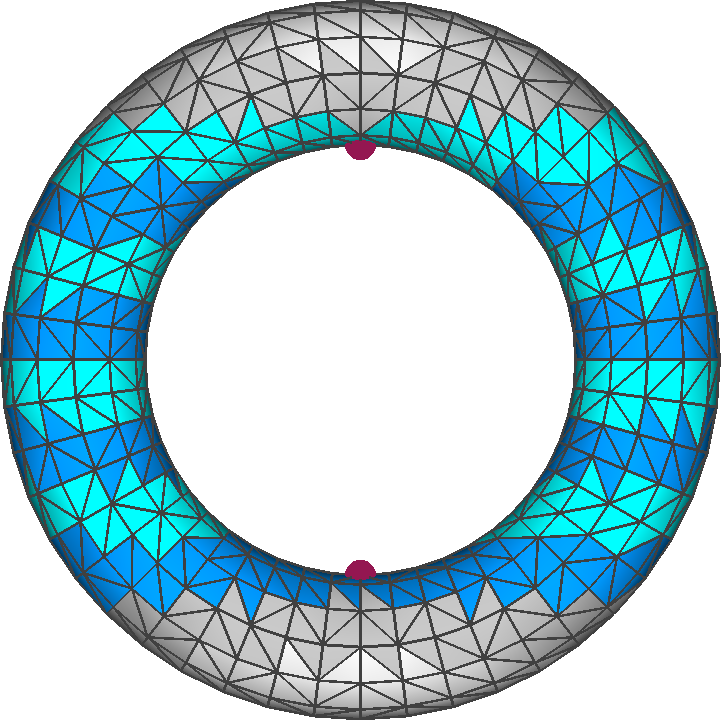} 
    \includegraphics[height=1.5cm]{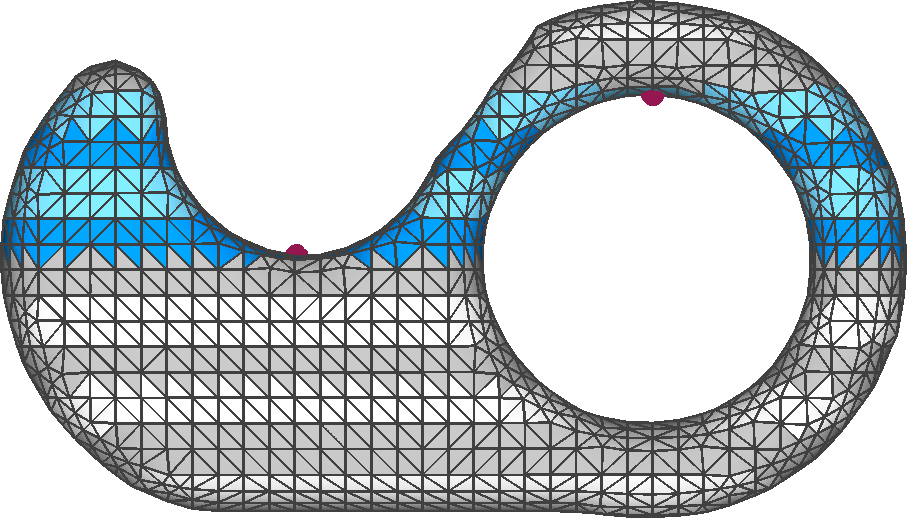}}
    
    \subfloat[Source Cut\label{fig:decomp:src}]{\includegraphics[height=1.5cm]{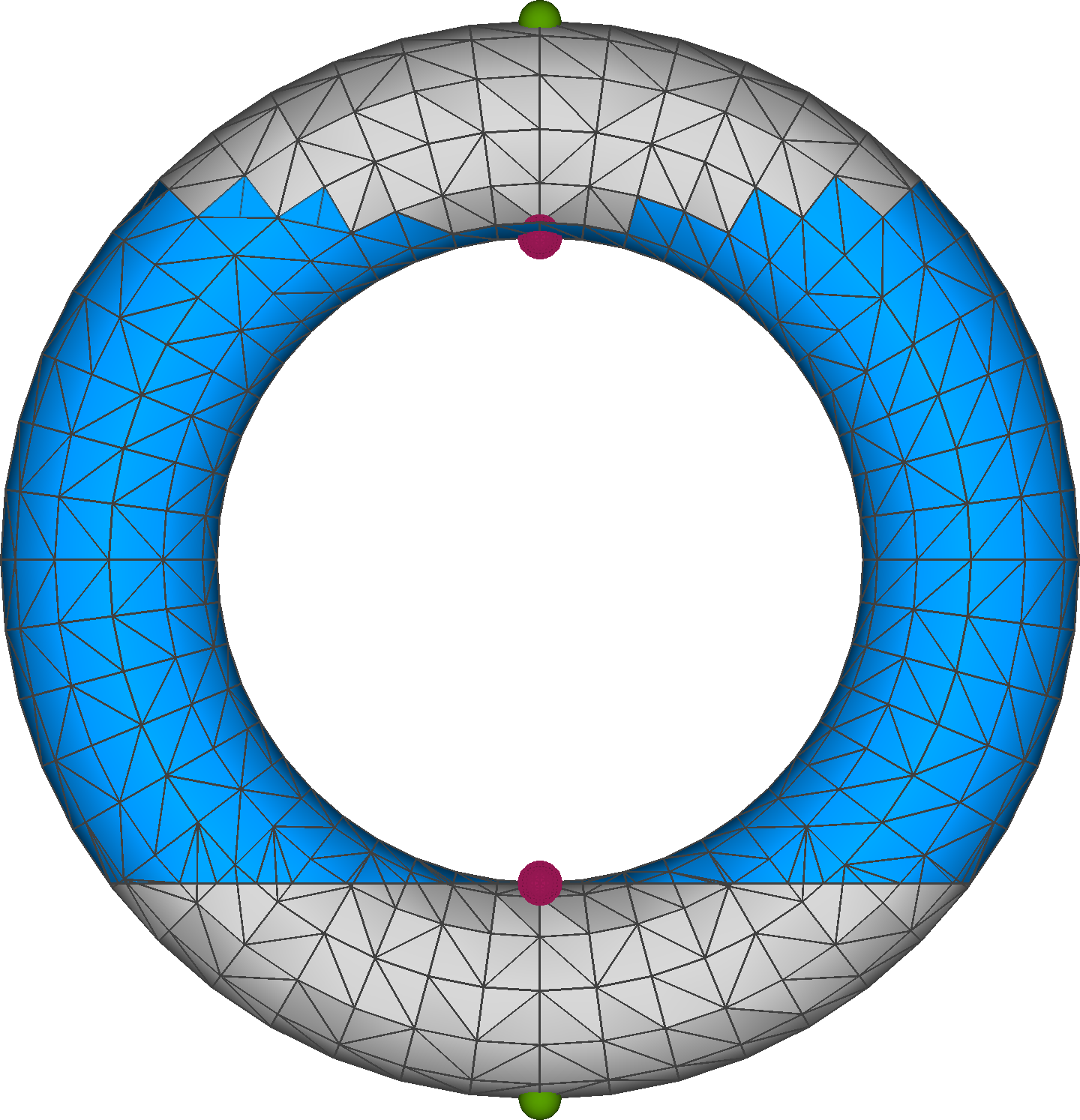} 
    \includegraphics[height=1.5cm]{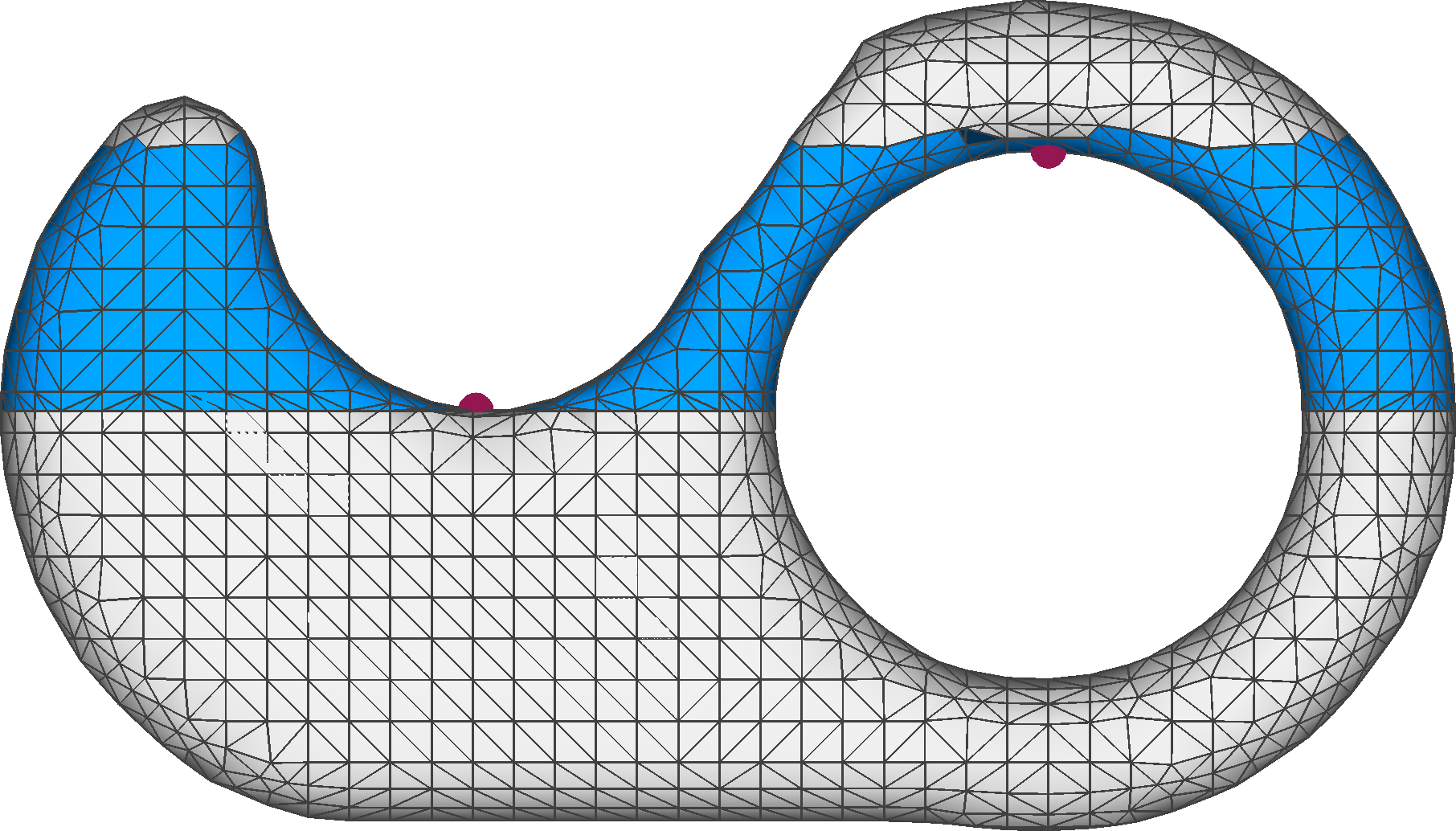}}
    \hfill
    \subfloat[Destination Cut\label{fig:decomp:dst}]{\includegraphics[height=1.5cm]{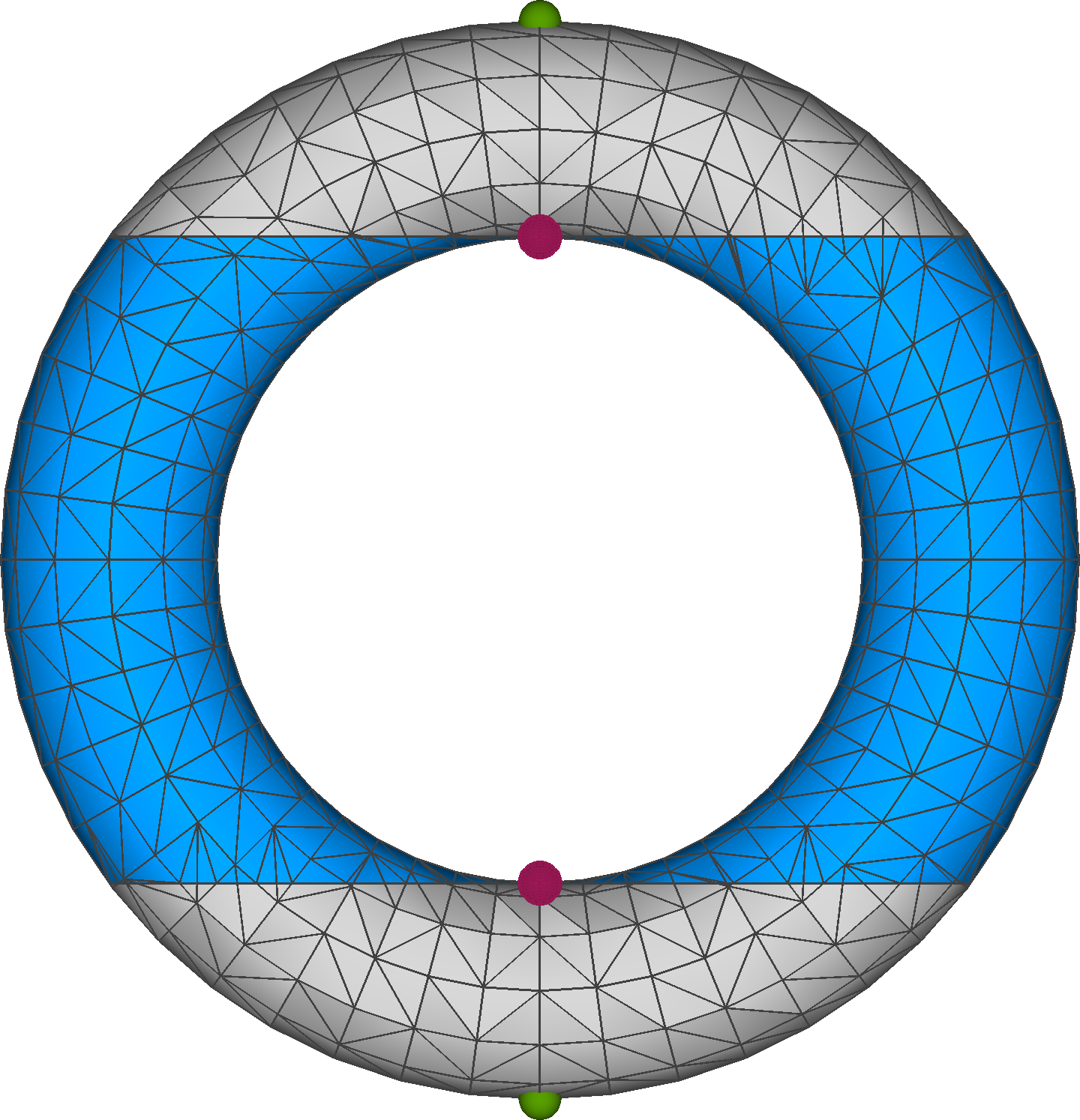} 
    \includegraphics[height=1.5cm]{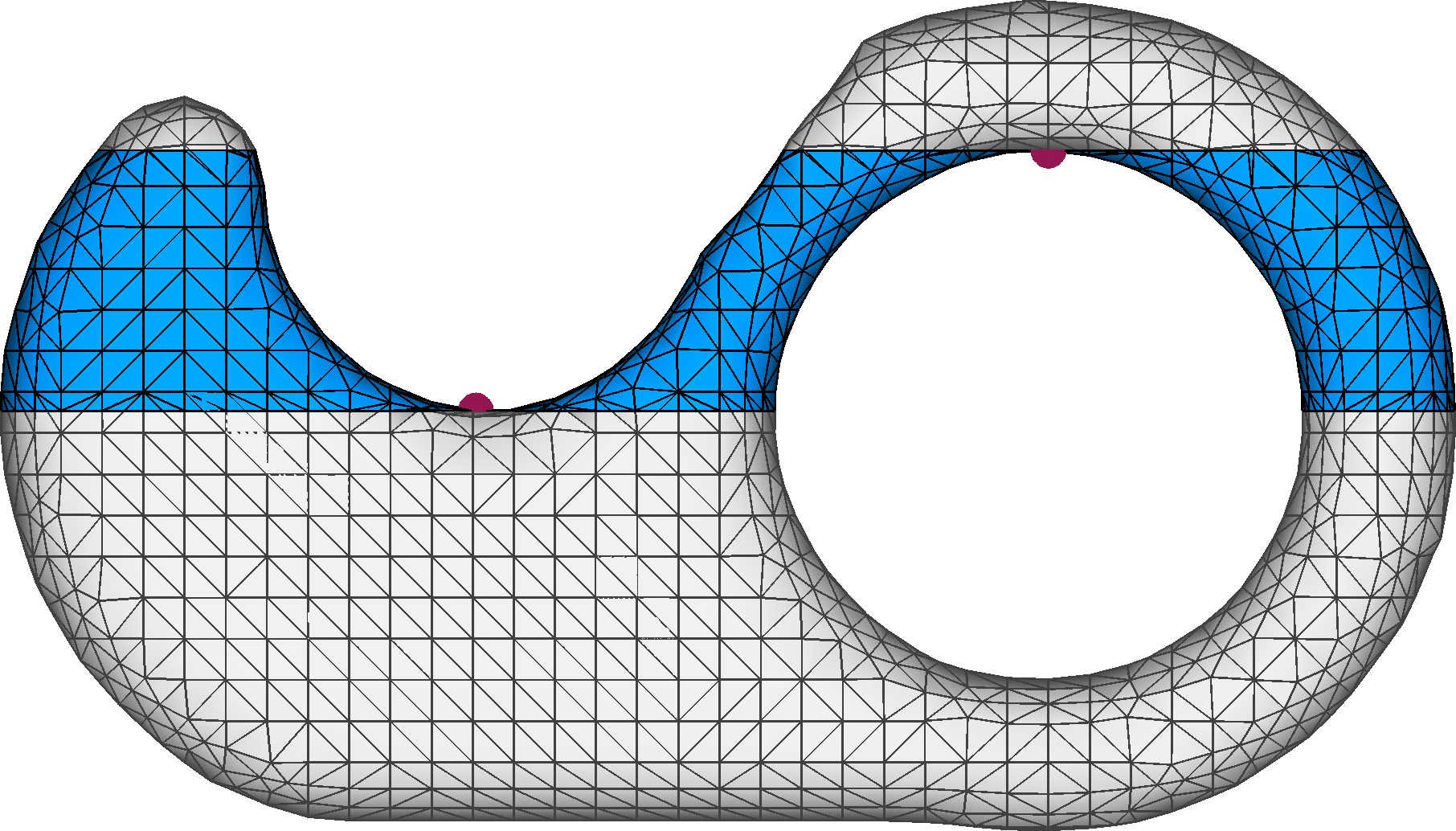}}

    { \color{gray}\rule{0.8\linewidth}{0.4pt} }

    \vspace{4pt}
    \begin{minipage}[b]{0.975\linewidth}
        \centering
        \subfloat[Does Not Contain Critical Point\label{fig:discarded-regions:connect}]{\includegraphics[height=1.7cm]{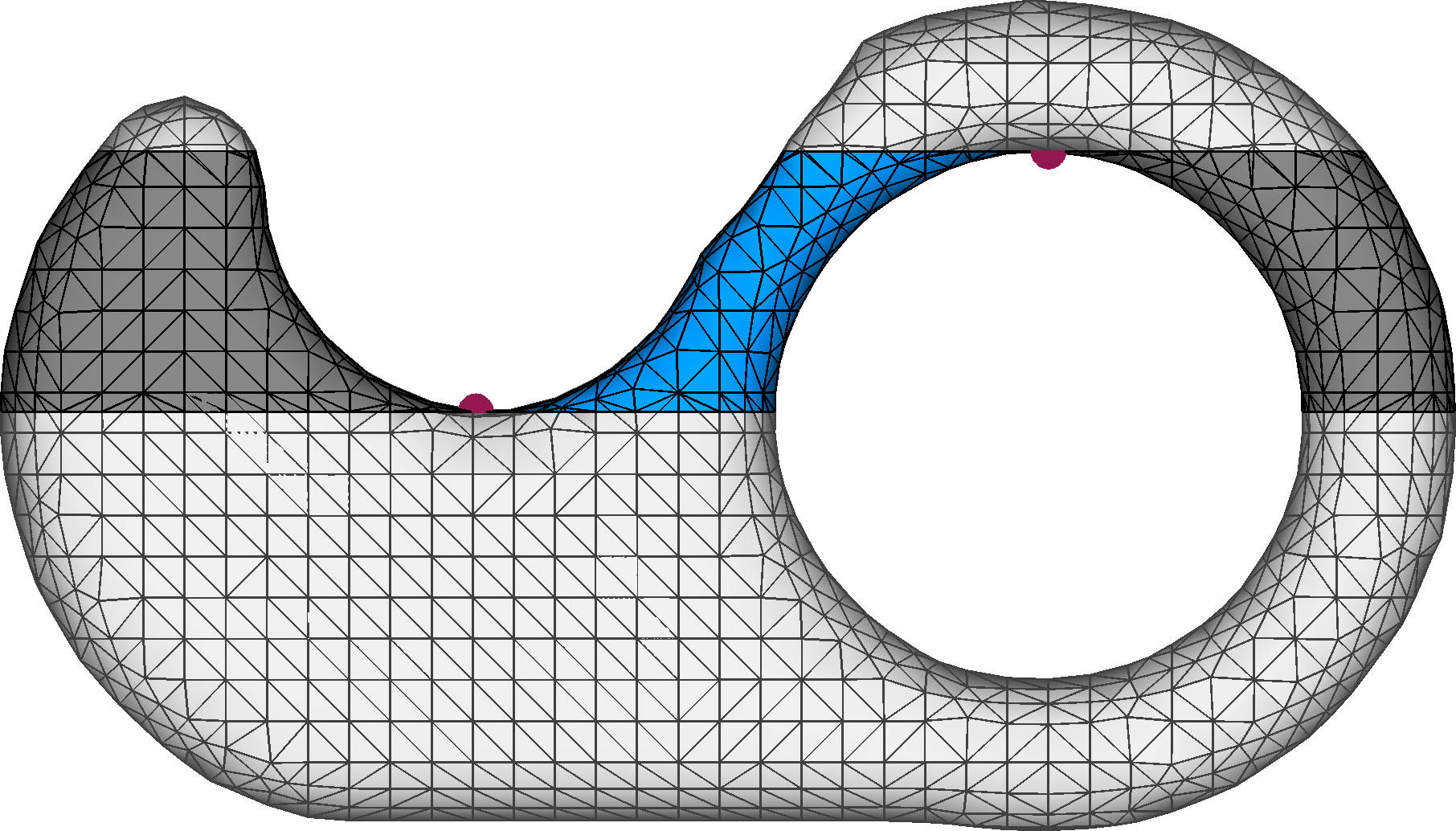}}
        \hfill
        \subfloat[Contains Extra Critical Point\label{fig:discarded-regions:cp}]{\includegraphics[height=1.7cm]{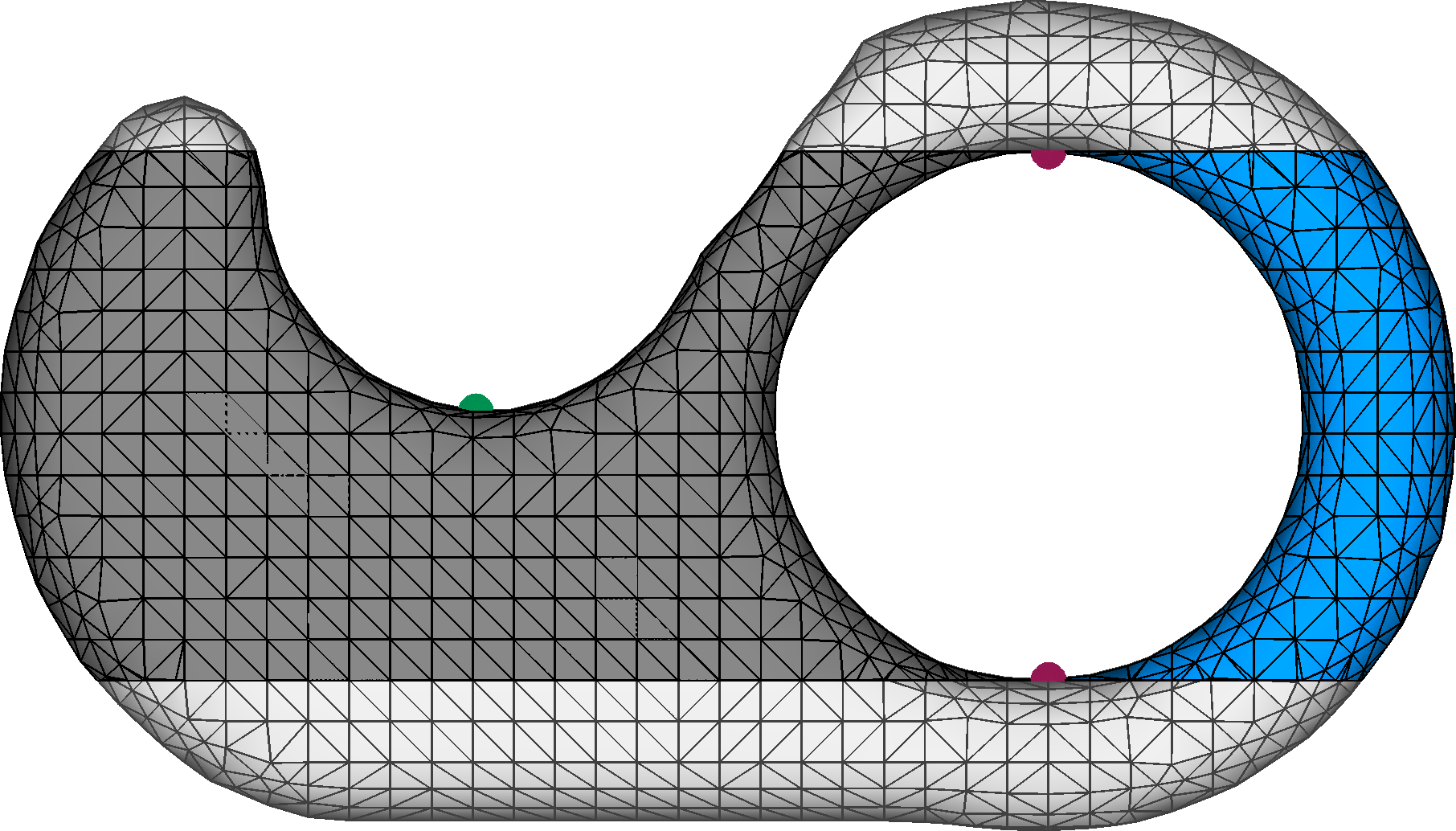}}
        \hfill
        \subfloat[Duplicity\label{fig:TwoPath}]{\includegraphics[height=1.71cm]{figs/implementation/torus/torus_dst_cut}}
    \end{minipage}
    \begin{minipage}[b]{1pt}
        \tiny
        \hspace{-29pt}$C_k$\hspace{-66pt}$C_k$\hspace{-105pt}$C_k$
        \vspace{-1pt}

        \hspace{-124pt}$C_l$\hspace{-104pt}$C_j$
        \vspace{9pt}
        
        \hspace{-29pt}$C_j$\hspace{-66pt}$C_j$ 
        \vspace{11pt}
    \end{minipage}
    \begin{minipage}[b]{1pt}
        \small
        \hspace{-57pt}\textcolor{blue}{$\overline{\Cspace}_1$}\hspace{40pt}\textcolor{blue}{$\overline{\Cspace}_2$}
        \hspace{-136pt}\textcolor{gray}{$\overline{\Cspace}_1$}\hspace{56pt}\textcolor{blue}{$\overline{\Cspace}_2$}
        \hspace{-171pt}\textcolor{gray}{$\overline{\Cspace}_1$}\hspace{9pt}\textcolor{blue}{$\overline{\Cspace}_2$}\hspace{39pt}\textcolor{gray}{$\overline{\Cspace}_3$}
        \vspace{35pt}

    \end{minipage}
    
    \caption{(a)~Torus (left) and modified torus (right) are (b)~binned by function value and have a rough cut in \textcolor{cyan}{\textit{cyan}} and \textcolor{blue}{\textit{blue}} performed between the critical points in \textcolor{purple}{\textit{red}}. Then, the triangles in \textcolor{blue}{\textit{blue}} have the (c)~source cut applied, (d)~followed by the destination cut.
    (e-g)~Examples of different candidate connected component validation criteria after a cut, where valid cylinders are \textcolor{blue}{\textit{blue}}, and invalid ones are \textcolor{gray}{\textit{gray}}.
    (e)~After a cut, a valid connected component will connect the two critical points ($c_j$ and $c_k$), e.g., \textcolor{blue}{$\overline{\Cspace}_2$}, whereas \textcolor{gray}{$\overline{\Cspace}_1$} and \textcolor{gray}{$\overline{\Cspace}_3$} connect to only one critical point, $c_j$ and $c_k$, respectively. (f)~After a cut, a valid connected component will not contain any other critical points, e.g., \textcolor{blue}{$\overline{\Cspace}_2$}, where \textcolor{gray}{$\overline{\Cspace}_1$} contains the extra critical point $c_l$. (g)~Special care must be taken for Reeb graph edges with duplicity as multiple valid topological cylinders will be identified (e.g., both \textcolor{blue}{$\overline{\Cspace}_1$} and \textcolor{blue}{$\overline{\Cspace}_2$}) and must be matched to those edges. 
    }
    \label{fig:decomp}
\end{figure}

\section{Implementation}

In this study, we utilized triangular meshes as our input, $\Mgroup$. To simplify the description, we show a principal direction as the height function, $f$, on the torus and modified torus models found in \Cref{fig:decomp}-\ref{fig:torus_final}. However, as we will describe, our algorithm is independent of this constraint, and our evaluation uses height and geodesic functions. To produce the Reeb graph, we utilized Recon~\cite{doraiswamy2012computing}. 

\subsection{Decomposition of the Triangle Mesh}
\label{sec:imp:decomp}

Decomposition is performed on a Reeb graph edge-by-edge basis.

\subsubsection{Splitting Between Critical Points}
Assume we have a Reeb graph edge, $E_i$, with critical points $C_j$ and $C_k$, with function values, $f(C_j)$ and $f(C_k)$, where $f(C_j)<f(C_k)$. 
We will apply two cut operations on the mesh at  $f(C_j)+\epsilon$, referred to as \textbf{\textit{source cut}}, and at $f(C_k)-\epsilon$, referred to as \textbf{\textit{destination cut}}.

Initially, to speed the process, a rough cut is performed to remove triangles outside the processing area. To do this, triangles are preprocessed into bins by their function value. Bins that would not contain any triangles between the critical points are excluded from the cutting process. \Cref{fig:decomp:rough} shows one such rough cut. Here, the models were divided into 12 bins\footnote{By default and in experiments, we use 20 bins, but the value is user-settable.}. The bins in \textcolor{cyan}{\textit{cyan}} and \textcolor{blue}{\textit{blue}} are selected for processing, while the \textcolor{lightgray}{\textit{white}} and \textcolor{gray}{\textit{gray}} bins are ignored.

\begin{figure}[!b]
    \centering

    \includegraphics[width=0.975\linewidth]{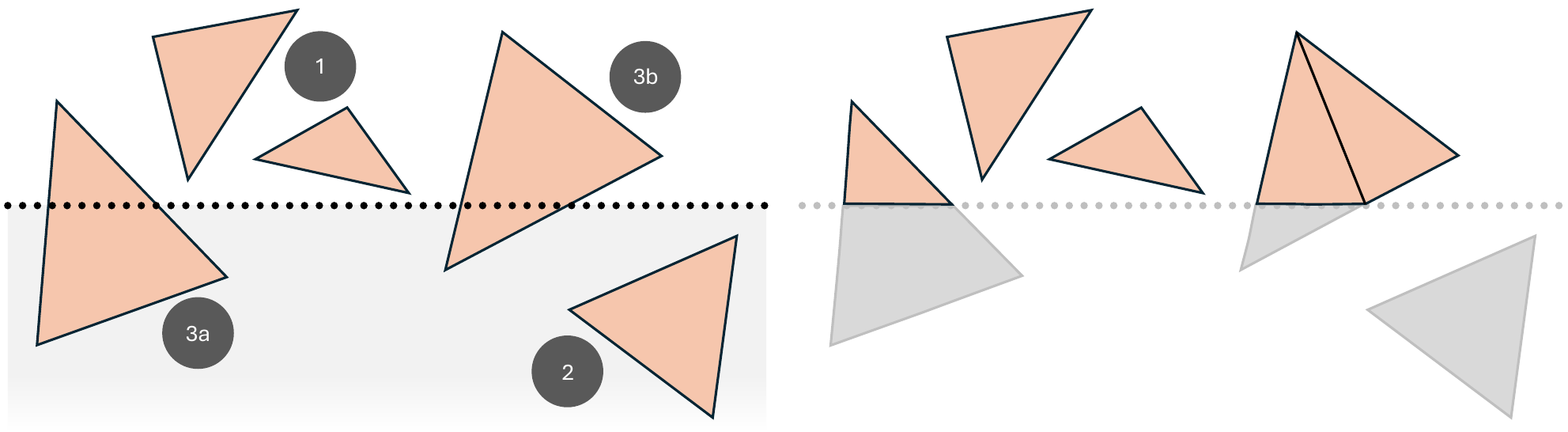}
    
    \vspace{-10pt}
    \subfloat[Before Source Cut\label{fig:cut:src_before}]{\hspace{0.475\linewidth}}\hfill
    \subfloat[After Source Cut\label{fig:cut:src_after}]{\hspace{0.475\linewidth}}

    \vspace{4pt}
    \hfill\includegraphics[width=0.975\linewidth]{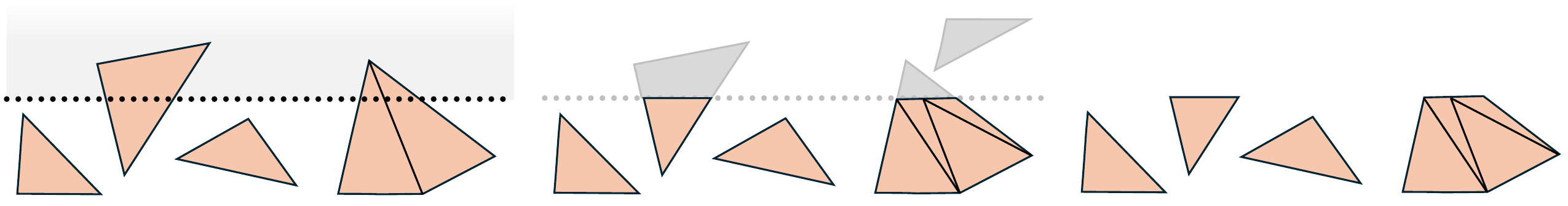}
    
    \vspace{-10pt}
    \subfloat[Before Destination Cut\label{fig:cut:dst_before}]{\hspace{0.34\linewidth}}
    \hspace{1pt}
    \subfloat[After Destination Cut\label{fig:cut:dst_after}]{\hspace{0.32\linewidth}}
    \subfloat[Final Triangles\label{fig:cut:final}]{\hspace{0.32\linewidth}}
    
    \caption{Examples of triangles (a/c)~before and (b/d)~after (a/b)~source and (c/d)~destination cuts. The four types of source cuts include: (1)~completely above, (2)~completely below, (3a)~partially above keeping a triangle, and (3b)~partially above keeping a quadrilateral. \textcolor{Peru}{\textit{Beige}} triangles are retained, while \textcolor{gray}{\textit{gray}} ones are discarded.}
    \label{fig:cut}
\end{figure}

The source and destination cut operations are then performed per triangle.  
For a source cut, triangles fall into one of three cases: (1)~completely above the cut value, which is retained as is (see \Cref{fig:cut:src_before}/\ref{fig:cut:src_after} type~1); (2)~completely below the cut value, which is excluded from further computation (see \Cref{fig:cut:src_before}/\ref{fig:cut:src_after} type~2); and (3)~crossing the cut value, in which case triangles are cut along that value. The portion above the cut line is retained in the form of a smaller triangle (see \Cref{fig:cut:src_before}/\ref{fig:cut:src_after} type~3a) or a quadrilateral, which is subdivided into two triangles (see \Cref{fig:cut:src_before}/\ref{fig:cut:src_after} type~3b). 
A similar process is repeated for the destination cut, shown in \Cref{fig:cut:dst_before}/\ref{fig:cut:dst_after}. \Cref{fig:decomp:src} and \Cref{fig:decomp:dst} show example source and destination cuts on models, respectively.

\subsubsection{Isolating the Topological Cylinder for a Reeb Graph Edge}
After the cut operations, one or more components will be extracted. To isolate the cylinder associated with the $E_i$, from the extracted component(s), the triangles are separated into candidate cylinders by calculating connected components. Candidate cylinders are discarded if: (1)~there is no path between $C_j$ and $C_k$ using the candidate cylinder triangles (see \Cref{fig:discarded-regions:connect}), or (2)~there is another critical point contained within the set of triangles from the candidate cylinder (see \Cref{fig:discarded-regions:cp}). If a candidate cylinder passes both tests, it is associated with the Reeb graph edge connecting the critical points. Special consideration needs to be made to simultaneously process Reeb graph edges that have duplicity (i.e., edges with the same critical points), such as in the torus (e.g., \Cref{fig:TwoPath}). This way, each of the multiple valid cylinders can be assigned to a unique Reeb graph edge.

\subsection{Reeb Graph Arc in a Topological Cylinder Mesh}
Once a topological cylinder is isolated, the Reeb graph arc is formed.

\subsubsection{Calculating Contours}
We use the marching triangles technique~\cite{hilton1996marching} to generate the contours on the object. 
The calculated number and spacing of contours follows the description in \Cref{sec:concept:arc}.

\subsubsection{Calculating Candidate Points}
\label{impl:candidate:points}
\begin{figure}[!t]
    \centering
    \begin{minipage}[b]{0.2\linewidth}
        \subfloat[3D Contour\label{fig:interiorMethod:before}]{\includegraphics[trim=630pt 100pt 300pt 150pt, clip, width=0.975\linewidth]{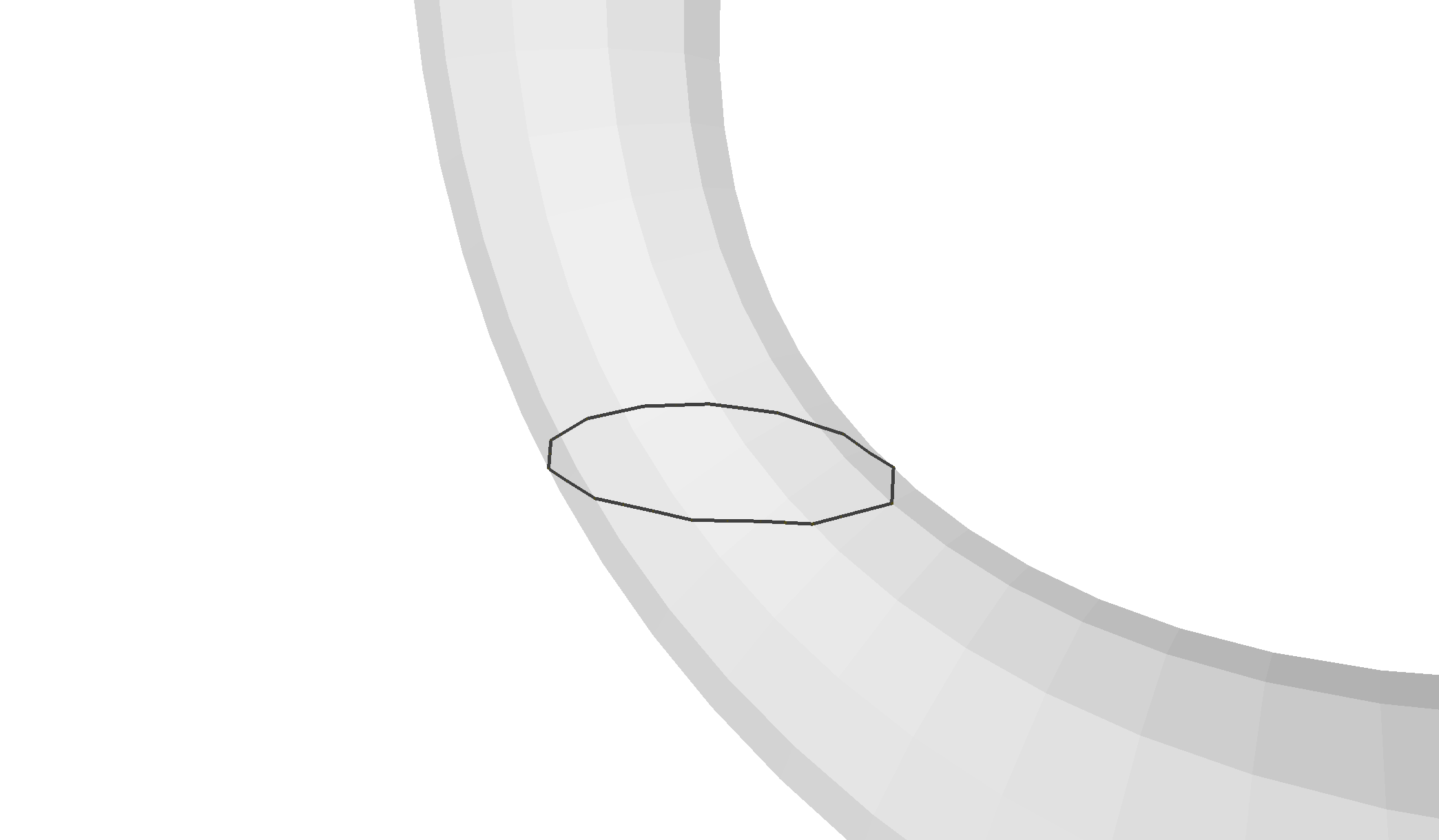}}
        
        \subfloat[2D Contour \& Points\label{fig:interiorMethod:2d}]{\hspace{5pt}\includegraphics[trim=600pt 290pt 600pt 290pt, clip, width=0.975\linewidth]{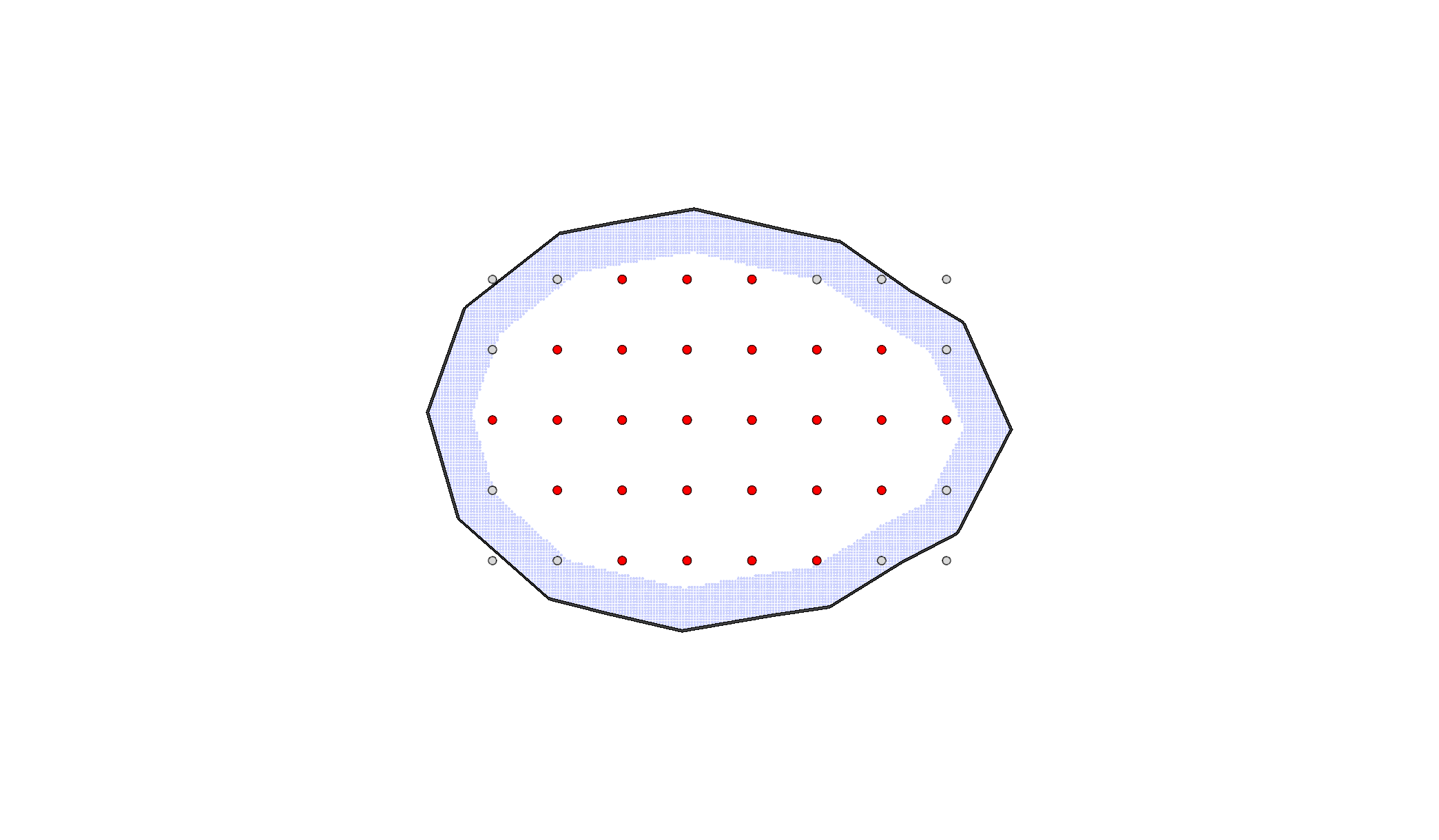}\hspace{5pt}}
        
        \subfloat[3D Contour \& Points\label{fig:interiorMethod:3d}]{\hspace{5pt}\includegraphics[trim=630pt 100pt 300pt 150pt, clip, width=0.975\linewidth]{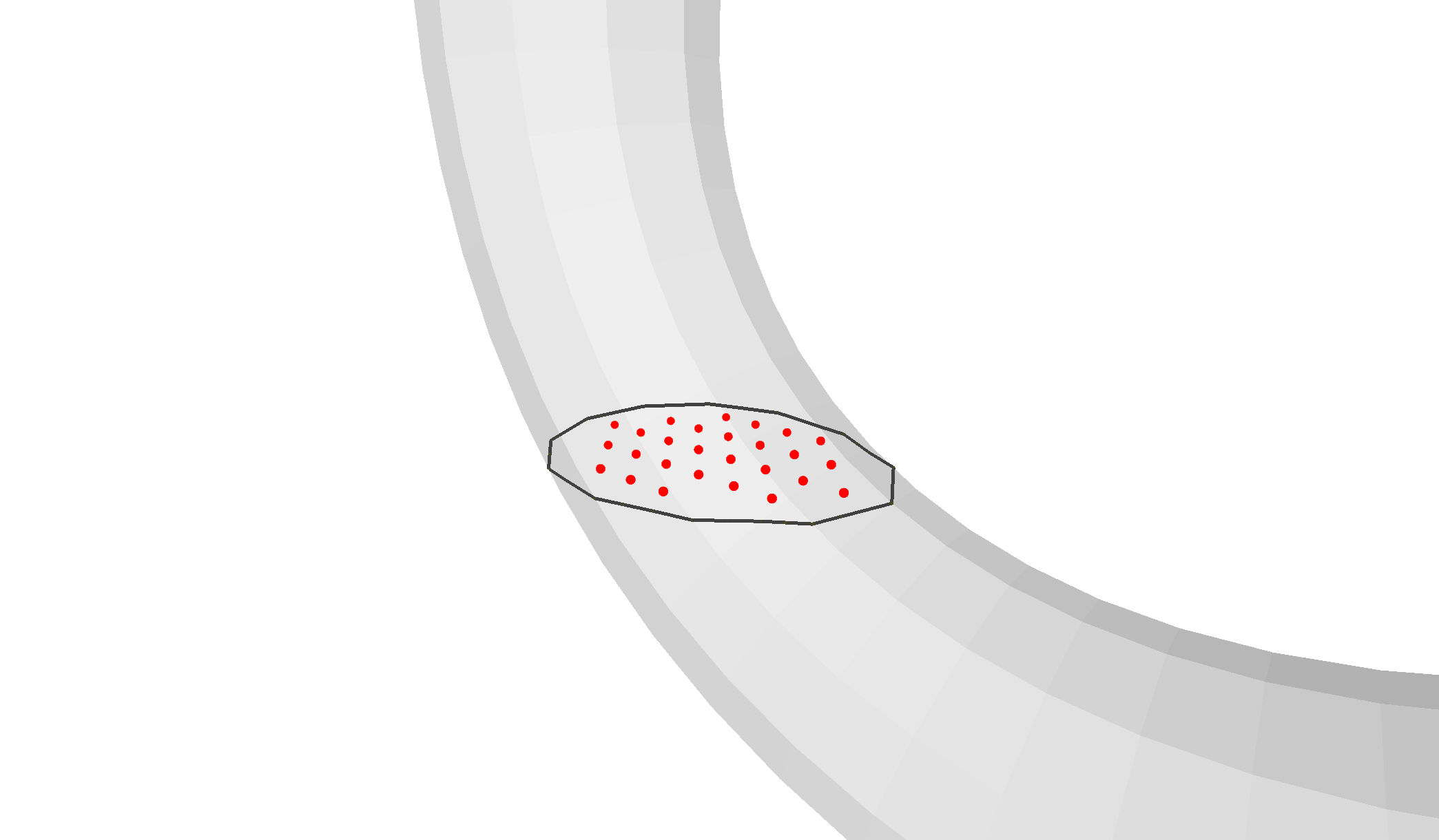}\hspace{5pt}}    
    \end{minipage}
    \hfill
    \begin{minipage}[b]{0.75\linewidth}
        \hfill
        \rotatebox{90}{\hspace{7pt} \tiny \textsf{Boundary Approach}}
        \subfloat[First Iteration\label{fig:contours:boundary}]{\includegraphics[width=0.31\linewidth]{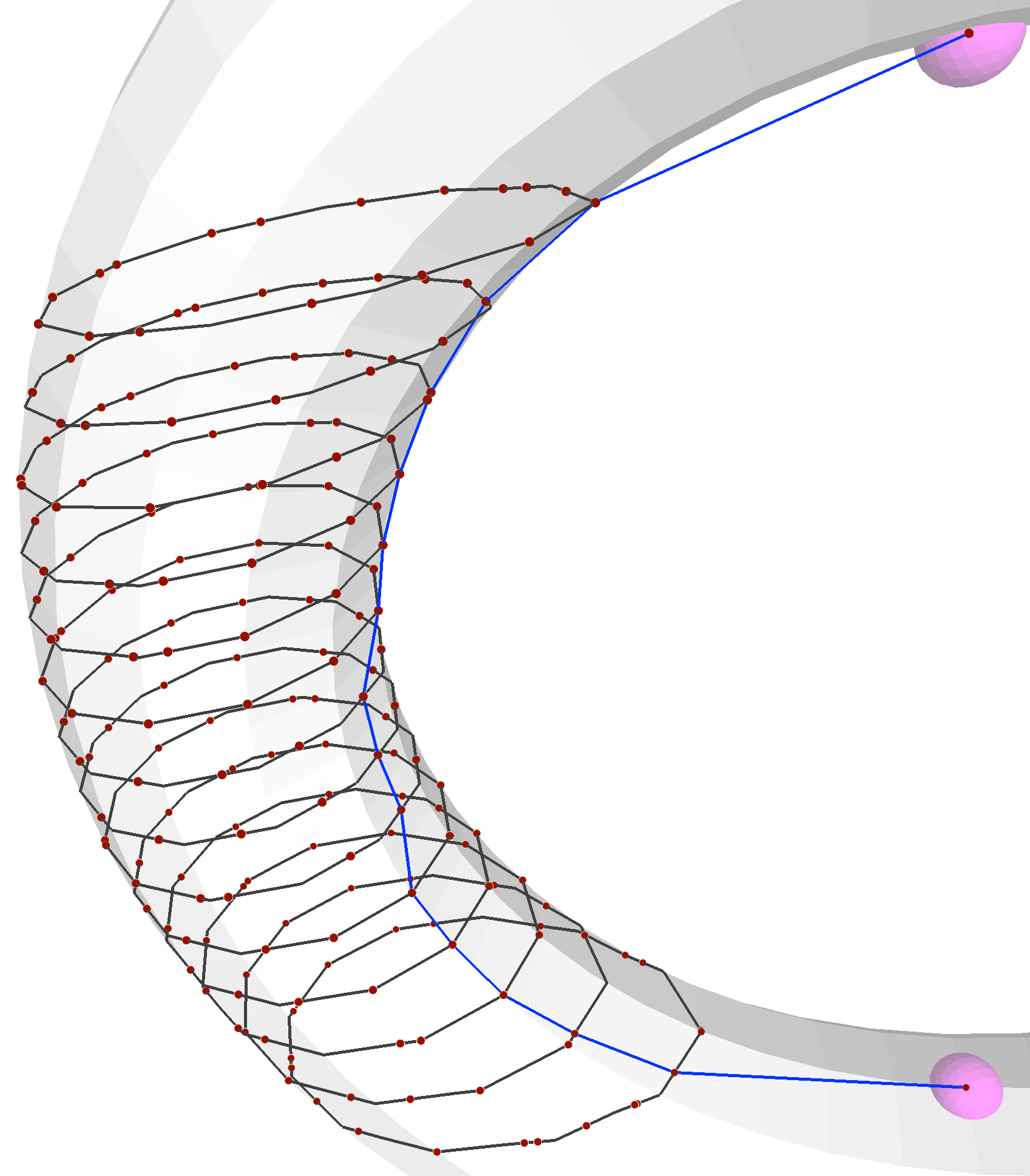}}
        \subfloat[Second Iteration\label{fig:contours:boundary:2nd}]{\includegraphics[width=0.31\linewidth]{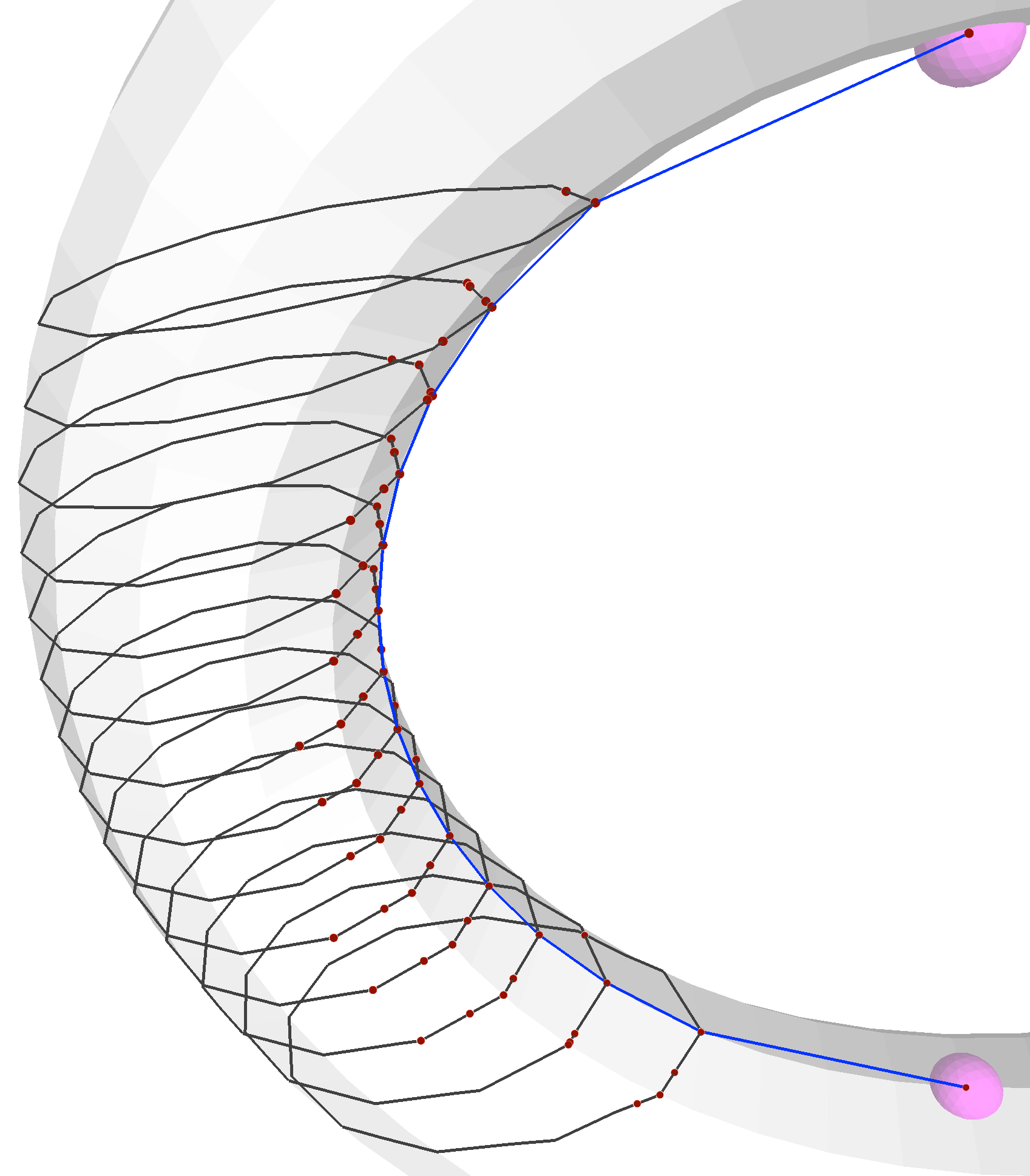}}
        \subfloat[Third Iteration\label{fig:contours:boundary:3rd}]{\includegraphics[width=0.31\linewidth]{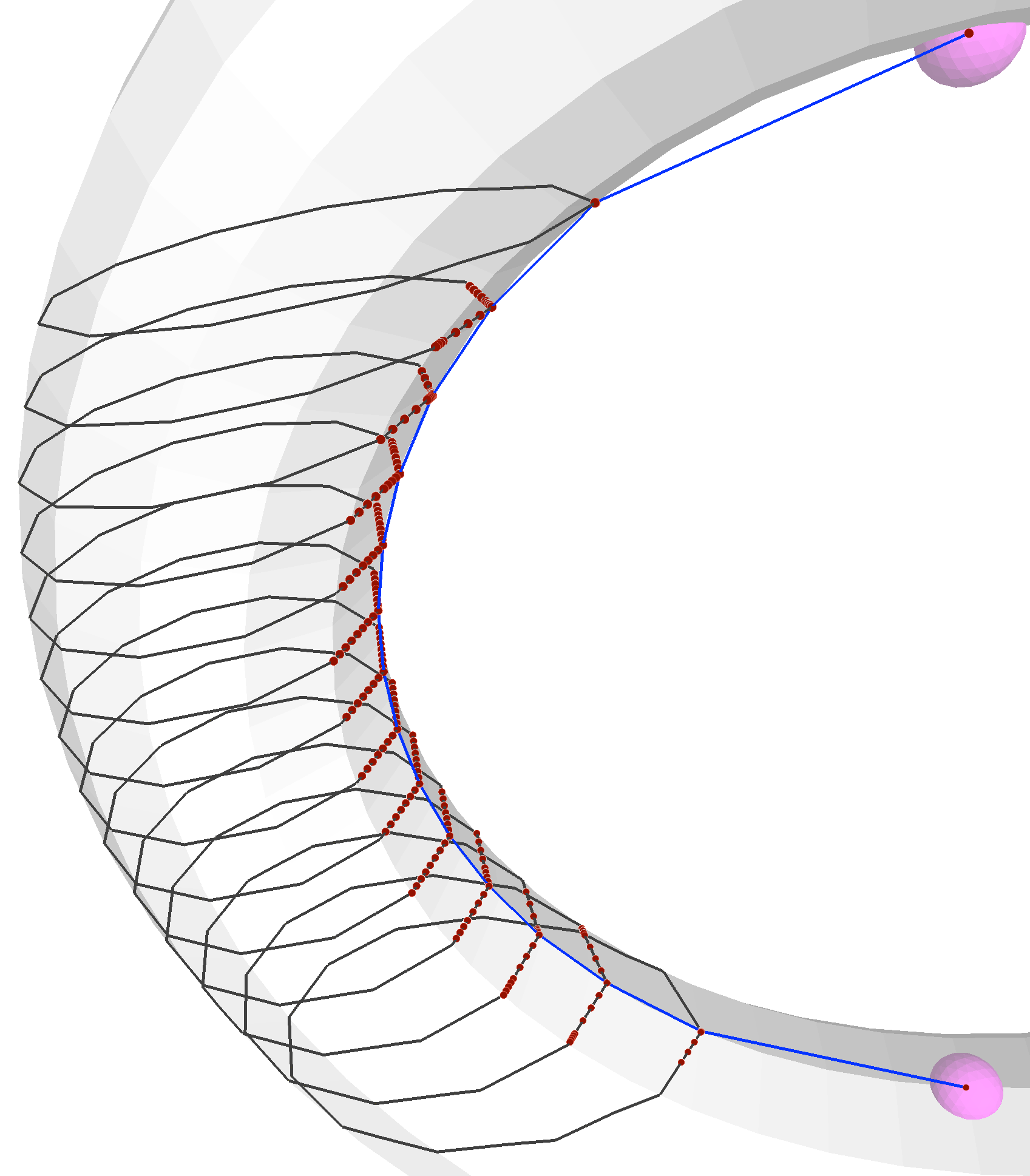}}

        \hfill
        \rotatebox{90}{\hspace{7pt} \tiny \textsf{Interior Approach}}
        \subfloat[First Iteration\label{fig:contours:interior}]{\includegraphics[width=0.31\linewidth]{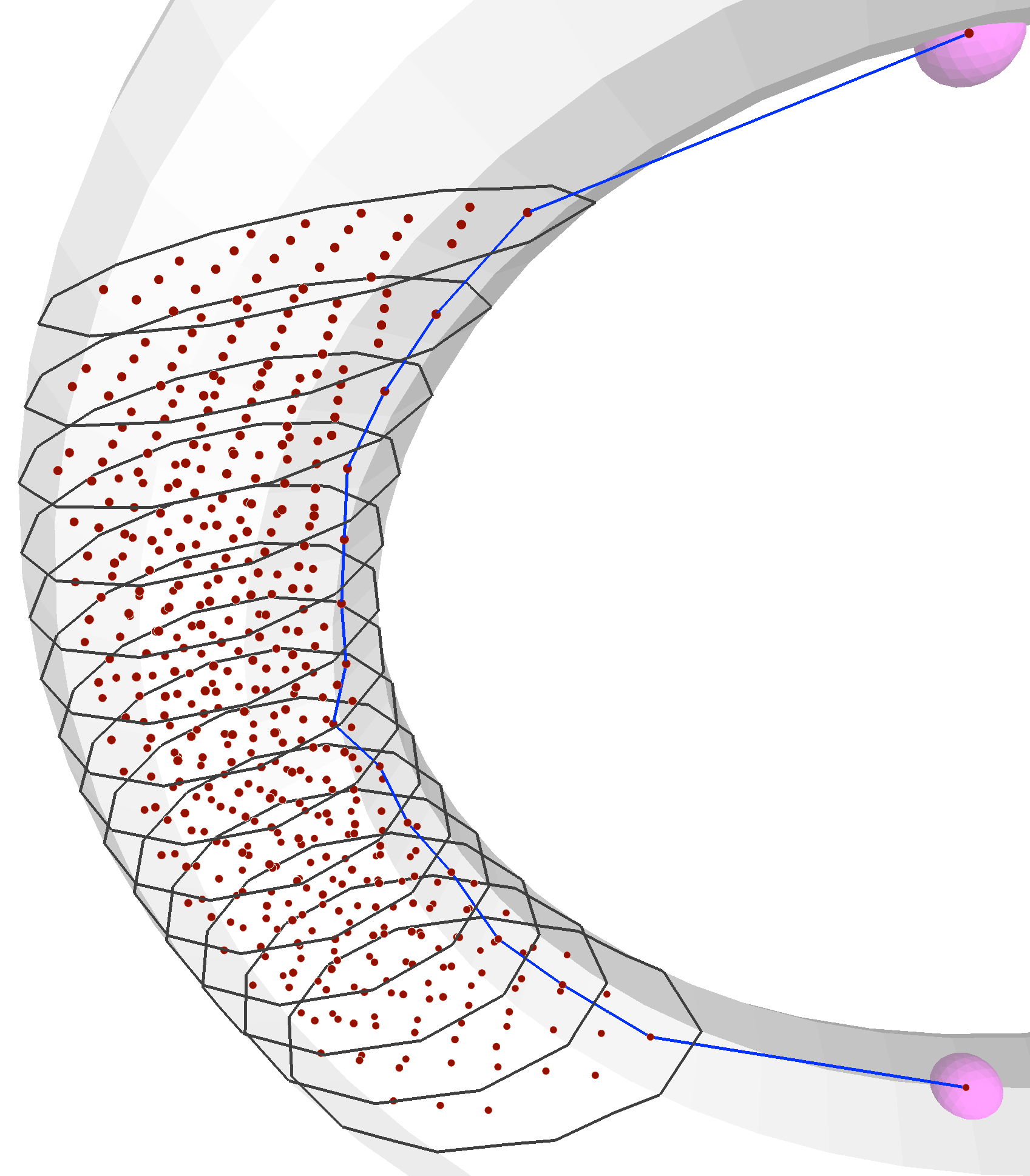}}
        \subfloat[Second Iteration\label{fig:contours:interior:2nd}]{\includegraphics[width=0.31\linewidth]{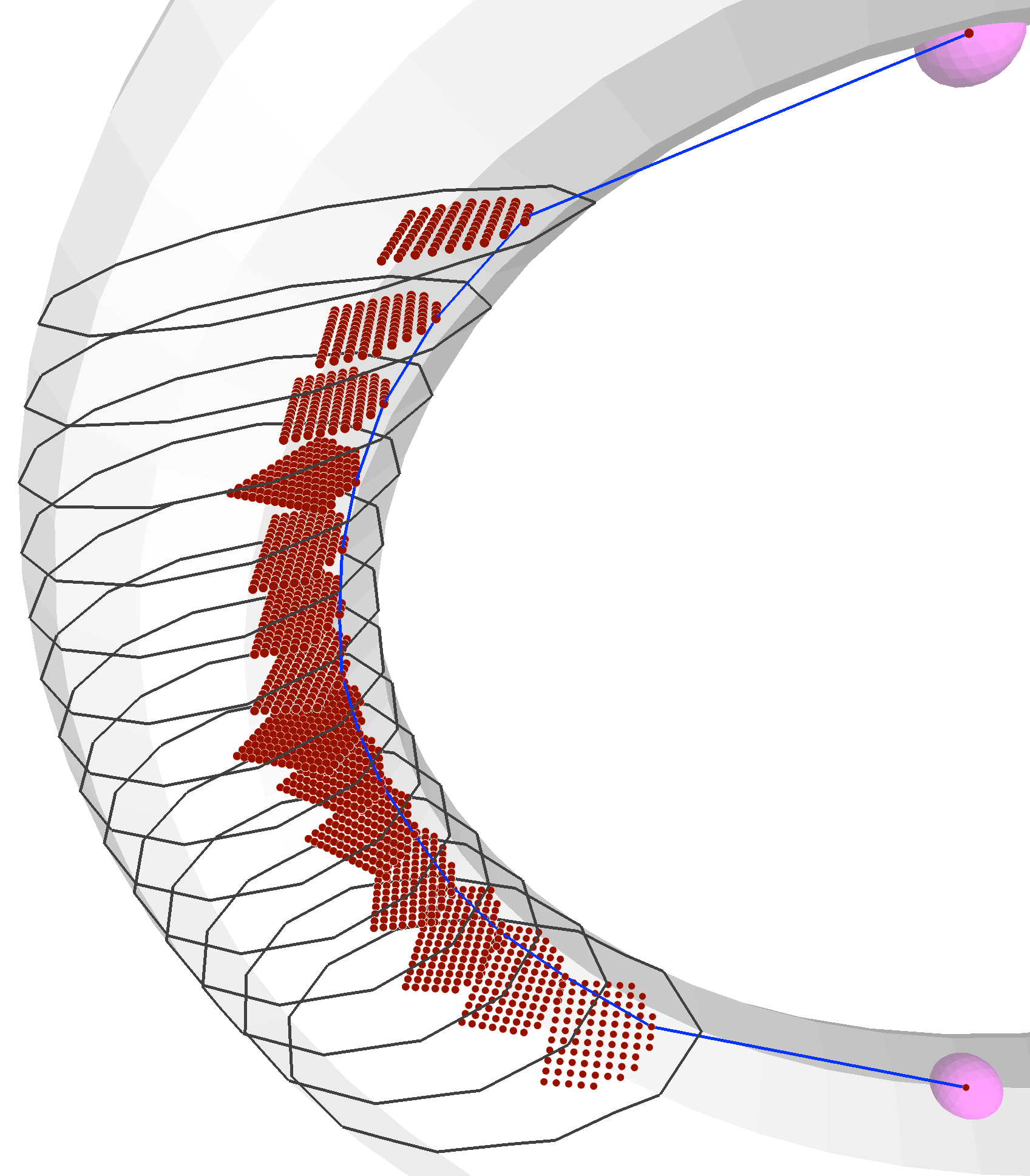}}
        \subfloat[Third Iteration\label{fig:contours:interior:3rd}]{\includegraphics[width=0.31\linewidth]{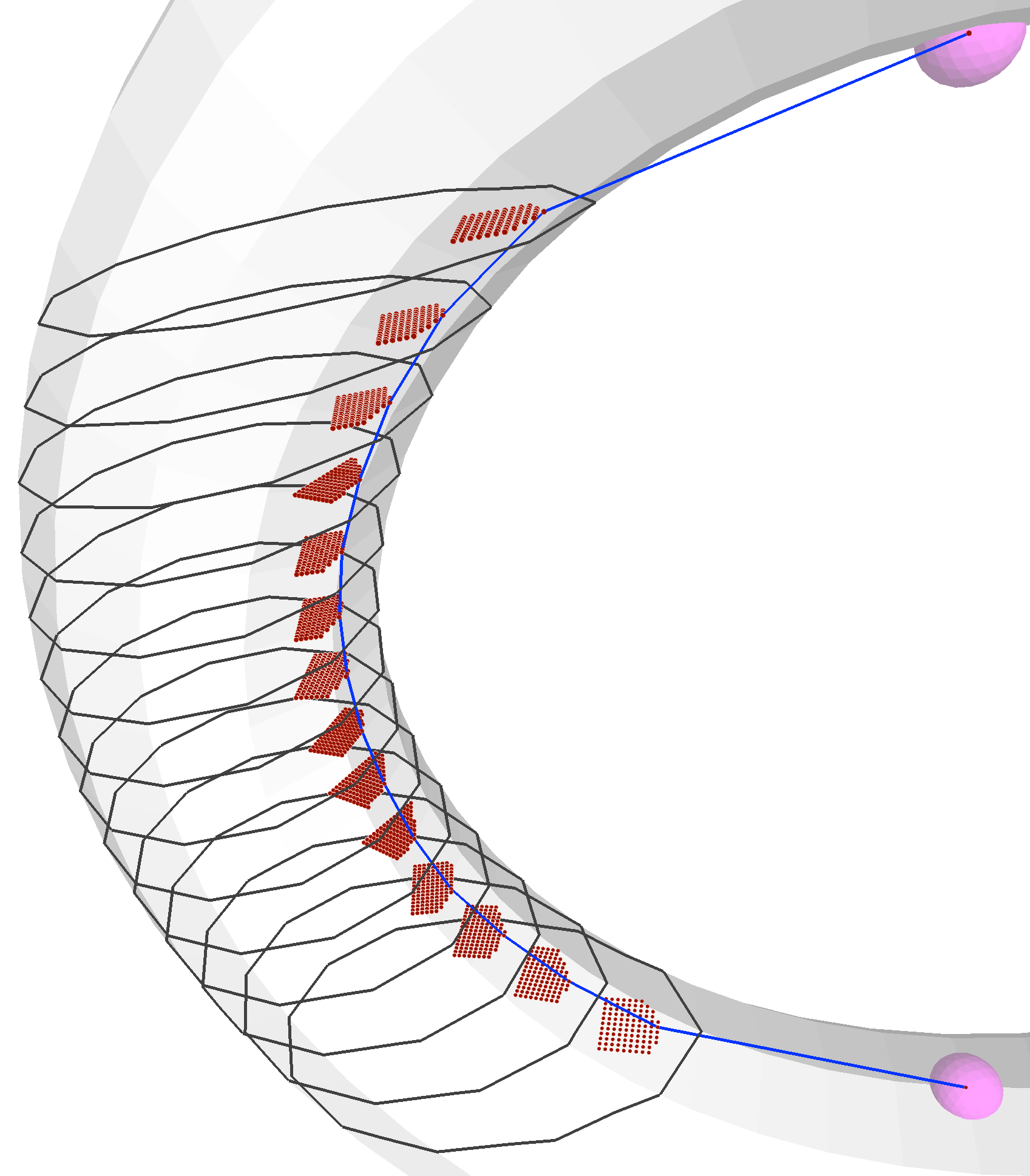}}
    \end{minipage}
    
    \caption{\textit{Left}: The process for the interior method (a)~starts with a contour in 3D. (b)~Principal component analysis is used to project the contour onto a plane, where a regular grid of points is set up. The points are checked to see if they are inside of the contour and no closer than a user-set distance from the contour (region in \textcolor{blue}{\textit{blue}}). (c)~Those points that pass both conditions (in \textcolor{red}{\textit{red}}) are projected back to 3D space. \textit{Right}: Example of the boundary approach (top) and interior method (bottom) candidate points in \textcolor{red}{\textit{red}} and the calculated path in \textcolor{blue}{\textit{blue}}. The process is iterative, (d,g)~starting with a coarse resolution, and refining it two times (e,h followed by f,i). The output path of the final step (f,i) is the Reeb graph arc used for the critical point pair.}
    \label{fig:interiorMethod_AND_contours_and_path}
\end{figure}

Next, the candidate points, $p$, for the Reeb graph arc path are calculated using one of two strategies, following \Cref{fig:interiorMethod_AND_contours_and_path}.

\paragraph{Boundary Approach} After the contours are generated (see \Cref{fig:interiorMethod:before}), the vertices coming from marching triangles are used as the vertices in the path graph (see \Cref{fig:contours:boundary}). 

\paragraph{Interior Approach} If the contours are approximately planar, the interior approach can identify candidate points inside of them (see \Cref{fig:interiorMethod:before}). It does this by first projecting the contour onto a 2D plane using principal component analysis (see \Cref{fig:interiorMethod:2d}). Then, a regular grid of candidate points is formed within the bounds of the contour. Those points are tested to see if they are inside the contour (point inside polygon test), and their distance from the contour is larger than a user-set buffer, $B$. Those points are projected back to 3D space (see \Cref{fig:interiorMethod:3d}) and used as the candidate points for the path graph (see \Cref{fig:contours:interior}).
The interior approach is more likely to keep Reeb graph arcs inside the object than the boundary approach. When the contour is planar or nearly planar, it can select suitable interior points. However, for non-planar contours, it may place some points outside the surface.

\subsubsection{Constructing the Path Graph and Path Extraction}
\label{impl:candidate:path}
Given the candidate points from the prior step, the path graph is formed next.
Candidate points serve as path graph vertices. Candidate points from adjacent contours have path graph edges added between them with weights set by their Euclidean distance. In addition, the contours adjacent to the critical points have path graph edges added between them and the critical points with weights set by their Euclidean distance.

After formulating the path graph, Dijkstra's algorithm is used to find the shortest path between the critical points, which represents the Reeb graph arc, as shown in \textcolor{blue}{\textit{blue}} in \Cref{fig:contours:boundary} and \Cref{fig:contours:interior} for the boundary and interior versions, respectively.

\subsubsection{Iterative Refinement} To improve the quality and speed of the operation, the process of calculating candidate points (see \cref{impl:candidate:points}) and constructing the path graph and extracting the path (see \cref{impl:candidate:path}) is done over multiple iterations at increasing resolutions. For both variations, the first iteration begins with a sparse selection of up to 40 candidate points per contour (see \Cref{fig:contours:boundary} and \Cref{fig:contours:interior}). Once an initial path is identified, a new set with a similar number of candidate points is selected at a finer granularity in the neighborhood of the initial path (see \Cref{fig:contours:boundary:2nd} and \Cref{fig:contours:interior:2nd}). Our implementation repeats this step one additional time (i.e., 2 refinement steps total, see \Cref{fig:contours:boundary:3rd} and \Cref{fig:contours:interior:3rd}). The benefit of this approach is that it limits the complexity of the path graph while providing a high fidelity in the output position.

\subsubsection{Thin Features}
\label{sec:thin}
For some topological cylinders, the functional distance between critical points is less than the contour spacing parameter (i.e., $f(C_k)-f(C_j)<S$). Under the process described above, these arcs would directly connect critical points, and the visualization would, therefore, be of low quality. Instead, we introduce a modified process for these \textit{thin features}, irrespective of whether the boundary or interior approach is selected.

\begin{figure}[!ht]
    \centering
    \rotatebox{90}{\footnotesize \hspace{3pt}\tiny \textsf{Example Pair 1}}
    \includegraphics[trim=175pt 400pt 500pt 700pt, clip, height=1.55cm]{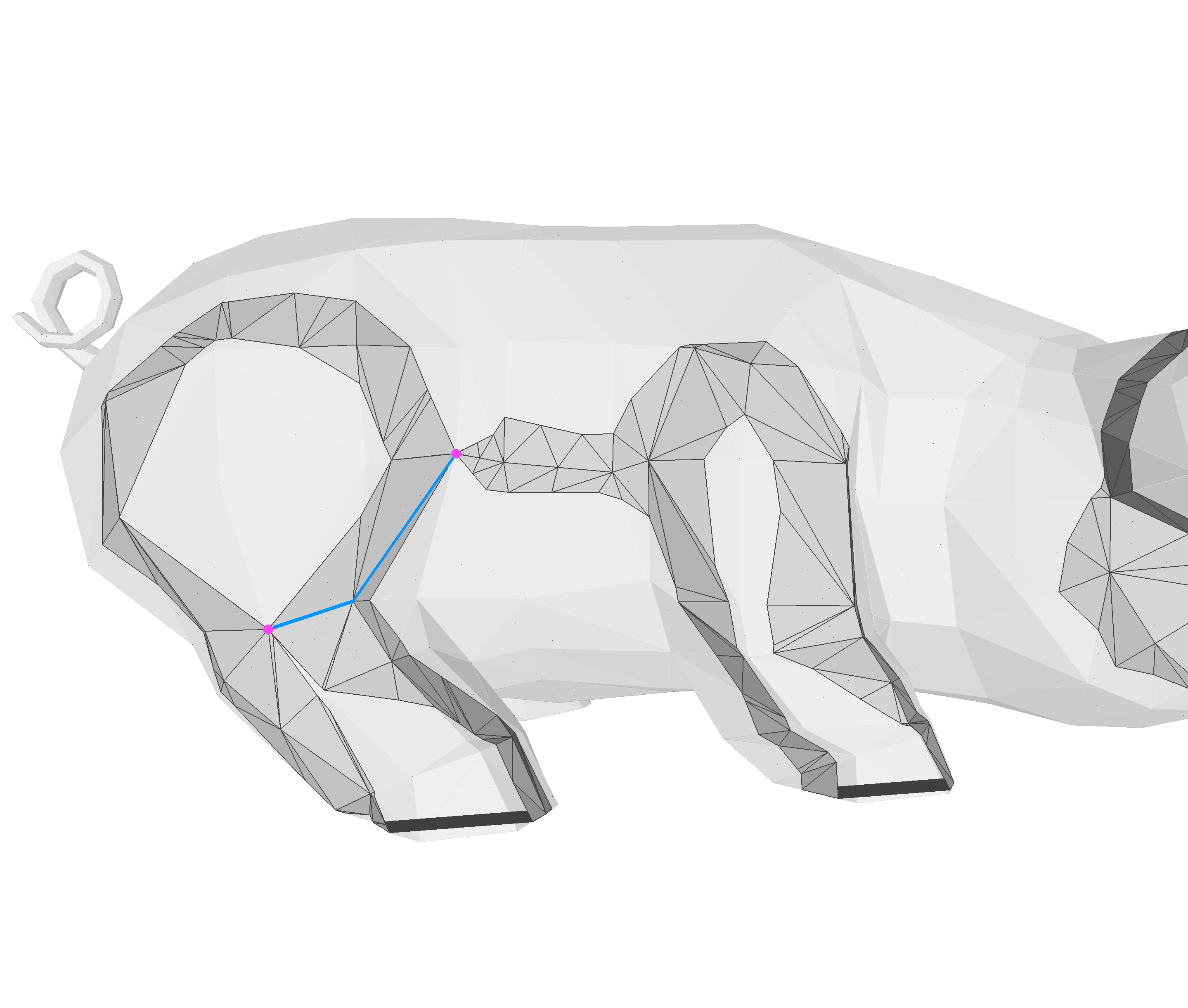}
    \hspace{1pt}
    \includegraphics[trim=150pt 650pt 1400pt 1000pt, clip,  height=1.55cm]{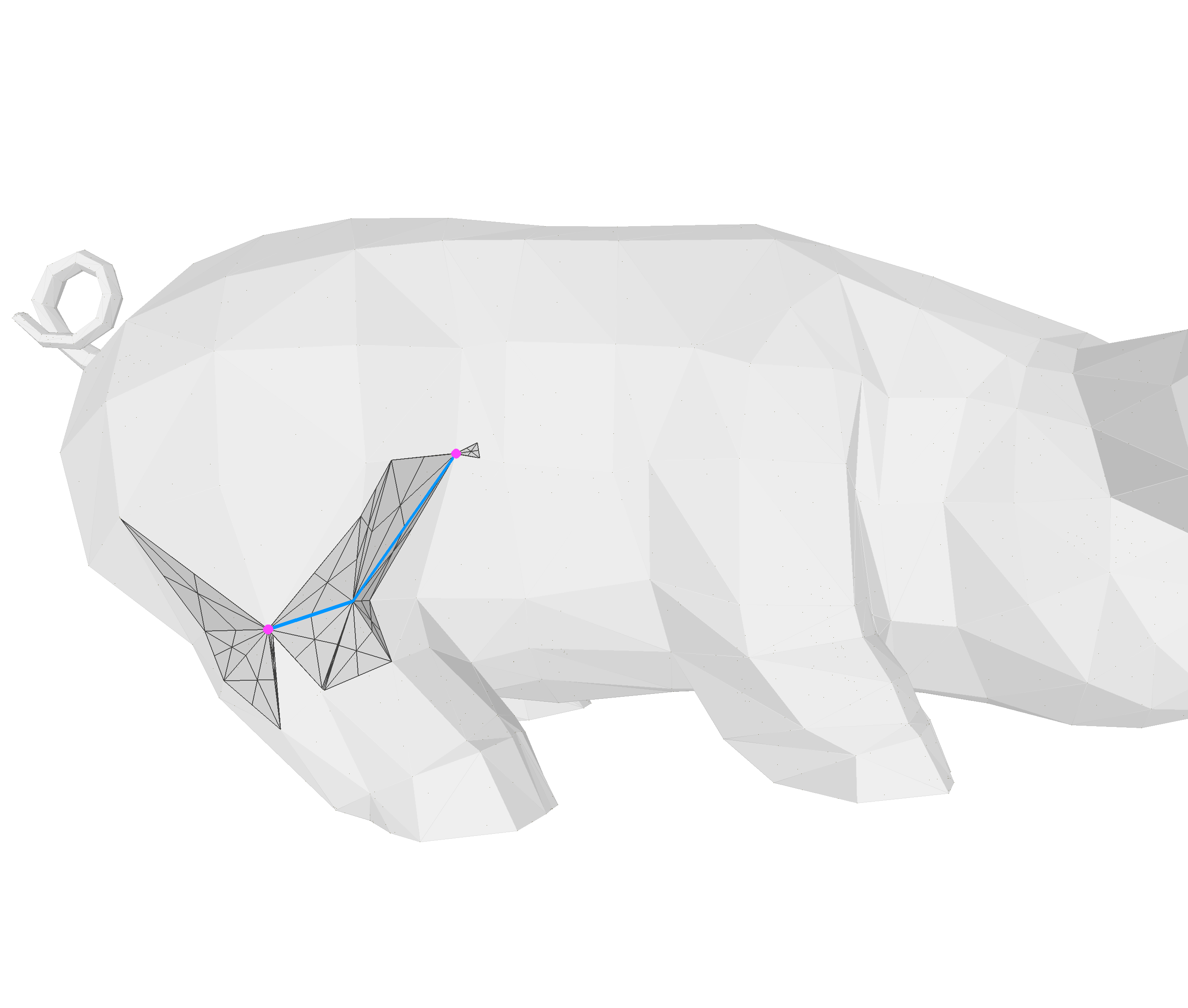}
    \hspace{1pt}
    \includegraphics[trim=150pt 650pt 1400pt 1000pt, clip, height=1.55cm]{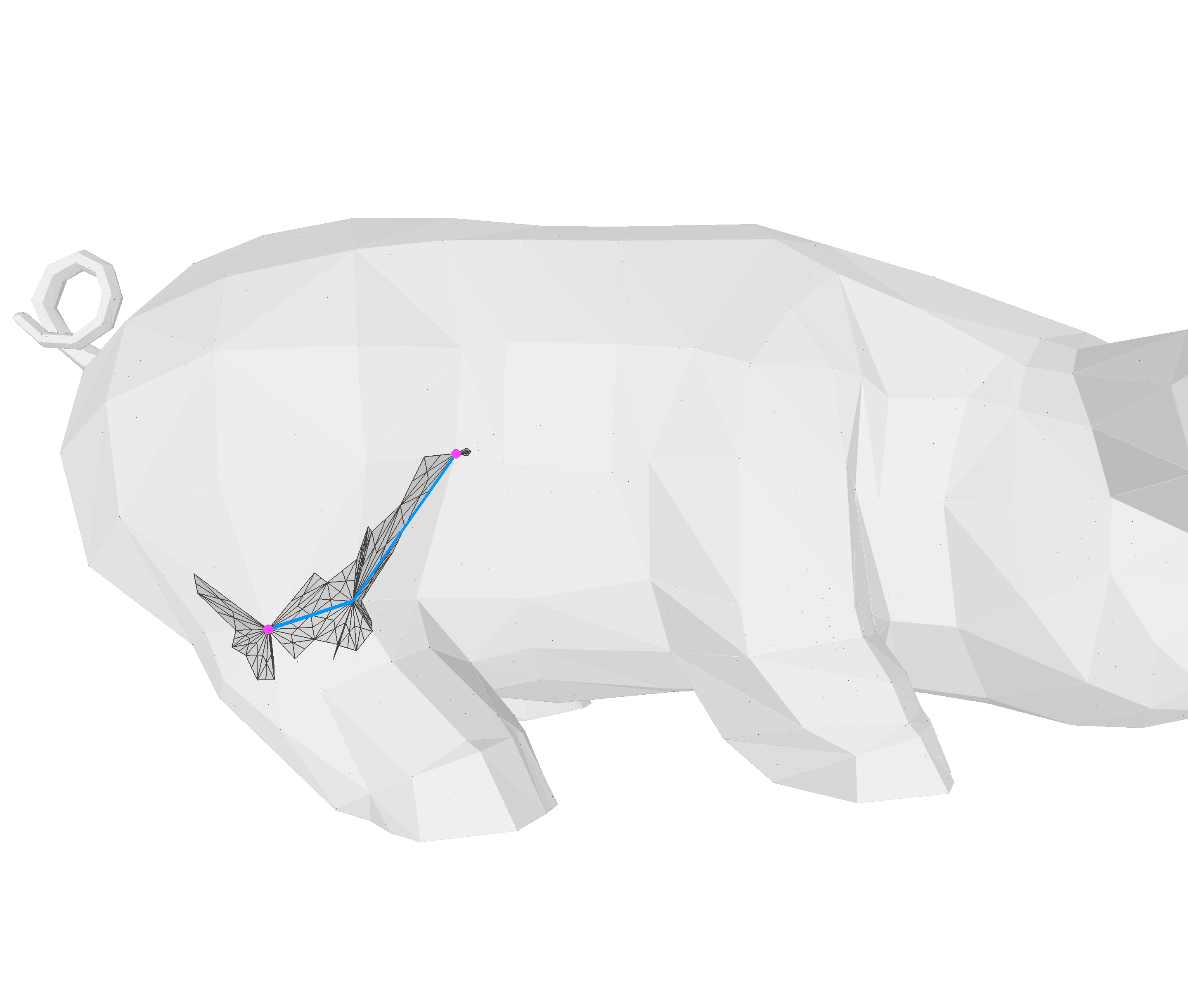}

    \vspace{3pt}
    \rotatebox{90}{\footnotesize \hspace{3pt} \tiny \textsf{Example Pair 2}}
    \subfloat[Initial Cut\label{fig:thin_ex:cut}]{\includegraphics[trim=875pt 800pt 170pt 550pt, clip, width=0.3\linewidth]{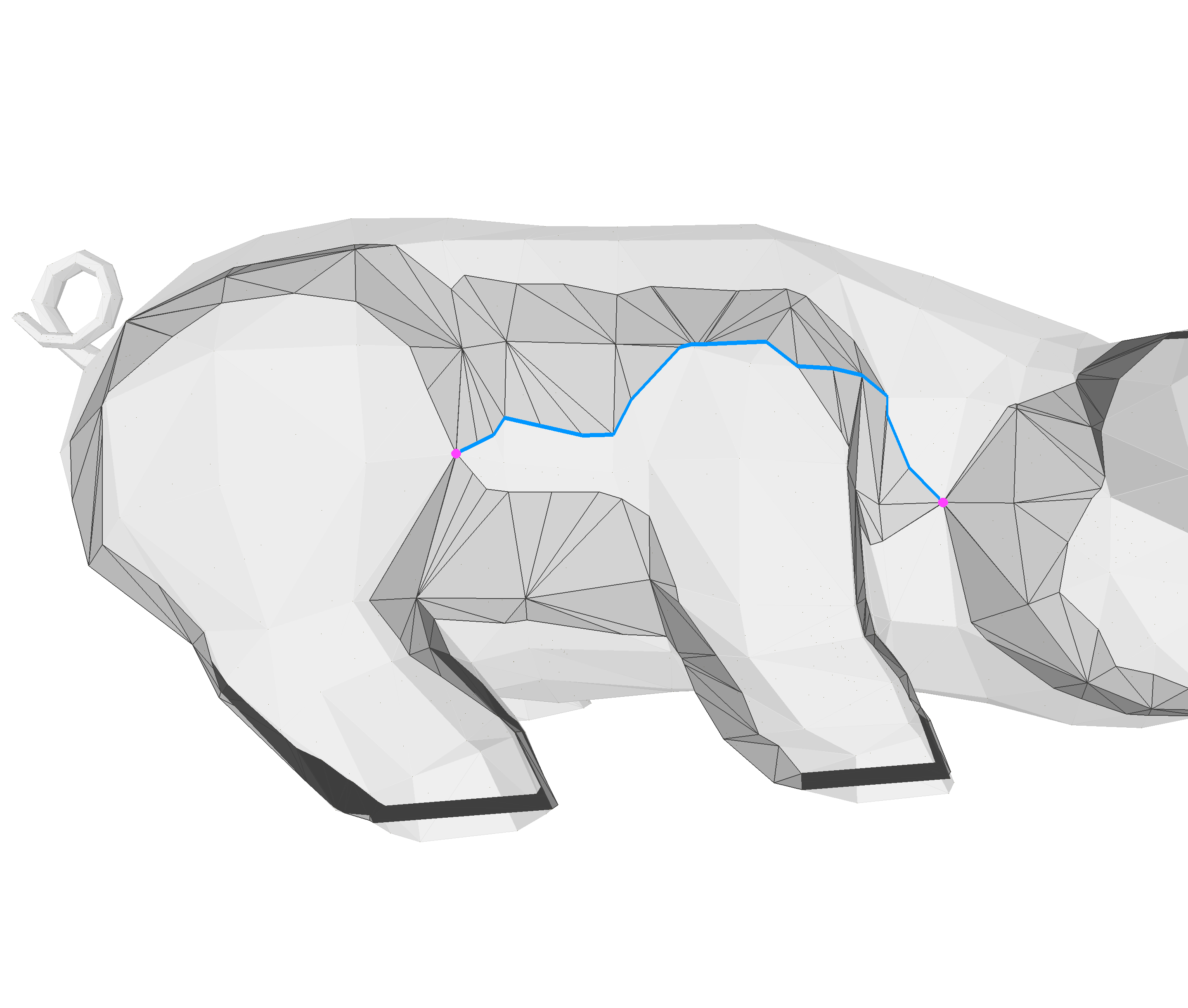}}
    \hspace{1pt}
    \subfloat[First Refinement\label{fig:thin_ex:first}]{\includegraphics[trim=875pt 800pt 170pt 550pt, clip, width=0.3\linewidth]{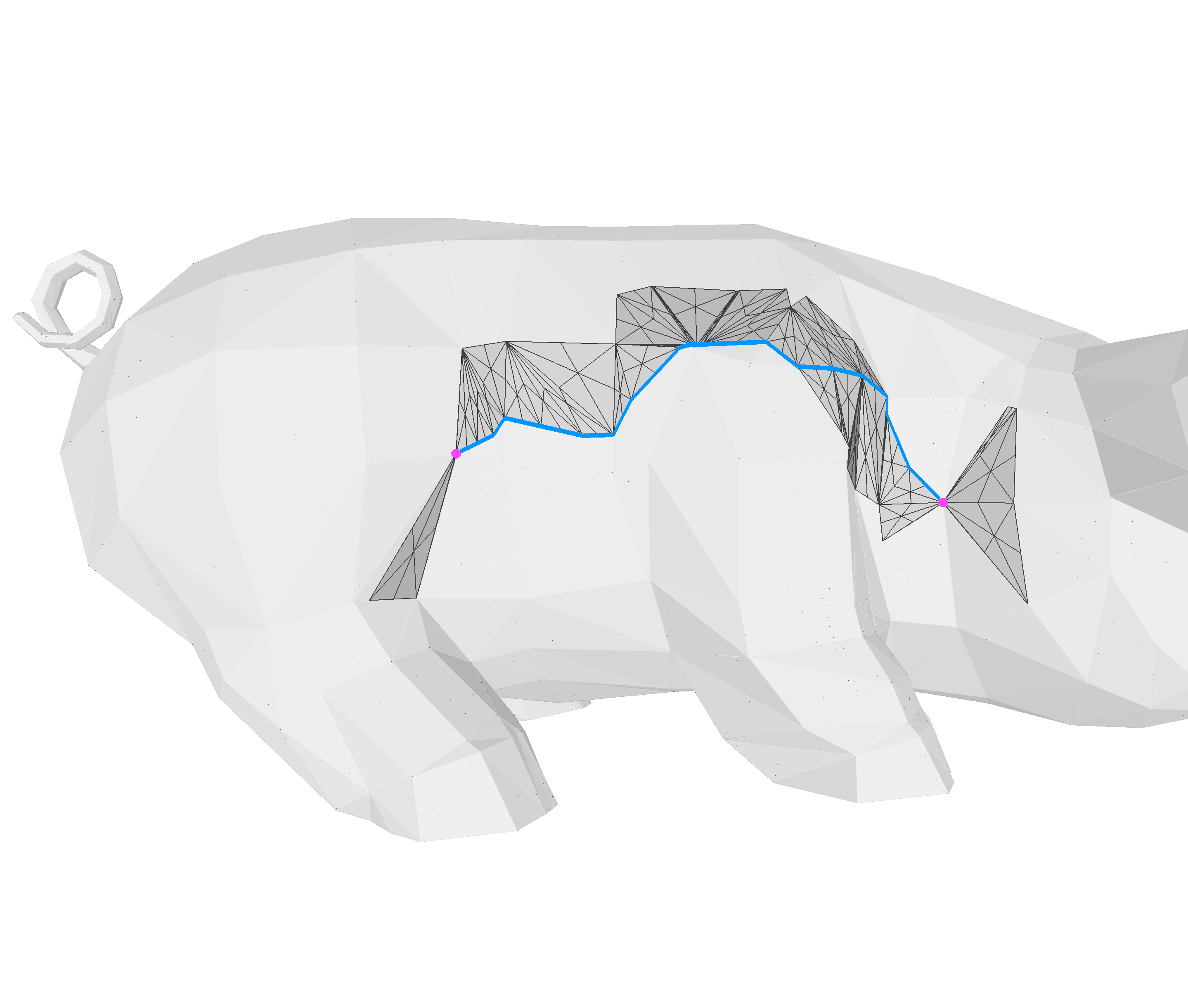}}
    \hspace{1pt}
    \subfloat[Second Refinement\label{fig:thin_ex:second}]{\includegraphics[trim=875pt 800pt 170pt 550pt, clip, width=0.3\linewidth]{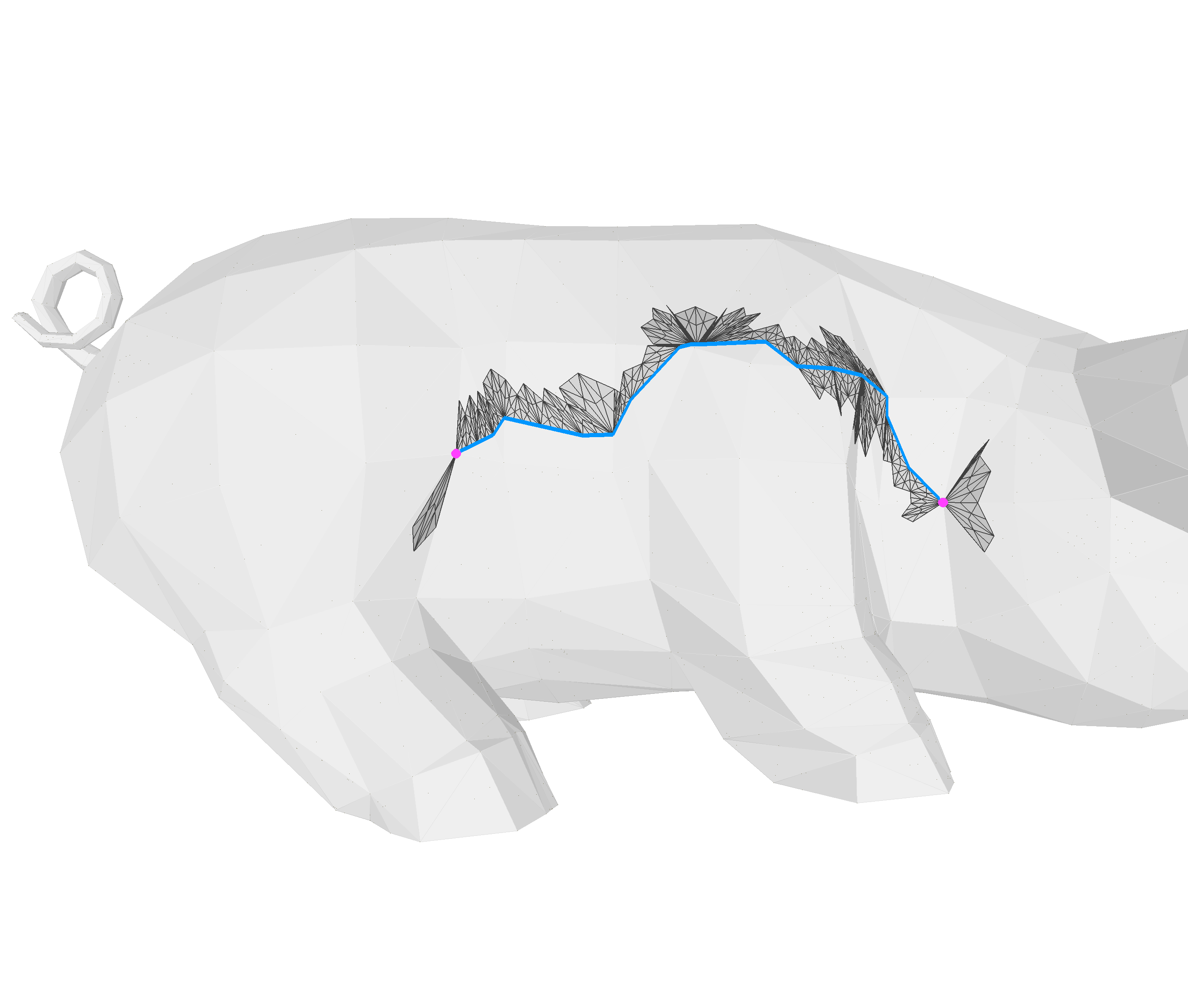}}
    
    \caption{Two examples of of thin features. (a)~Initially, a cut is performed between the two critical points in \textcolor{Magenta}{\textit{pink}} and the shortest path, in \textcolor{blue}{\textit{blue}}, along the triangle edges is found. (b)~Triangles adjacent to the initial path are subdivided and the path is recomputed. (c)~The process is repeated until the path no longer changes or a maximum of five iterations.}
    \label{fig:thin_ex}
\end{figure}

\paragraph{Decomposition} The decomposition process is identical in all regards, except the cuts occur at $f(C_j)$ and $f(C_k)$, in other words excluding the $\epsilon$ offset (see \Cref{fig:thin_ex:cut}).

\paragraph{Path Graph and Path Extraction} The path graph for thin features uses the edges of the extracted triangle mesh. The shortest path between critical points is found using Dijkstra's algorithm (see \Cref{fig:thin_ex:cut}).

\paragraph{Iterative Refinement} To produce a smoother arc, triangles that are adjacent to the initial path are subdivided and the path is re-computed (see \Cref{fig:thin_ex:first}). This step is performed up to five times or when the refinement no longer changes the path (see \Cref{fig:thin_ex:second}).

\paragraph{Discussion}
Compared with drawing Reeb graph arcs from contours, the thin feature ensures that the arcs stay on the surface. However, applying this approach to the entire model introduces some drawbacks. First, the arcs that are generated are \textit{not guaranteed} to align with the function gradients. Additionally, because it operates directly on mesh triangles, the smoothness of the arcs depends on the input mesh resolution. Therefore, we employ the thin feature only when it is necessary.

\subsection{Reeb Graph Assembly}
\begin{figure}[!ht]
    \centering

    \subfloat[Torus (Boundary)]{\hspace{5pt}\includegraphics[height=1.55cm]{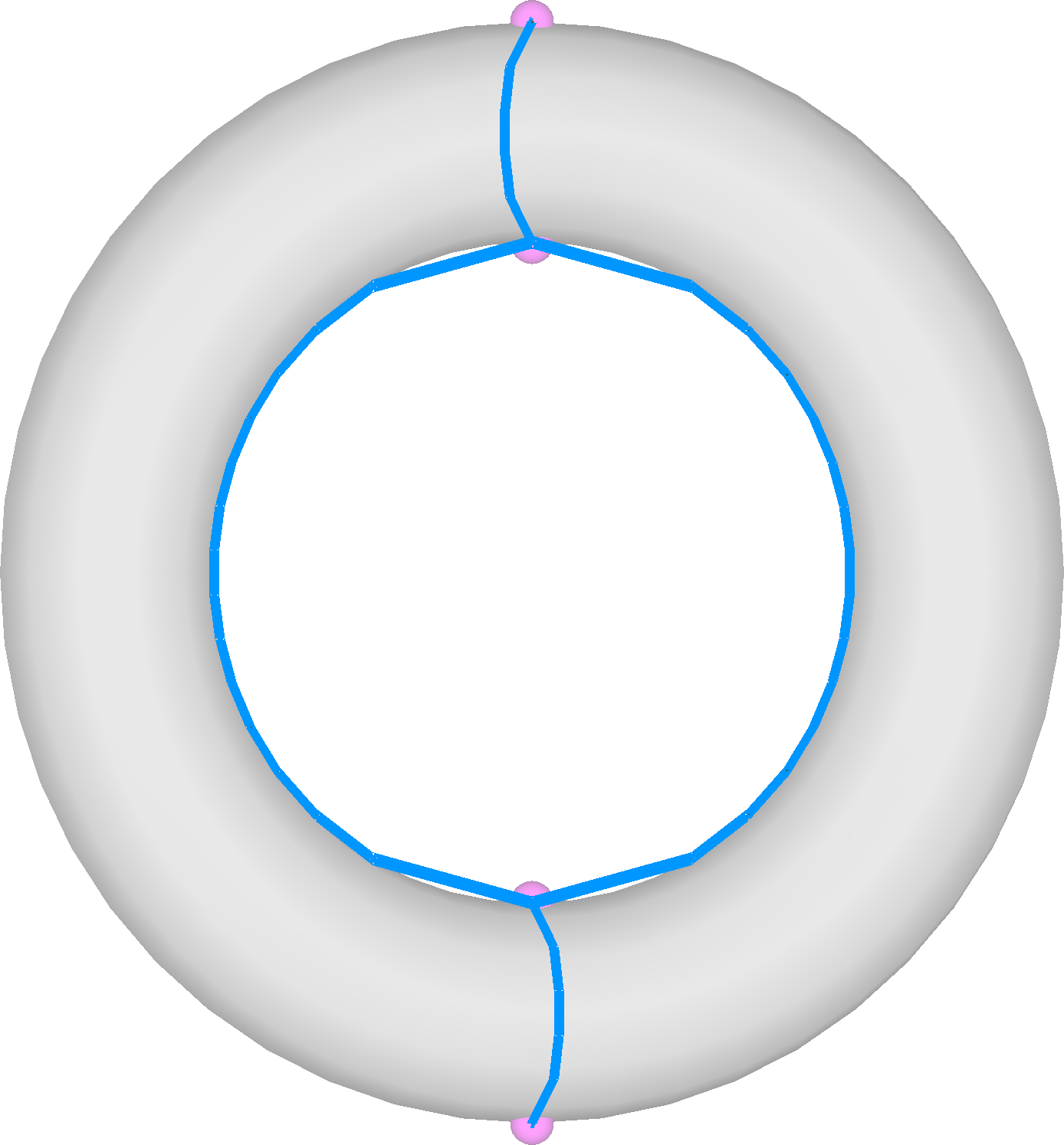}\hspace{5pt}}
    \hfill
    \subfloat[Torus (Interior)]{
    \includegraphics[height=1.55cm]{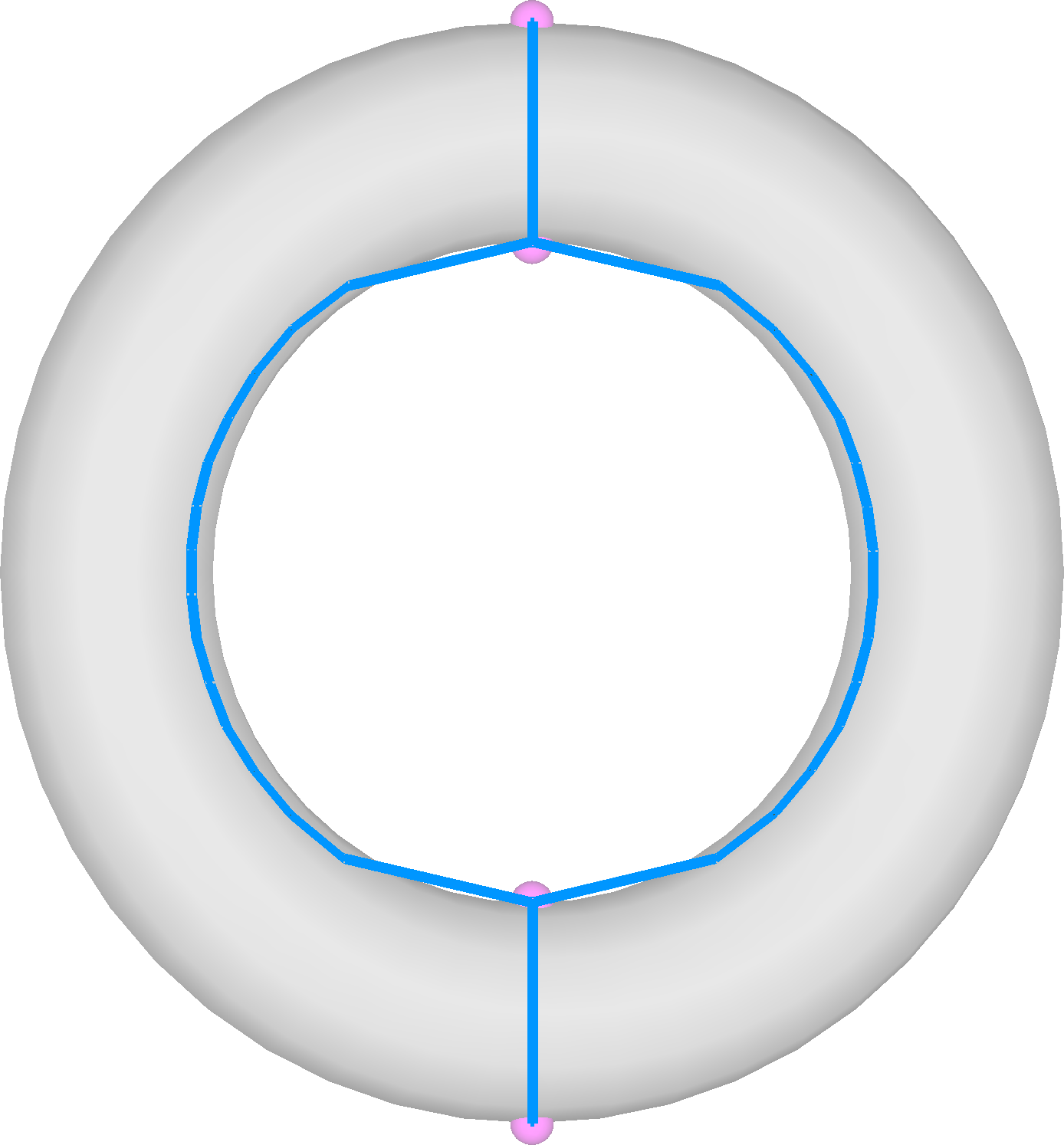}}
    \hfill
    \subfloat[Modified Torus (Boundary)\label{fig:torus_final:mt_boundary}]{\includegraphics[height=1.55cm]{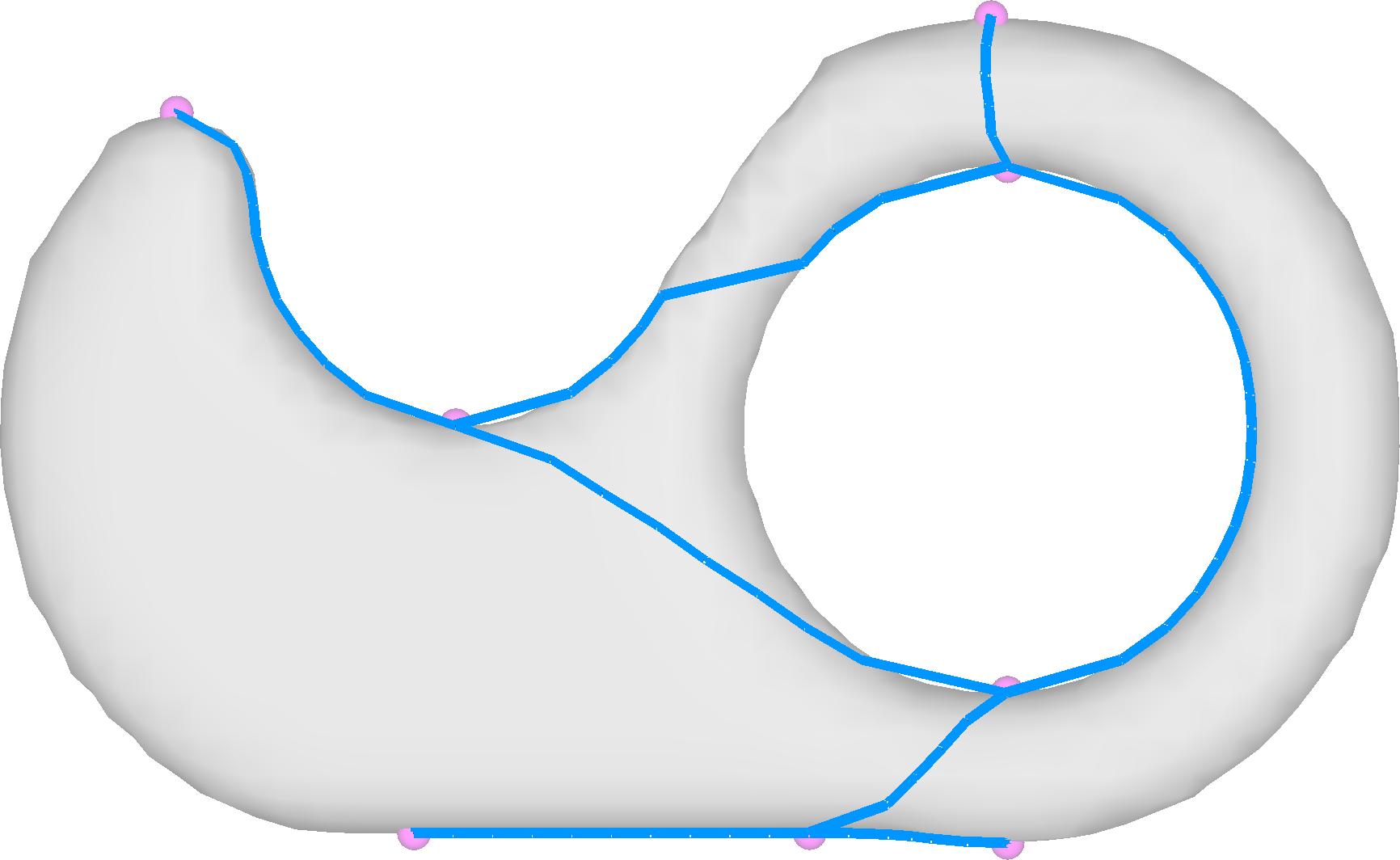}}
    \hfill
    \subfloat[Modified Torus (Interior)\label{fig:torus_final:mt_interior}]{
    \includegraphics[height=1.55cm]{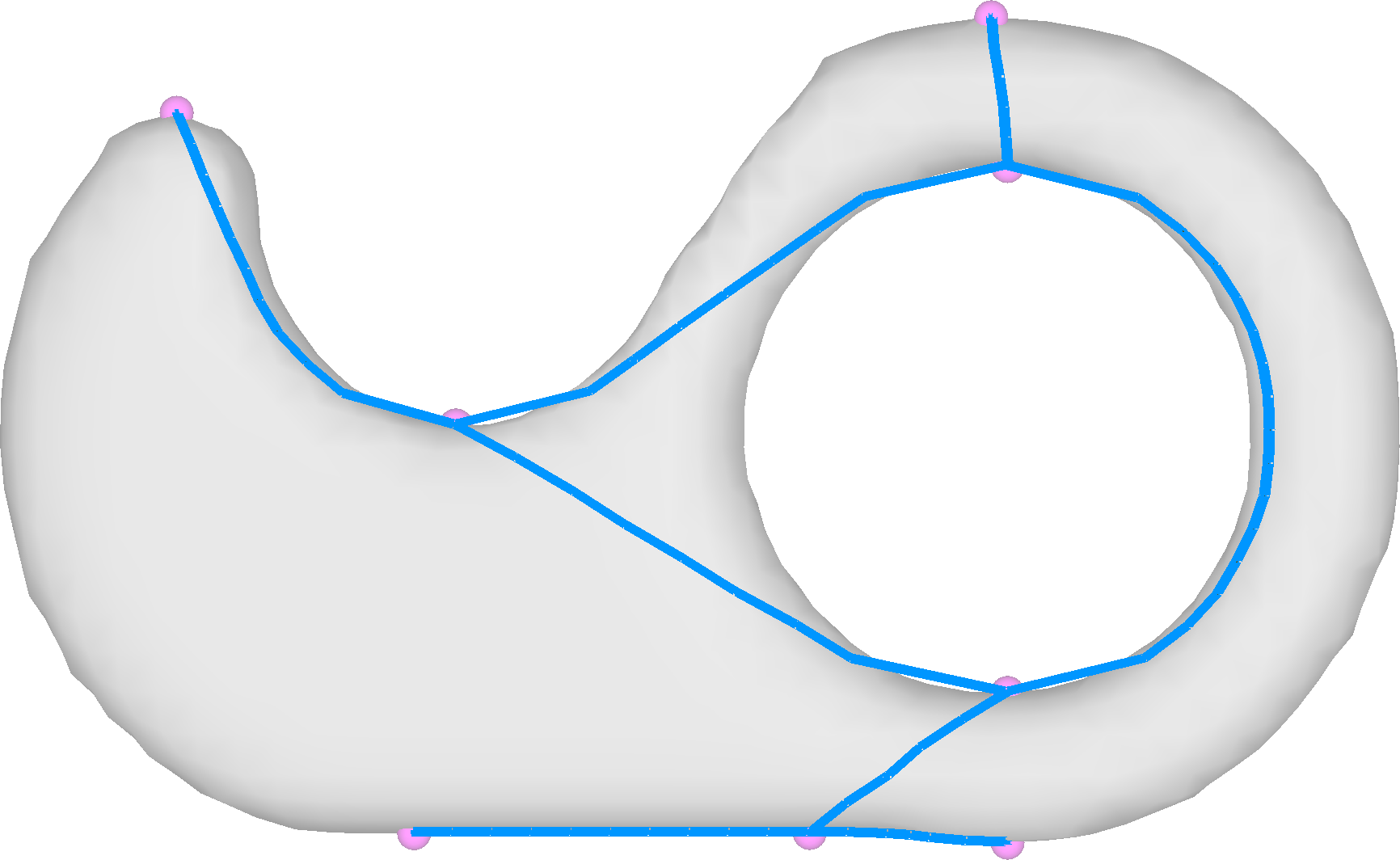}}
    
    \caption{Final Reeb graph visualizations for the (a,b)~torus and (c,d)~modified torus models with the (a,c)~boundary and (b,d)~interior approaches.}
    \label{fig:torus_final}
\end{figure}

For each pair of connected critical points, we generated the arcs of the Reeb graph as described above. Combining all the arcs, the complete Reeb graph is realized. \Cref{fig:torus_final} shows examples of the torus and modified torus with both the boundary and interior approach.

\section{Evaluation}

We perform a mixed-method evaluation of our approach and compare it to Geometric Barycenter (GB) algorithm, as implemented in TTK. 
We primarily frame our evaluation around the three essential properties of a Reeb graph that impact the faithfulness of its representation (see \cref{eval:issue1}-\cref{eval:issue3}) and one optional property (see \cref{eval:aesthetic}). Through this detailed comparative analysis, we highlighted the improvements offered by GASP. We also demonstrate the impact of the user-set parameters of our approach have on the quality of the output Reeb graphs. Finally, we discuss the computational complexity of our algorithm (see \cref{eval:comp}).

\begin{figure}[!b]
    \centering

    \begin{minipage}[m]{0.2\linewidth}
        \centering
        \includegraphics[width=\linewidth]{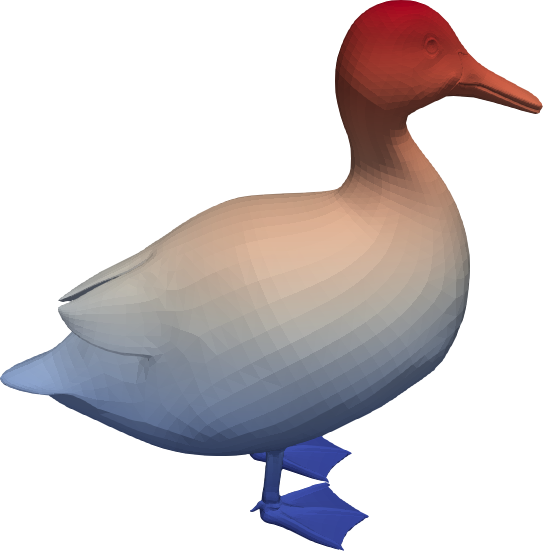} \\
        \tiny \texttt{duck} \\ 
        $y$ \textsf{function}
    \end{minipage}
    \hspace{5pt}
    \begin{minipage}[m]{0.7\linewidth}
        \rotatebox{90}{\hspace{5pt} \tiny \textsf{Boundary Approach}}
        \hspace{3pt}
        \subfloat[Spacing $S\mathbin{=}0.025$\label{fig:duck-c-0.025}]{\includegraphics[width=0.3\linewidth]{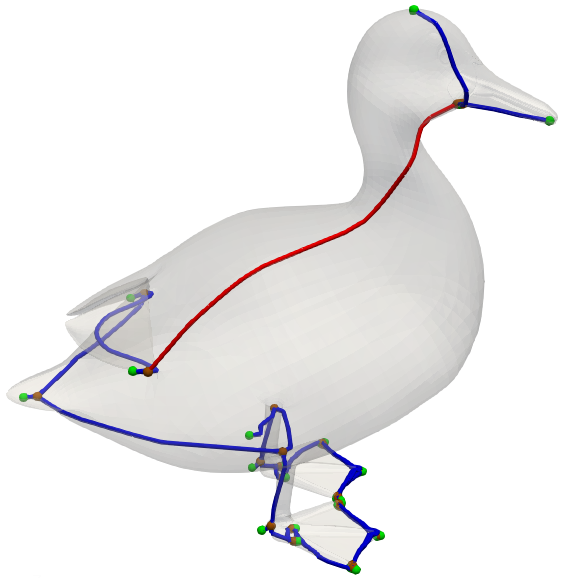}}     
        \hfill
        \subfloat[Spacing $S\mathbin{=}0.05$\label{fig:duck-c-0.05}]{\includegraphics[width=0.3\linewidth]{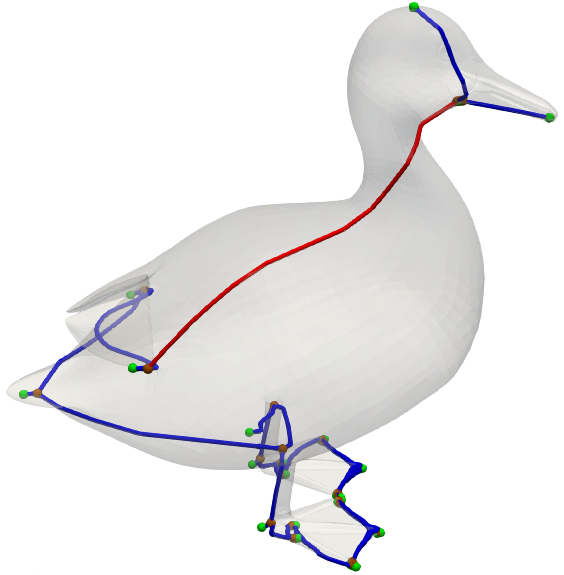}}
        \hfill
        \subfloat[Spacing $S\mathbin{=}0.1$\label{fig:duck-c-0.1}]{\includegraphics[width=0.3\linewidth]{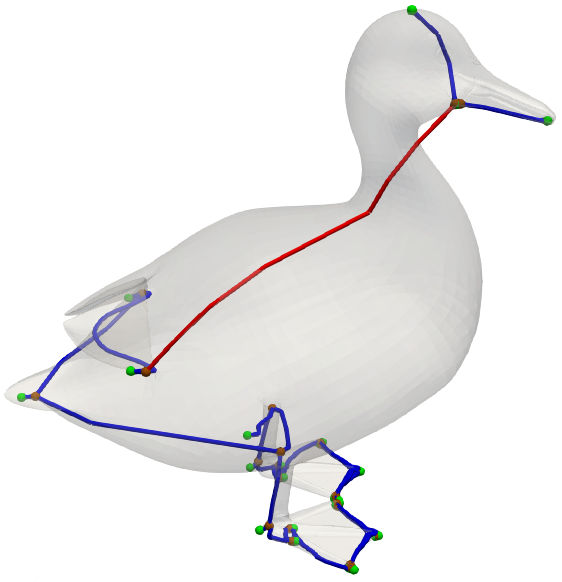}}        
        \hfill

        \rotatebox{90}{\hspace{5pt} \tiny \textsf{Interior Approach}}
        \hspace{3pt}
        \subfloat[Buffer $B\mathbin{=}0.0$\label{fig:duck-i-0.00}]{\includegraphics[width=0.3\linewidth]{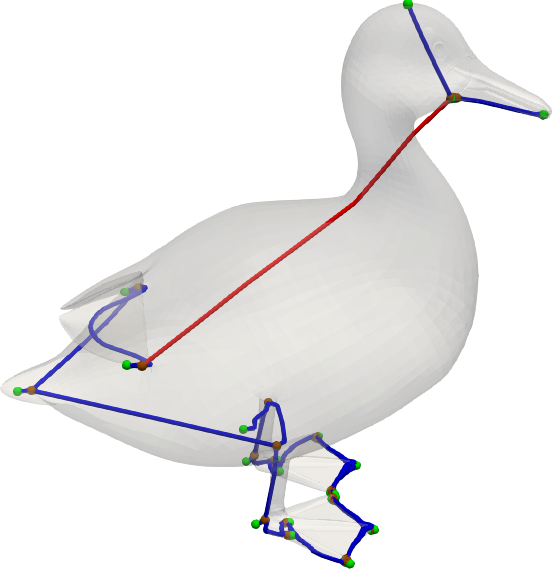}}
        \hfill
        \subfloat[Buffer $B\mathbin{=}0.025$\label{fig:duck-i-0.025}]{\includegraphics[width=0.3\linewidth]{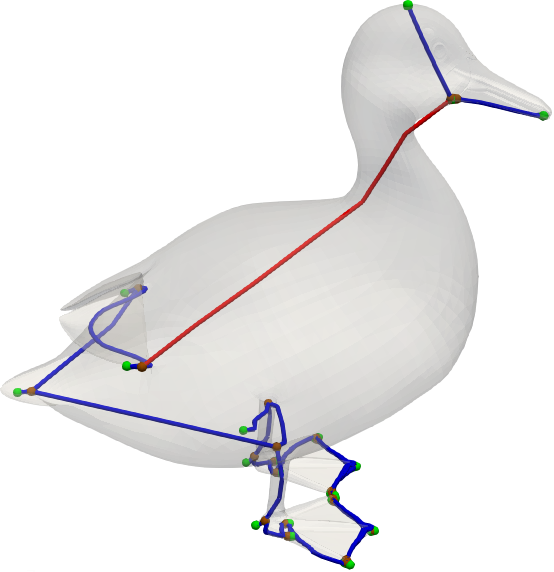}}
        \hfill
        \subfloat[Buffer $B\mathbin{=}0.05$\label{fig:duck-i-0.05}]{\includegraphics[width=0.3\linewidth]{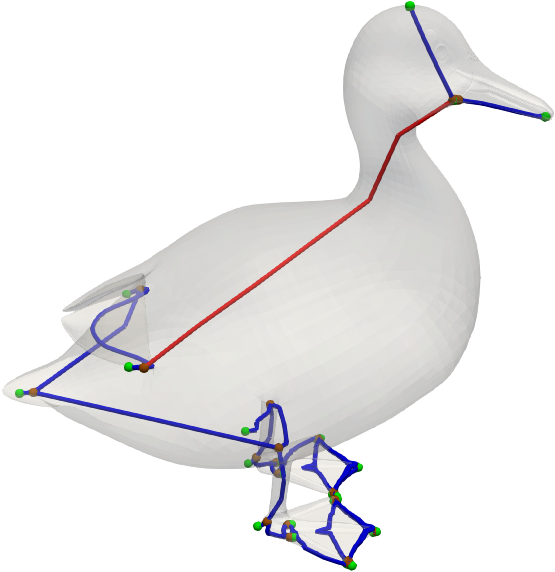}}        
    \end{minipage}

    \caption{Illustration of the effects of varying the (a-c)~\textit{contour spacing} for the boundary approach and the (d-f)~\textit{buffer size} for the interior approach with fixed contour spacing ($S\mathbin{=}0.1$). Complete Reeb graphs include \textit{\textcolor{red}{red}} and \textit{\textcolor{blue}{blue}} arcs; \textit{\textcolor{red}{red}} arcs highlight comparisons.}
    \label{fig:varying_spacing_and_buffer}
\end{figure}

\paragraph{Data and Visualizations} We have utilized 30 triangle meshes from~\cite{TA:2024:topoSensitivity}. All models were normalized such that their largest dimension range was $[-1,1]$. For each model, we generated Reeb graphs for height functions in all three principal directions (i.e., $x$, $y$, and $z$) and geodesic distance functions that emanate from the most extreme point in each principal direction (i.e., $top$, $bottom$, $left$, $right$, $back$, and $front$). While the Reeb graph arc paths are calculated using our approach, all visualizations are rendered using \textit{ParaView}~\cite{TA:2005:paraview}. Only a subset of the datasets is discussed in the paper, but the analysis of all datasets, source code, and a video of our approach compared to GB (TTK) can be found in the supplemental materials.

\paragraph{GASP Parameters}
As outlined in \Cref{sec:candidate-points}, our approach features two main variations: \textit{boundary} and \textit{interior}. Both approaches incorporate a \textit{contour spacing} parameter, $S$, while the interior approach has an additional \textit{buffer size} parameter, $B$. The smaller the contour spacing, the finer the detail in the output. Throughout the evaluation, we utilize three levels of contour spacing with the largest being 0.1 (see \Cref{fig:duck-c-0.1}), representing 5\% of the $[-1,1]$ domain, and 0.05 (see \Cref{fig:duck-c-0.05}) and 0.025  (see \Cref{fig:duck-c-0.025}) representing $2 \times$ and $4 \times$ the resolution, respectively. The buffer spacing determines the minimum distance an arc should be from the contour in the interior method. Again, three levels are utilized from 0 (see \Cref{fig:duck-i-0.00}), which allows arcs to touch the contour, to 0.025 (see \Cref{fig:duck-i-0.025}), to 0.05 (see \Cref{fig:duck-i-0.05}) representing 2.5\% of the domain. 
Parameters for testing were selected such that at the lowest level of detail, the output quality would be good but noticeably different from the highest level of detail. We selected the highest level of detail such that additional detail would not be noticeable.

\begin{figure}[!ht]

    \begin{minipage}[m]{0.175\linewidth}
        \centering
        \includegraphics[width=\linewidth]{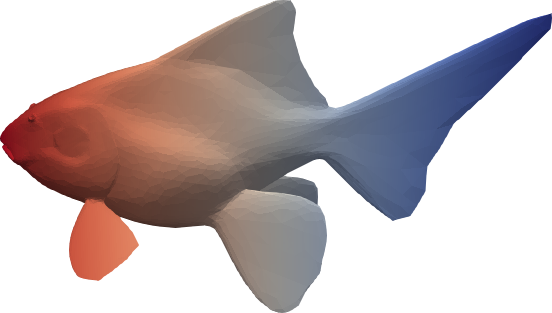} \\
        \tiny \texttt{fish} \\
        $z$ \textsf{function}
    \end{minipage}
    \hfill
    \begin{minipage}[m]{0.8\linewidth}
        \rotatebox{90}{\hspace{4pt} \tiny \textsf{Fixed}}
        \rotatebox{90}{\tiny \textsf{Smoothing 15}}
        \subfloat[Sampling 5 (TTK 5/15)\label{fig:TTK-ArcSampling-5}]{{\includegraphics[width=0.31\linewidth]{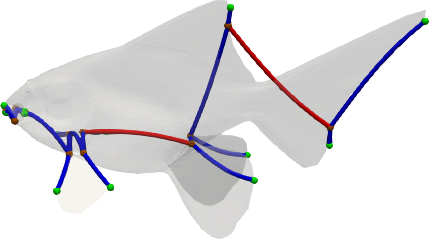}}}
        \hfill
        \subfloat[Sampling 10\label{fig:TTK-ArcSampling-10}]{\includegraphics[width=0.31\linewidth]{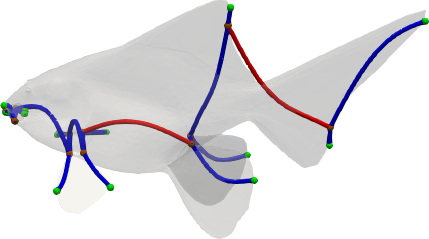}}
        \hfill
        \subfloat[Sampling 15 (TTK 15/15)\label{fig:TTK-ArcSampling-15}]{\includegraphics[width=0.31\linewidth]{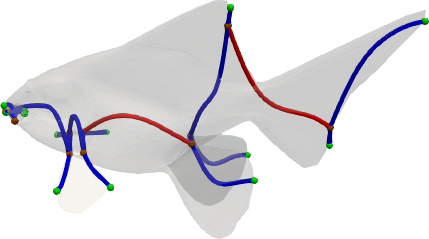}}

        \rotatebox{90}{\hspace{4pt} \tiny \textsf{Fixed}}
        \rotatebox{90}{\tiny \textsf{Sampling 15}}
        \subfloat[Smoothing 5\label{fig:TTK-iteration-5}]{\includegraphics[width=0.31\linewidth]{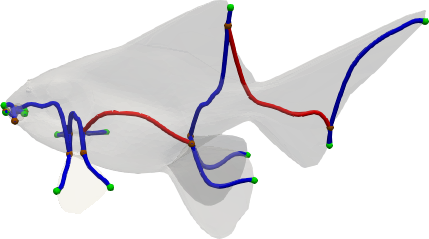}} 
        \hfill
        \subfloat[Smoothing 10\label{fig:TTK-iteration-10}]{\includegraphics[width=0.31\linewidth]{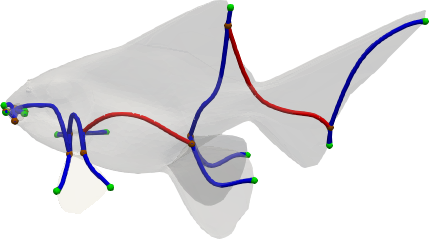}}
        \hfill
        \subfloat[Smoothing 15\label{fig:TTK-iteration-15}]{\includegraphics[width=0.31\linewidth]{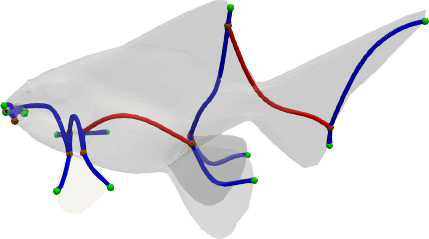}}
    \end{minipage}
    
    \caption{Illustration of varying TTK parameters: (a-c)~arc sampling and (d-f)~arc smoothing. As arc sampling increases, arcs get longer. As smoothing increases, the output arcs take smoother, more aesthetically appealing routes. Complete Reeb graphs include \textit{\textcolor{red}{red}} and \textit{\textcolor{blue}{blue}} arcs; \textit{\textcolor{red}{red}} arcs highlight parameter variations.}
    \label{fig:TTK-parameters}
\end{figure}

\paragraph{TTK Parameters} 
As briefly outlined in \Cref{sec:reeb-vis}, TTK's approach to drawing Reeb graphs utilizing GB subdivides a straight edge into multiple arc segments based on a user-defined \textit{arc sampling} parameter. It then applies \textit{geometric smoothing} over several iterations, controlled by another user-defined parameter. \Cref{fig:TTK-parameters} illustrates the effect of varying both TTK parameters. For the evaluation, we consider two versions of TTK, both with smoothing set to 15 iterations and arc sampling set to 5 (referred to as \textit{TTK 5/15}), representing a less smooth variant (see \cref{fig:TTK-ArcSampling-5}), and arc sampling 15 (referred to as \textit{TTK 15/15}), representing a smoother, aesthetically more pleasing version (see \cref{fig:TTK-ArcSampling-15}).

\subsection{Property 1: Arcs Confined to the Model}
\label{eval:issue1}

\begin{figure}[!b]
    \centering
    \begin{minipage}[m]{0.125\linewidth}
        \centering
        \includegraphics[width=\linewidth]{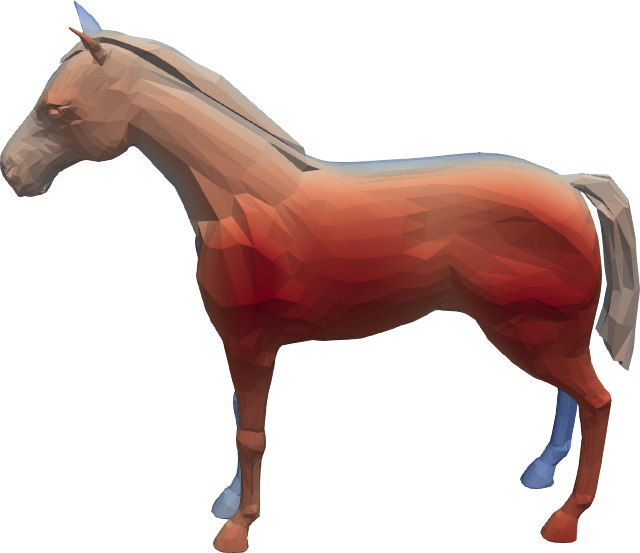} \\
        \tiny \texttt{horse} \\
        $x$ \textsf{function}
    \end{minipage}
    \hspace{5pt}
    \begin{minipage}[m]{0.825\linewidth}
        \subfloat[GASP Boundary\label{fig:i1-BvC:ours:boundary}]{\includegraphics[width=0.24\linewidth]{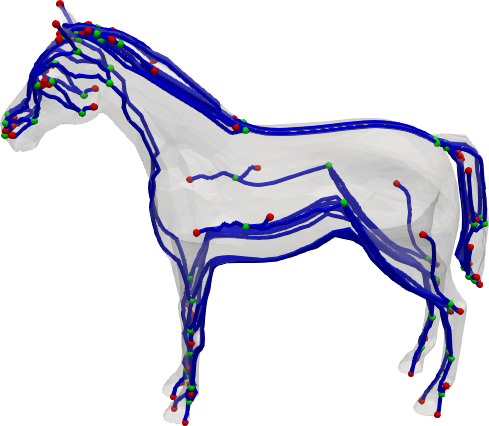}}
        \hfill
        \subfloat[GASP Interior\label{fig:i1-BvC:ours:interior}]{\includegraphics[width=0.24\linewidth]{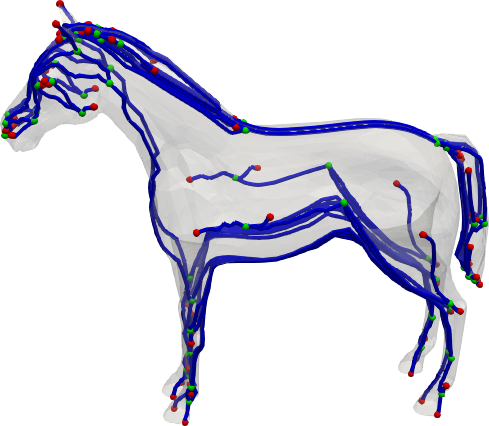}}
        \hfill
        \subfloat[GB (TTK 5/15)\label{fig:i1-BvC:ttk:10}]{\includegraphics[width=0.24\linewidth]{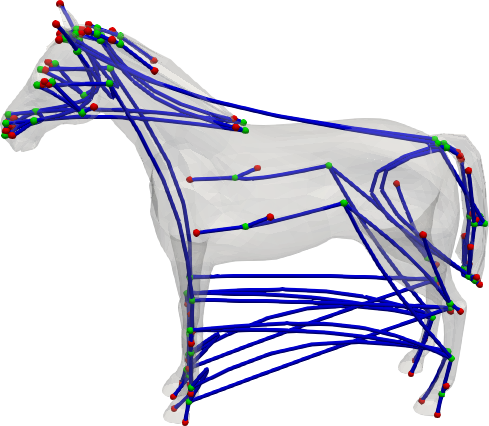}}
        \hfill
        \subfloat[GB (TTK 15/15)\label{fig:i1-BvC:ttk:15}]{\includegraphics[width=0.24\linewidth]{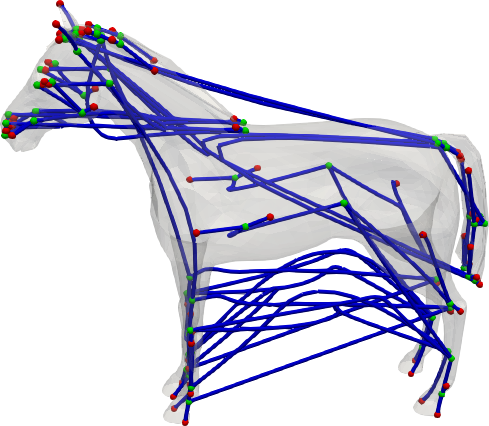}}
    \end{minipage}
    \begin{minipage}[m]{1pt}
        \hspace{-123pt} \scalebox{0.75}{\tiny $S\mathbin{=}0.05$}

        \vspace{-6pt}
        \hspace{-180pt} \scalebox{0.75}{\tiny $S\mathbin{=}0.05$}    
        \hspace{40pt} \scalebox{0.75}{\tiny $B\mathbin{=}0.05$}
        \vspace{46pt}
    \end{minipage}

    { \color{gray}\rule{0.8\linewidth}{0.4pt} }
    
    \vspace{5pt}
    \begin{minipage}[m]{0.675\linewidth}
        \begin{minipage}[m]{0.11\linewidth} 
            \centering 
            \includegraphics[width=\linewidth]{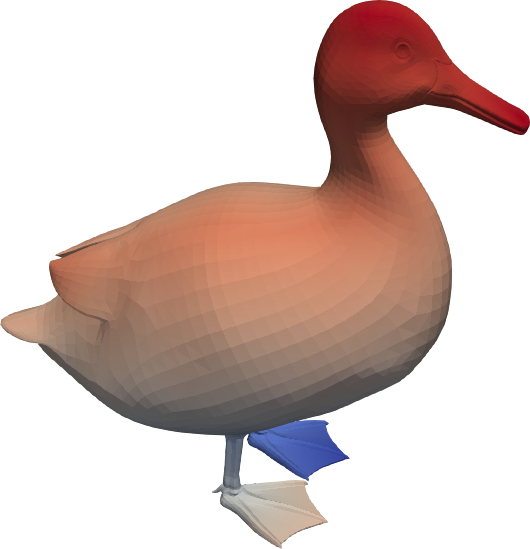} \\ 
            \rotatebox{90}{\tiny \hspace{2pt} \texttt{duck}}
            \rotatebox{90}{\tiny \hspace{0pt} $bottom$}
            \rotatebox{90}{\tiny \hspace{2pt} \textsf{func}}
            \vspace{5pt}
        \end{minipage}%
        \hfill
        \begin{minipage}[m]{0.85\linewidth}
            \hspace{0pt}
            \subfloat[Spac.~$S\mathbin{=}0.025$\label{fig:i1:cs-duck-0.025}]
            {\hspace{5pt}\includegraphics[width=0.265\linewidth]{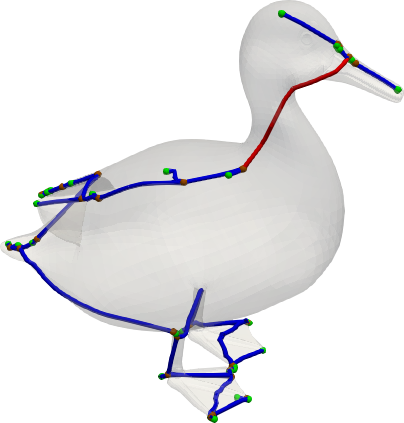}\hspace{5pt}}
            \hfill
            \subfloat[Spac.~$S\mathbin{=}0.05$\label{fig:i1:cs-duck-0.05}]
            {\hspace{5pt}\includegraphics[width=0.265\linewidth]{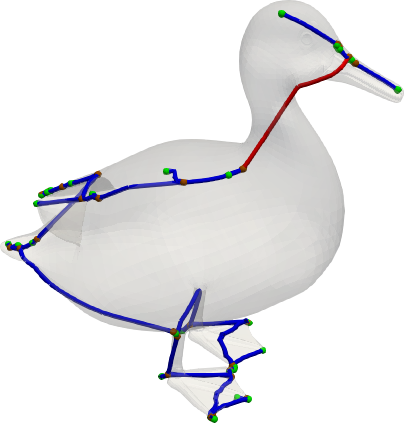}\hspace{5pt}}
            \hfill
            \subfloat[Spac.~$S\mathbin{=}0.1$\label{fig:i1:cs-duck-0.1}]
            {\hspace{2pt}\includegraphics[width=0.265\linewidth]{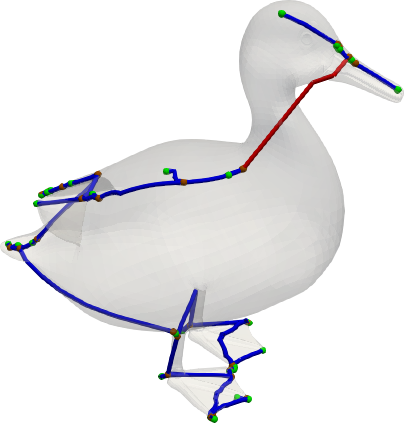}\hspace{2pt}}
            \hfill
        \end{minipage}
    
        \begin{minipage}[m]{0.11\linewidth}
            \centering
            \includegraphics[width=1\linewidth]{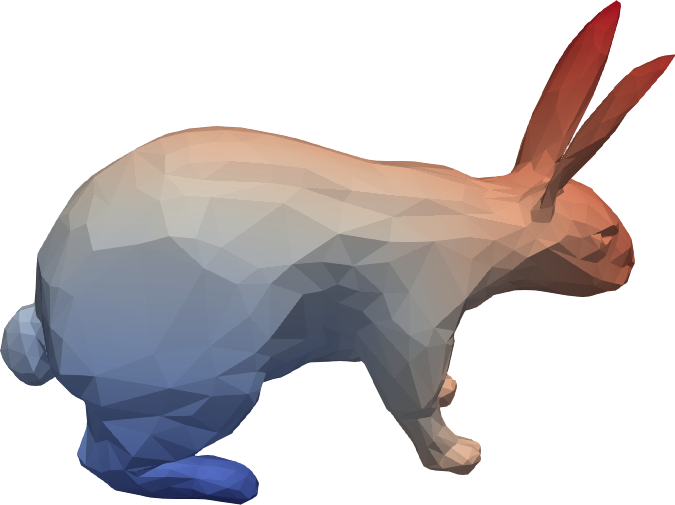} \\
            \rotatebox{90}{\tiny \hspace{0pt} \texttt{rabbit}}
            \rotatebox{90}{\tiny \hspace{1pt} $bottom$}
            \rotatebox{90}{\tiny \hspace{3pt} \textsf{func}}
        \end{minipage}
        \hfill
        \begin{minipage}[m]{0.85\linewidth}
            \hspace{0pt}
            \subfloat[Buffer $B\mathbin{=}0.0$\label{fig:i1:buffer-0}]{\includegraphics[width=0.32\linewidth]{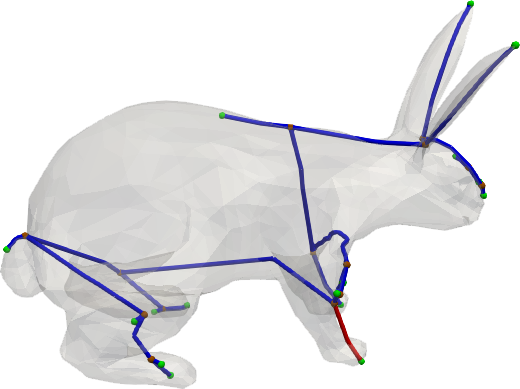}}
            \hfill
            \subfloat[Buffer $B\mathbin{=}0.025$\label{fig:i1:buffer-0.025}]{\includegraphics[width=0.32\linewidth]{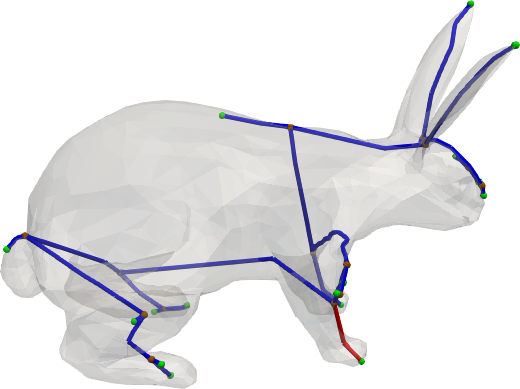}}
            \hfill
            \subfloat[Buffer $B\mathbin{=}0.05$\label{fig:i1:buffer-0.05}]{\includegraphics[width=0.32\linewidth]{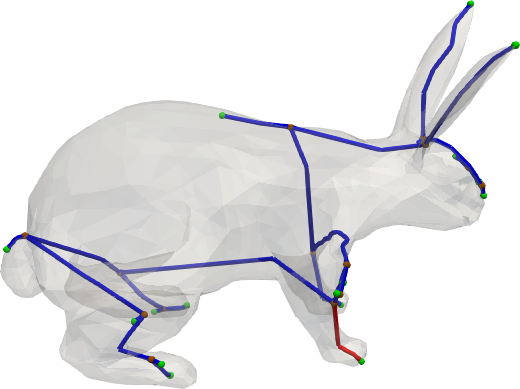}}
        \end{minipage}
        \begin{minipage}[m]{1pt}
            \hspace{-20pt} \scalebox{0.75}{\tiny $S\mathbin{=}0.1$}
            \vspace{-20pt}
        \end{minipage}        
    \end{minipage}
    \hfill
    \begin{minipage}[m]{0.275\linewidth}

        \begin{minipage}[m]{0.195\linewidth}
            \includegraphics[width=1.3\linewidth]{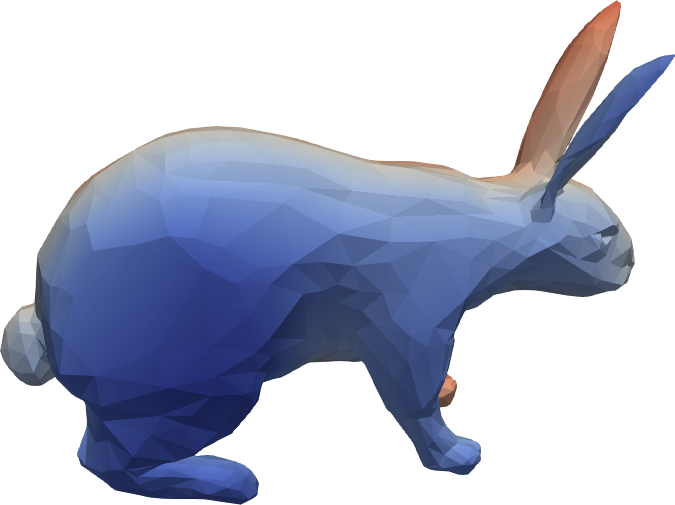}
            \rotatebox{90}{\tiny \hspace{0pt} \texttt{rabbit}}
            \rotatebox{90}{\tiny \hspace{1pt} $x$ \textsf{func}}
            \vspace{60pt}
        \end{minipage}
        \hfill
        \begin{minipage}[m]{0.75\linewidth}
            \hspace{0pt}
            \subfloat[Spacing $S\mathbin{=}0.025$\label{fig:i1:cs-rabbit-0.025}]
            {\includegraphics[width=\linewidth]{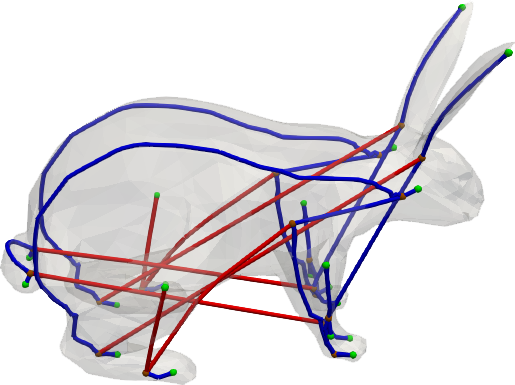}}

            \hspace{0pt}
            \subfloat[Spacing $S\mathbin{=}0.05$\label{fig:i1:cs-rabbit-0.05}]
            {\includegraphics[width=\linewidth]{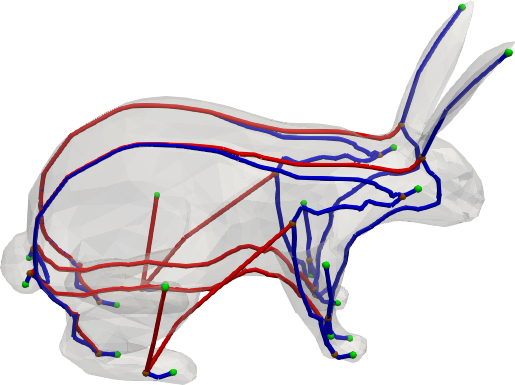}}     
        \end{minipage}
        \vspace{0pt}
    \end{minipage}

    \caption{Examples for Property 1, arcs confined to the model.
    \textit{Top}: Examples comparing (a-b)~GASP arcs that remain confined to the model's boundary, while (c-d)~some GB arcs go far beyond it.
    \textit{Bottom}:
    (e-f)~At lower contour spacing, the Reeb graph is more likely to remain inside the model, (g)~while at larger contour spacing values, arcs are more likely to extend outside the model (e.g., near the neck).
    (h)~When buffer size is $0$, arcs around the front legs are partially outside the boundary (not easily visible in the figures). (i-j)~When buffer size increases to $0.025$ and $0.05$, arcs move further inside the model.
    (k)~One GASP limitation is that for critical points with small function distance and large Euclidean distance, long arcs can appear. (l)~Interestingly and somewhat counterintuitively, as contour spacing increases (i.e., fewer contours are used), these long arcs disappear because topological cylinders switch from regular features to thin features.
    Complete Reeb graphs include \textit{\textcolor{red}{red}} and \textit{\textcolor{blue}{blue}} arcs; \textit{\textcolor{red}{red}} arcs highlight comparisons. Additional examples in supplement.
    }

    \label{fig:issue1:Boundary_vs_contour:Horse-x}

\end{figure}

The first property for visualizing Reeb graphs is that arcs should ideally be contained within the model. For example, in \Cref{fig:i1-BvC:ttk:10}-\ref{fig:i1-BvC:ttk:15}, the GB Reeb graph generated by TTK has several arcs outside of the \texttt{horse} mesh. The position of these arcs outside of the mesh is important, as it may misrepresent the function, which travels on the surface of the mesh, and causes unnecessary visual clutter.

\begin{figure}[!t]
    \centering

    \begin{minipage}[m]{0.025\linewidth}
        \rotatebox{90}{\hspace{10pt}\tiny \textsf{Mean Arc Outside Ratio}}  %
    \end{minipage}
    \begin{minipage}[m]{0.5\linewidth}    
    \subfloat[Comparison of GASP and GB (\textit{x} function)\label{fig:vis-i1-CwT-x}]{\includegraphics[width=\linewidth]{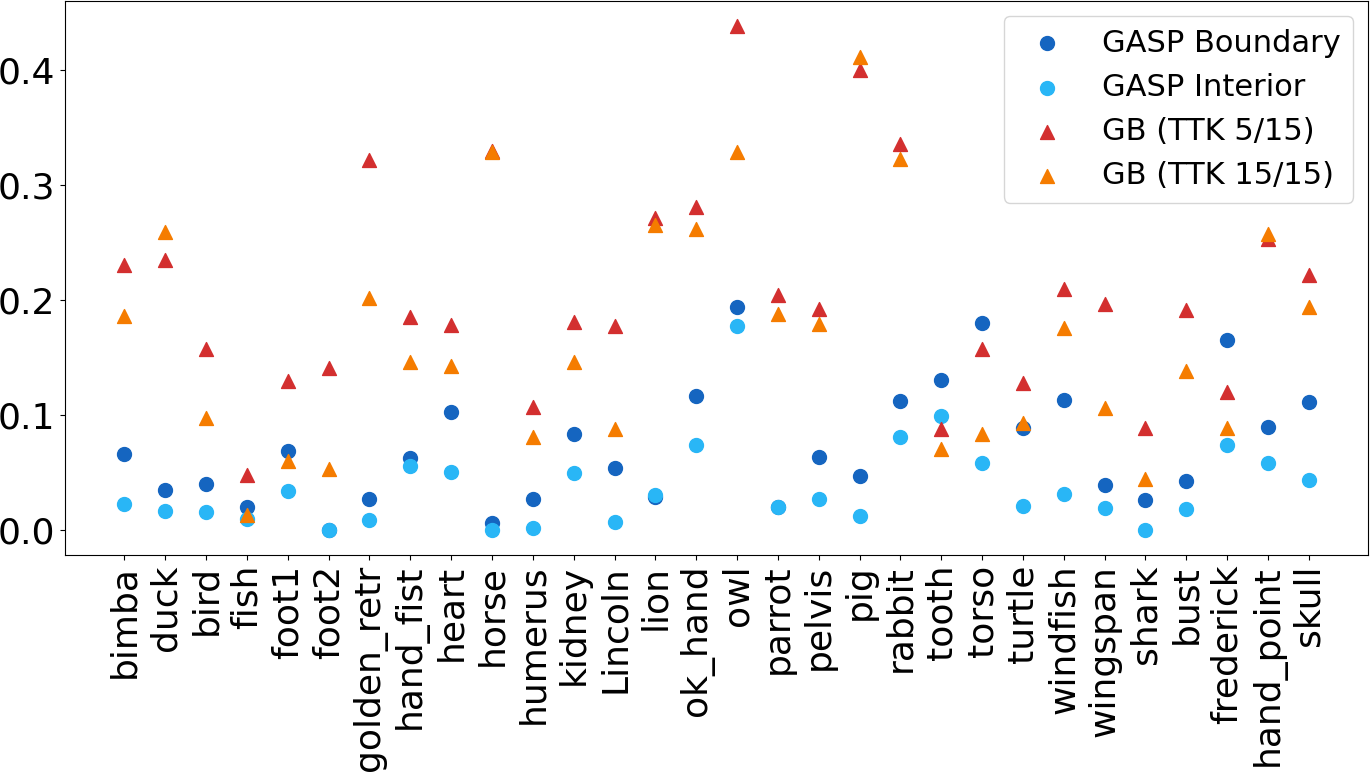}}
    \end{minipage}
    \hfill
    \begin{minipage}[m]{0.29\linewidth}
        \subfloat[Contour Spacing (\textit{bottom} func)\label{fig:vis-i1-cs-bottom}]{\includegraphics[width=\linewidth]{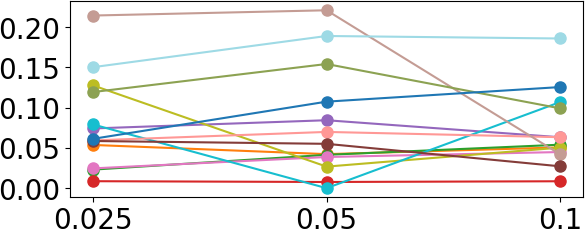}}

        \subfloat[Buffer Size (\textit{bottom} func)\label{fig:issue1:buffer_size}]{\includegraphics[width=\linewidth]{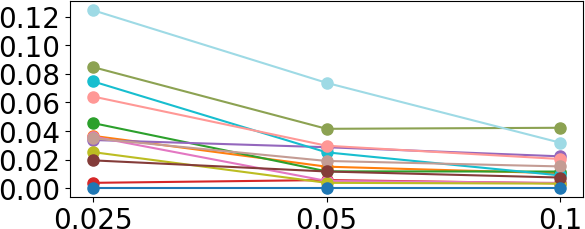}}        
    \end{minipage}
    \hfill
    \begin{minipage}[m]{0.125\linewidth}
        \includegraphics[width=\linewidth]{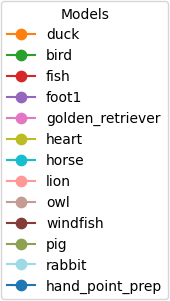}
        \vspace{7pt}
    \end{minipage}

    \begin{minipage}[m]{0.025\linewidth}
        \rotatebox{90}{\hspace{10pt}\tiny \textsf{Arc Area Outside}}  %
    \end{minipage}
    \begin{minipage}[m]{0.5\linewidth}    
    \subfloat[Comparison of GASP and GB (\textit{x} function)\label{fig:vis-i1-area-CwT-x}]{\includegraphics[width=\linewidth]{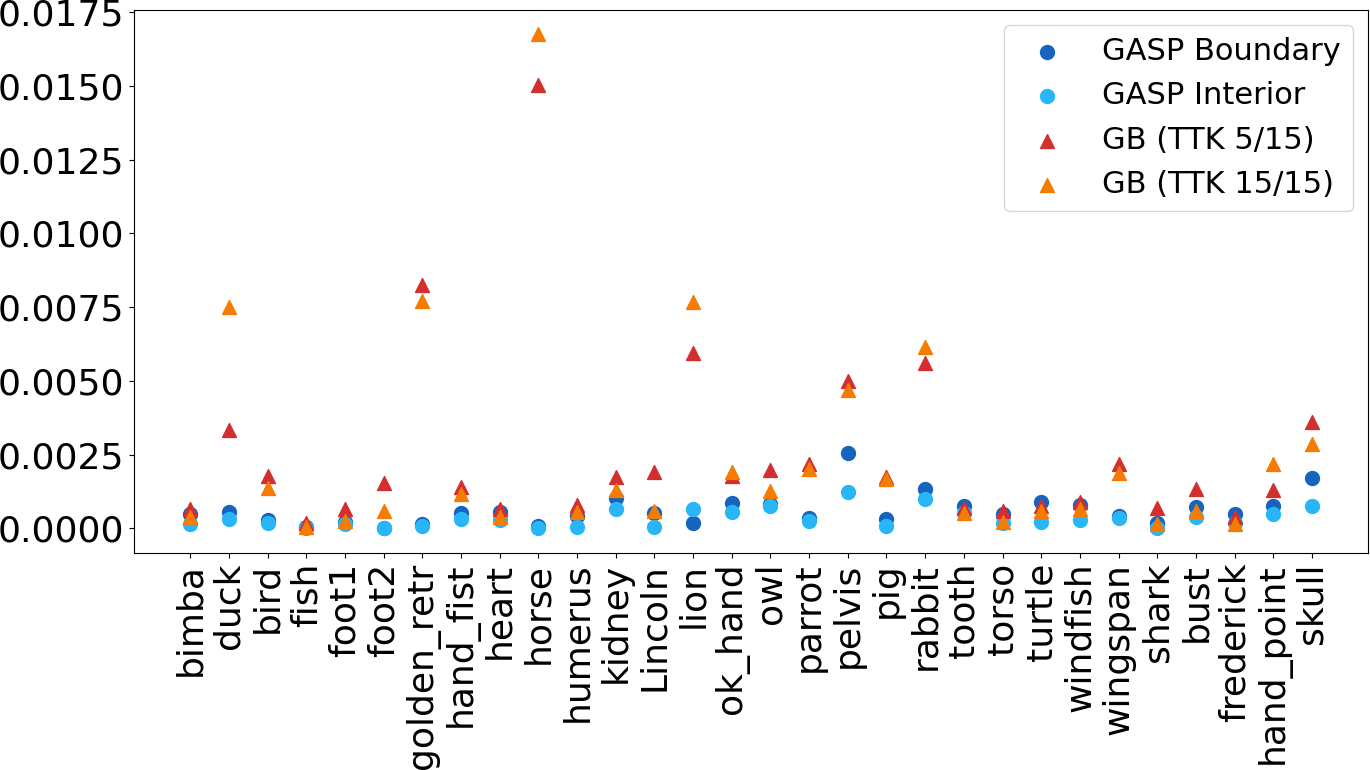}}
    \end{minipage}
    \hfill
    \begin{minipage}[m]{0.29\linewidth}
        \subfloat[Contour Spacing (\textit{bottom} func)\label{fig:vis-i1-area-cs-bottom}]{\includegraphics[width=\linewidth]{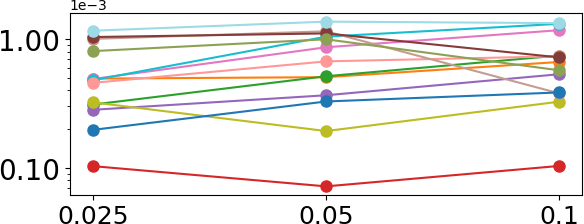}}

        \subfloat[Buffer Size (\textit{bottom} func)\label{fig:issue1:area-buffer_size}]{\includegraphics[width=\linewidth]{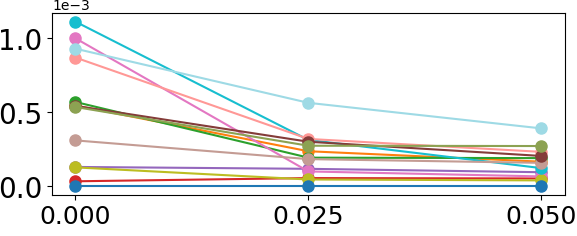}} 
    \end{minipage}
    \hfill
    \begin{minipage}[m]{0.125\linewidth} \hfill
    \end{minipage}

    \caption{
    Quantitative analysis for Property 1, arc confined to the model, using the (top) mean ratio and (bottom) total area of arcs outside the model (lower is better). 
    (a/d)~Comparison of GASP and GB across all models shows that for the majority, both configurations of GASP outperformed GB. 
    (b/e)~Charts illustrating the influence of the contour spacing on GASP for models used in the paper. Generally, as contour spacing increases, slightly more of the arcs end up outside of the model. The models with irregular patterns arise from variations in the number of thin and regular features as contour spacing changes.
    (c/f)~Charts illustrating the influence of the buffer size on GASP for models in the paper. Generally, buffer size increases result in a gradual decrease in arcs outside of the model, suggesting that larger buffer sizes help constrain arcs within the model's surface.
    Additional examples in supplement.
    }
    \label{fig:issue1:BvC-visualization}

\end{figure}

\paragraph{Comparison with GB}
GASP uses strategies that try to constrain the Reeb graph to the mesh surface using the boundary approach (see \Cref{fig:i1-BvC:ours:boundary}) or inside the mesh when using the interior approach (see \Cref{fig:i1-BvC:ours:interior}), while GB Reeb graphs extend beyond the boundary (see \Cref{fig:i1-BvC:ttk:10}-\ref{fig:i1-BvC:ttk:15}). 
We quantitatively evaluate the effectiveness of GASP and GB by measuring the ratio of arc $Length_{outside} / Length_{Total}$ ($0$ means the arc is entirely interior, while $1$ indicates the arc is entirely outside the mesh). The ratio is calculated per arc, and we present the average values per Reeb graph. 
To measure how far an arc extends beyond the model, we calculate the sum of the areas of all arc segments outside the mesh surface. In particular, if an arc segment of length $l$ whose endpoints are $d_1$ and $d_2$ units away from the boundary of the mesh, respectively, the area is $a = {l} \cdot \frac{d_1 + d_2}{2}$.
Lower area values indicate the extended arcs are closer to the mesh.
The scatter plots in \Cref{fig:vis-i1-CwT-x} (ratio of arcs outside the model) and \Cref{fig:vis-i1-area-CwT-x} (area of arcs outside the model) show the improved performance of GASP compared to the 
two TTK variants of GB across different models for the same function (scatter plots for additional functions are provided in the supplement). Furthermore, the GASP interior method is generally more effective than the boundary approach.

\paragraph{Effects of Contour Spacing and Buffer Size}
\uline{Contour Spacing:}
With GASP, the Reeb graph arc can go outside the model in two situations. It occurs most frequently when the Reeb graph arc covers non-convex portions of mesh, but it can also occur when the Reeb graph arc is traveling through thinner parts of the model (see \cref{fig:i1:cs-duck-0.1}). 
Reducing contour spacing (i.e., increasing the number of contours per arc) can mitigate these issues (see \cref{fig:i1:cs-duck-0.025}-\ref{fig:i1:cs-duck-0.05}).
The quantitative results in \Cref{fig:vis-i1-cs-bottom} and \Cref{fig:vis-i1-area-cs-bottom} show that across a variety of models, larger contour spacing generally results in a slight increase in the arcs being outside of the model. The outliers will be discussed in the forthcoming limitations.
\uline{Buffer Size:}
The buffer causes candidate points to be positioned deeper inside the mesh (see \cref{impl:candidate:points}), causing the Reeb graph to shift inward. Consequently, a larger buffer size confines more of the Reeb graph within the model compared to a smaller buffer size.
This effect is illustrated in \Cref{fig:i1:buffer-0.025}-\ref{fig:i1:buffer-0.05}. With a small buffer size (see \cref{fig:i1:buffer-0}), some arcs extend outside the model. As the buffer size increases (see \cref{fig:i1:buffer-0.025}-\ref{fig:i1:buffer-0.05}), the arcs shift inward. \Cref{fig:issue1:buffer_size} and \Cref{fig:issue1:area-buffer_size} further quantify this negative correlation, showing that a larger buffer size reduces the mean percentage of arcs extending outside the model.

\paragraph{Summary and Limitations}
Overall, GASP outperformed the TTK implementation of GB in confining Reeb graph arcs within the model. 
However, GASP can struggle when a large arc segment passes through a non-convex region. This can happen when a pair of critical points are close in function distance but far in Euclidean distance (see \Cref{fig:i1:cs-rabbit-0.025}). This problem can be mitigated by either decreasing the contour spacing, which adds more contours, or by increasing the contour spacing, which causes those features to switch from regular features to thin features (see \Cref{fig:i1:cs-rabbit-0.05}). It would, of course, be possible to heuristically force long edges to switch to thin features, but we chose not to make that a core feature of our implementation. These edges are also responsible for the few irregular patterns observed in \Cref{fig:vis-i1-cs-bottom}.

\begin{figure}[!b]
    \centering
    \begin{minipage}[m]{0.11\linewidth}
        \centering
        \includegraphics[width=\linewidth]{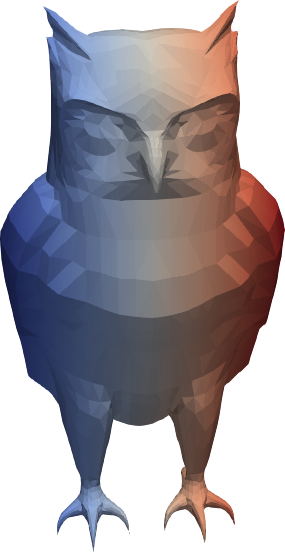} \\
        \tiny \texttt{owl} \\
        $x$ \textsf{function}        
    \end{minipage}%
    \hspace{5pt}
    \begin{minipage}[m]{0.8\linewidth}
        \subfloat[GASP Boundary\label{fig:i2:BvC:ours:boundary}]{\hspace{3pt}\includegraphics[trim=185pt 95pt 185pt 75pt, clip, width=0.21\linewidth]{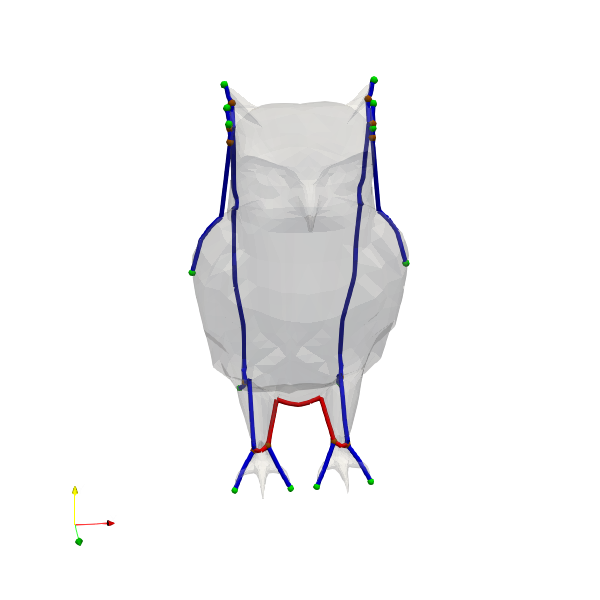}\hspace{3pt}}
        \hfill
        \subfloat[GASP Interior\label{fig:i2:BvC:ours:interior}]{\includegraphics[trim=185pt 95pt 185pt 75pt, clip, width=0.21\linewidth]{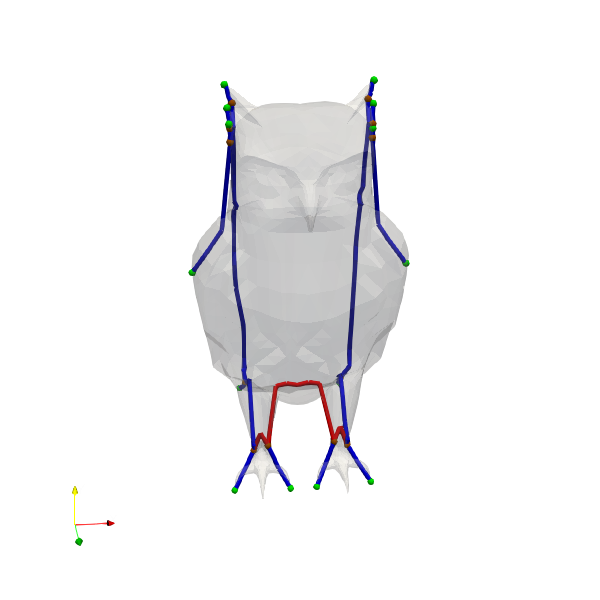}}
        \hfill
        \subfloat[GB (TTK 5/15)\label{fig:i2:BvC:ttk:10}]{\includegraphics[trim=185pt 95pt 185pt 75pt, clip, width=0.21\linewidth]{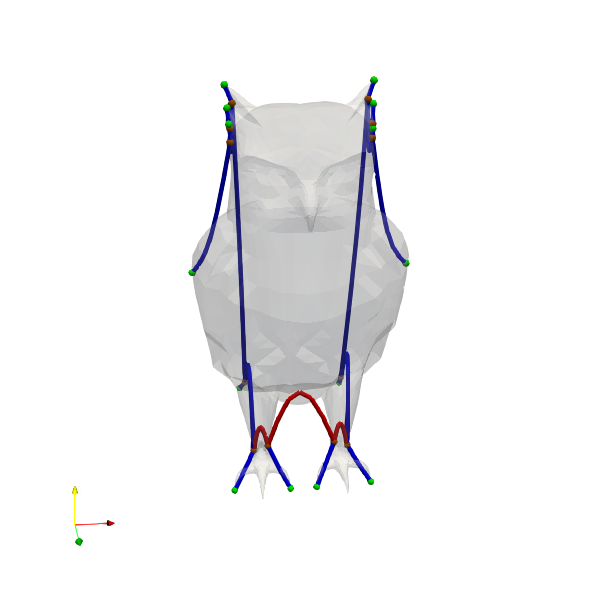}}
        \hfill
        \subfloat[GB (TTK 15/15)\label{fig:i2:BvC:ttk:15}]{\includegraphics[trim=185pt 95pt 185pt 75pt, clip, width=0.21\linewidth]{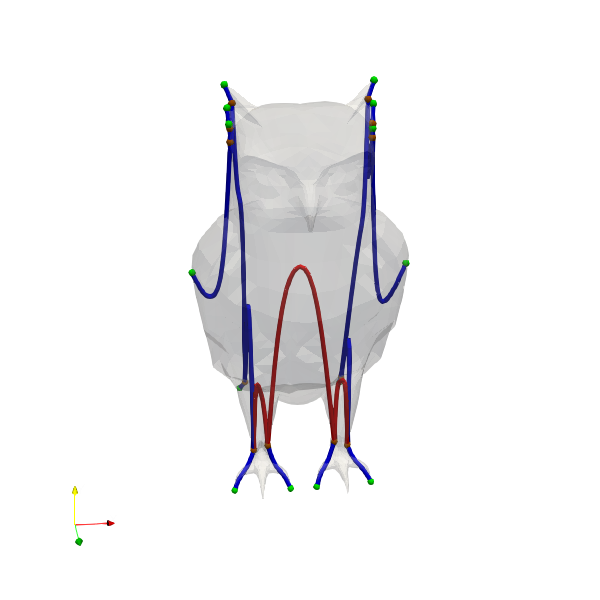}}
    \end{minipage}%
    \begin{minipage}[m]{1pt}
        \vspace{5pt}
        \hspace{-158pt} \rotatebox{90}{\scalebox{0.75}{\tiny $S\mathbin{=}0.05$}}

        \vspace{-14pt}
        \hspace{-103pt} \rotatebox{90}{\scalebox{0.75}{\tiny $S\mathbin{=}0.05$, $B\mathbin{=}0.05$}}
        \vspace{60pt}
    \end{minipage}    

    \vspace{-3pt}
    { \color{gray}\rule{0.8\linewidth}{0.4pt} }
    
    \vspace{2pt}
    \begin{minipage}[m]{0.675\linewidth}
        \begin{minipage}[m]{0.11\linewidth}
            \centering
            \includegraphics[width=1\linewidth]{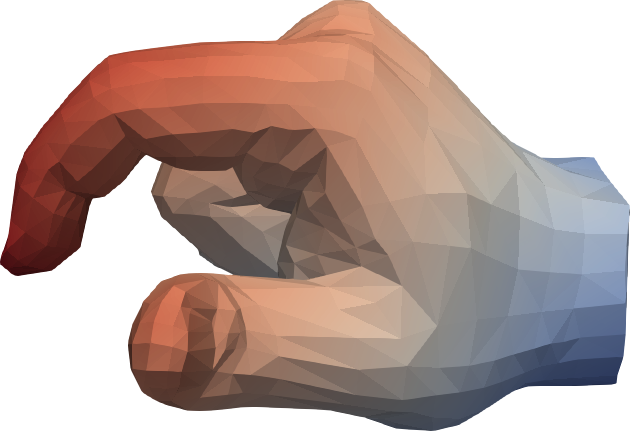}
            \rotatebox{90}{\tiny \hspace{0pt} \texttt{hand point}}
            \rotatebox{90}{\tiny \hspace{2pt} $front$ \textsf{func}}
            \vspace{8pt}
        \end{minipage}
        \hfill
        \begin{minipage}[m]{0.85\linewidth}
            \subfloat[Spac.~$S\mathbin{=}0.025$\label{fig:i2:cs-hand-point-prep-0.025}]{\hspace{2pt}\includegraphics[width=0.3\linewidth]{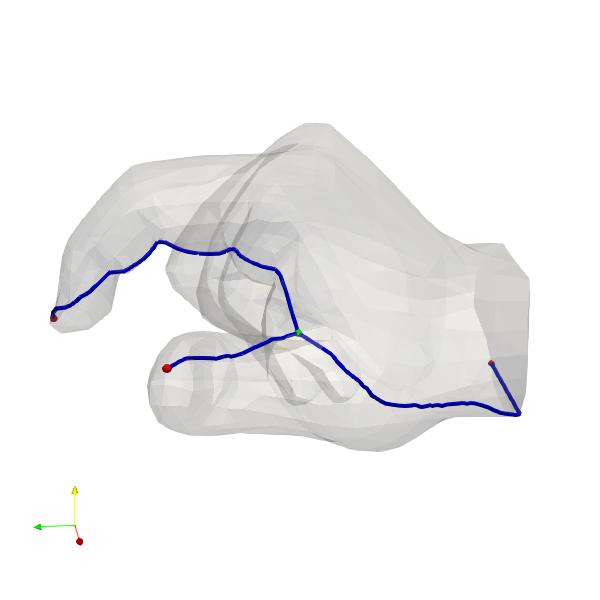}\hspace{2pt}}
            \hfill
            \subfloat[Spac.~$S\mathbin{=}0.05$\label{fig:i2:cs-hand-point-prep-0.05}]{\includegraphics[width=0.3\linewidth]{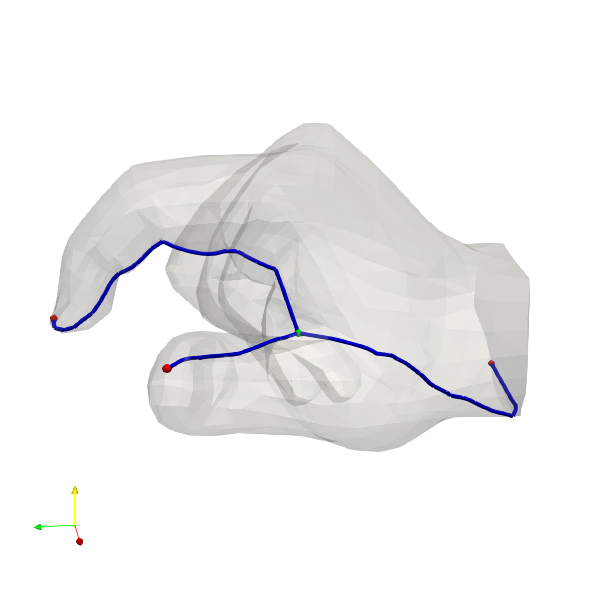}}
            \hfill
            \subfloat[Spac.~$S\mathbin{=}0.1$\label{fig:i2:cs-hand-point-prep-0.1}]{\includegraphics[width=0.3\linewidth]{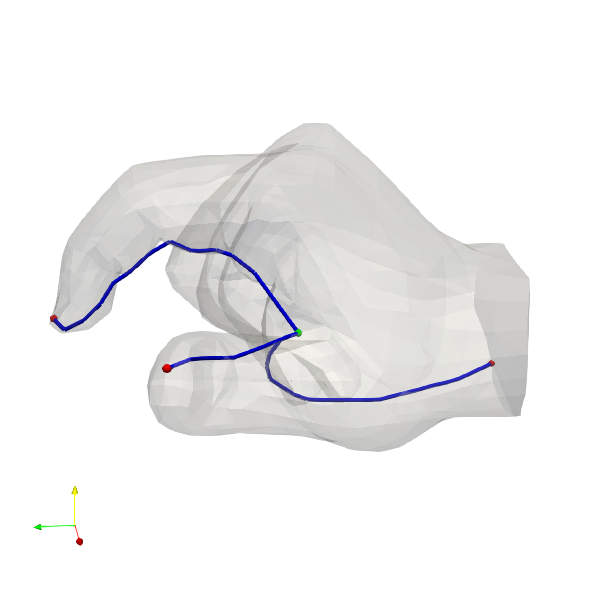}}
            \hfill
        \end{minipage}
    
        \begin{minipage}[m]{0.11\linewidth}
            \centering
            \includegraphics[width=1\linewidth]{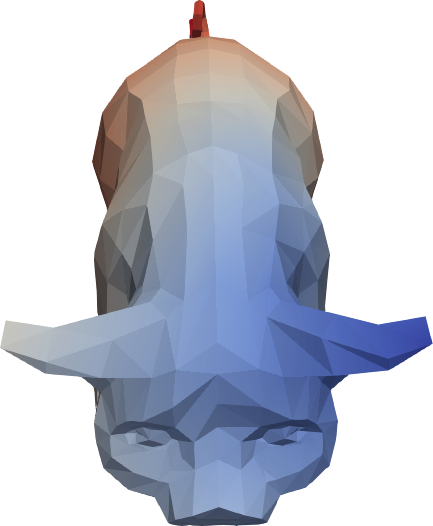}
            \rotatebox{90}{\tiny \hspace{1pt} \texttt{pig}}
            \rotatebox{90}{\tiny \hspace{0pt} $right$}
            \rotatebox{90}{\tiny \hspace{1pt} \textsf{func}}            
            \vspace{35pt}
        \end{minipage}
        \begin{minipage}[m]{0.85\linewidth}
            \subfloat[Buffer $B\mathbin{=}0.0$\label{fig:i2:buffer-0}]{\includegraphics[width=0.32\linewidth]{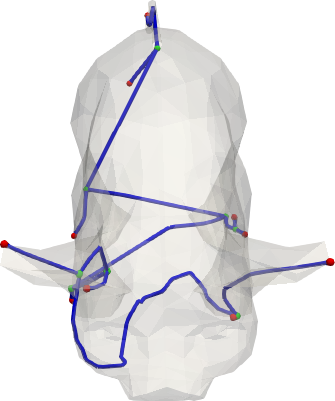}}
            \hfill
            \subfloat[Buffer $B\mathbin{=}0.025$\label{fig:i2:buffer-0.025}]{\includegraphics[width=0.32\linewidth]{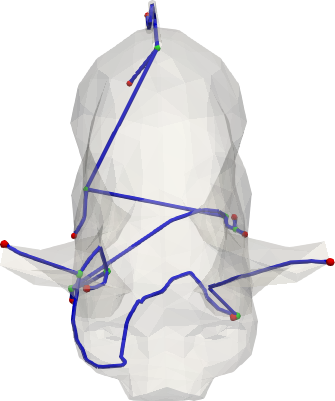}}
            \hfill
            \subfloat[Buffer $B\mathbin{=}0.05$\label{fig:i2:buffer-0.05}]{\includegraphics[width=0.32\linewidth]{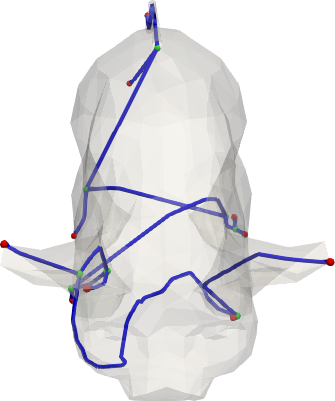}}
        \end{minipage}
        \begin{minipage}[m]{1pt}
            \hspace{-20pt} \scalebox{0.75}{\tiny $S\mathbin{=}0.1$}
            \vspace{-35pt}
        \end{minipage}        
    \end{minipage}
    \hfill
    \begin{minipage}[m]{0.275\linewidth}
        \begin{minipage}[m]{0.225\linewidth}
            \centering
            \includegraphics[width=1\linewidth]{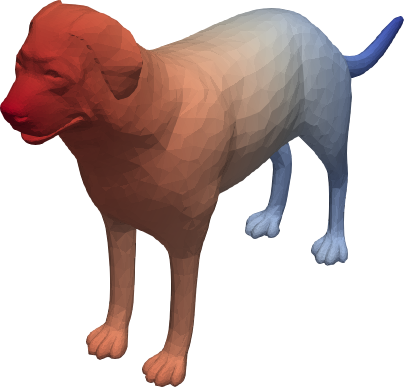}
            \rotatebox{90}{\tiny \hspace{5pt} \texttt{golden}}
            \rotatebox{90}{\tiny \hspace{0pt} \texttt{retriever}}
            \rotatebox{90}{\tiny \hspace{6pt} $z$ \textsf{func}}
            \vspace{60pt}
        \end{minipage}%
        \hfill
        \begin{minipage}[m]{0.71\linewidth}
            \subfloat[GASP Boundary\label{fig:i2:Limitation:ours-boundary}]{\includegraphics[width=\linewidth]{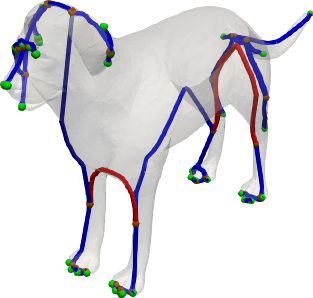}}

            \subfloat[GB (TTK 15/15)\label{fig:i2:Limitation:TTK}]{\includegraphics[width=\linewidth]{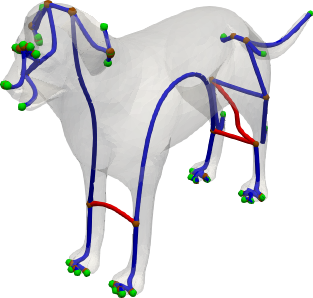}}
        \end{minipage}
    \end{minipage}

    \caption{
    Examples for Property 2, compact arc.
    \textit{Top}:
    Examples comparing GASP and GB show that (c-d) GB can produce Reeb graph arcs that are unnecessarily longer, while GASP arcs cover a minimal distance.
    \textit{Bottom}:
    (e-g)~Compared to smaller contour spacing, e.g., (e)~$0.025$, (f)~$0.05$, larger contour spacing, e.g., (g)~$0.1$, causes arcs to be more subdivided, resulting in slightly shorter arc lengths.
    (h-j)~Increases in buffer size from (h)~$0.0$ to (i)~$0.025$ to (j)~$0.05$ increases arc length by pushing arcs toward the interior of the model. 
    (k-l)~One limitation of (k)~GASP is that it can produce long arcs in non-convex areas of high curvature compared to (g)~GB. However, the GB arcs are only shorter because they go outside of the mesh, highlighting an important trade-off.
    Complete Reeb graphs include \textit{\textcolor{red}{red}} and \textit{\textcolor{blue}{blue}} arcs; \textit{\textcolor{red}{red}} arcs highlight comparisons. For additional models and functions, see supplement.
    }
    \label{fig:issue2:Boundary_vs_Contour:Owl}
\end{figure}

\subsection{Property 2: Compact Arcs}
\label{eval:issue2}

To reduce visual clutter and make arcs easier to track, a Reeb graph should be as compact as possible by minimizing arc length.

\begin{figure}[!t]
    \centering
    \scriptsize

    \begin{minipage}[m]{0.025\linewidth}
        \rotatebox{90}{\hspace{10pt}{\tiny \textsf{Mean Arc Length Ratio}}}  %
    \end{minipage}
    \begin{minipage}[m]{0.5\linewidth}
        \subfloat[Comparison of GASP and GB (\textit{x} function)\label{fig:issue2:BvC}]{\includegraphics[width=\linewidth]{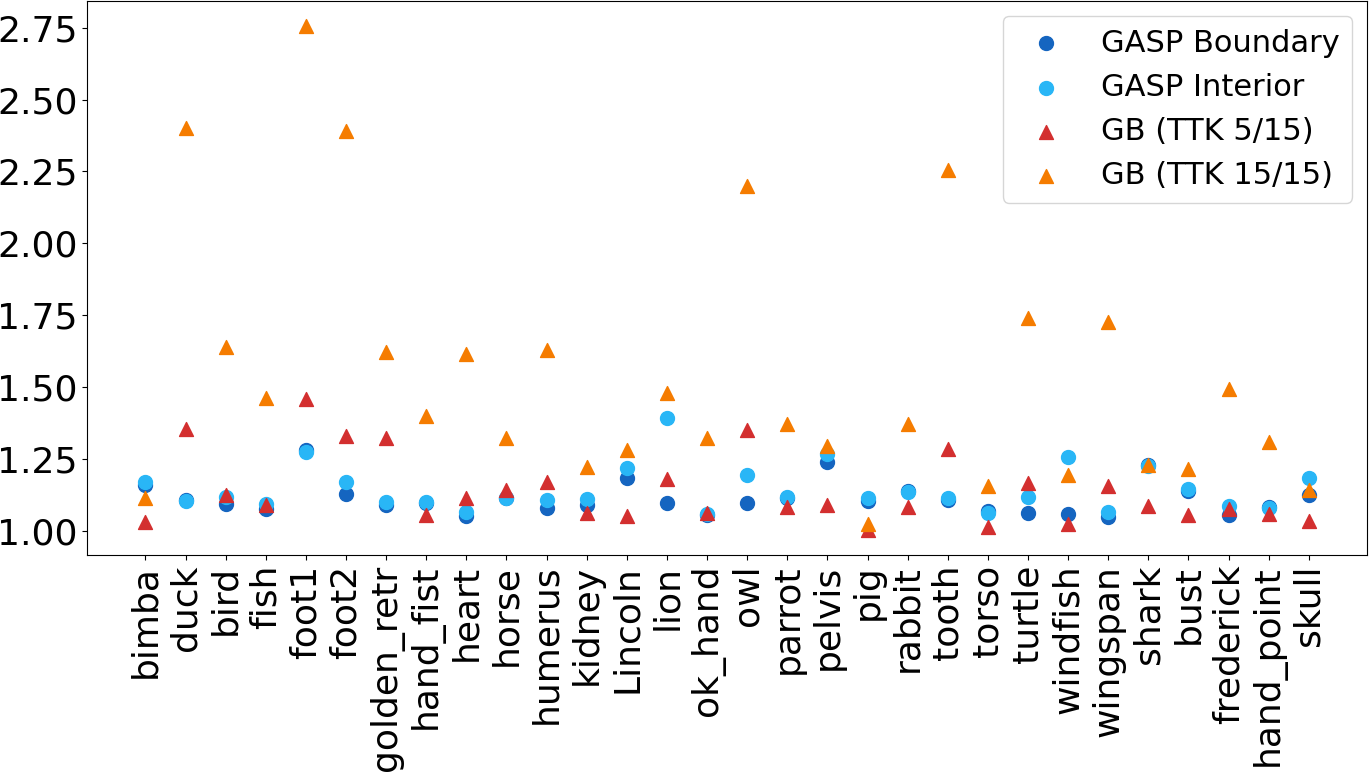}}
    \end{minipage}
    \hfill
    \begin{minipage}[m]{0.29\linewidth}
        \subfloat[Contour Spacing (\textit{front} func)\label{fig:issue2:CS:Vis:all}]{\includegraphics[width=\linewidth]{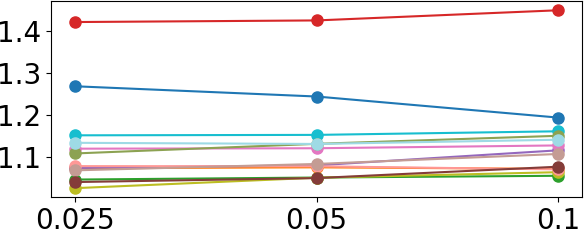}}
        
        \subfloat[Buffer Size (\textit{right} func)\label{fig:issue2:buffer_size:all}]{\includegraphics[width=\linewidth]{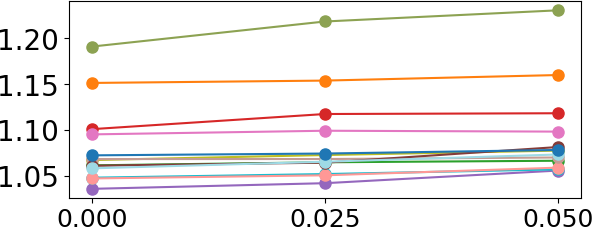}}        
    \end{minipage}
    \hfill
    \begin{minipage}[m]{0.125\linewidth}
        \includegraphics[width=\linewidth]{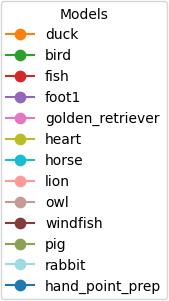}
        \vspace{10pt}
    \end{minipage}

    \caption{Quantitative analysis for Property 2, compact arc, using the mean ratio of the arc length to the distance between critical points  (lower is better).
    (a)~Comparison of GASP and GB across all models shows that for the majority of configurations, GASP outperformed GB. 
    (b)~Chart illustrating the influence of contour spacing on GASP for the subset of models used in the paper. Increases in spacing result in increases in arc length, which is mostly attributed to an increase in thin features, which tend to be longer than their regular counterparts.
    (c)~Chart illustrates the influence of buffer size on GASP for the same subset of models. 
    A slight increase in arc length is observed. As buffer size increases, arcs are pushed further inside of the model, causing their length to increase.
    Additional examples are in the supplement.
    }

\end{figure}

\paragraph{Comparison with GB}
When TTK creates a Reeb graph of the model using GB, it transforms the straight lines into arcs by utilizing arc sampling and smoothing. The length of the arcs increases as the arc sampling parameter increases. Moreover, in some cases, adjusting the iteration parameter allows the arc to position itself within the model at the cost of increased length. \cref{fig:TTK-parameters} shows an example where increasing the sampling causes the arcs to respect the model boundary while also getting longer. 
In contrast, GASP employs the shortest path to ensure that the path taken is minimal while also respecting the boundary (up to the limits of contour spacing). An example is illustrated in \cref{fig:i2:BvC:ours:boundary}-\ref{fig:i2:BvC:ttk:15}. GB results in several unnecessarily long arcs (highlighted in \textcolor{red}{\textit{red}}), with the less smooth version (see \cref{fig:i2:BvC:ttk:10}) producing shorter arcs than the smoother version (see \cref{fig:i2:BvC:ttk:15}).   
However, both variations of GASP, respect the boundary of the model while keeping their length minimal (see \cref{fig:i2:BvC:ours:boundary}-\ref{fig:i2:BvC:ours:interior}). 
To quantitatively measure how well a Reeb graph minimizes arc length, we measure the total length of each Reeb graph arc and compare it to the Euclidean distance between the associated critical points. The ratio $Length_{Total} / Distance_{CP}$ is calculated per arc, and smaller is generally better ($1$ is a straight edge). We report the average values per Reeb graph. \Cref{fig:issue2:BvC} shows that for almost all cases, GASP arcs are significantly shorter than GB arcs and that the GASP interior approach arcs are shorter than boundary approach arcs.

\paragraph{Contour Spacing and Buffer Size}
\uline{Contour Spacing:}
In principle, increases in the contour spacing should result in a decrease in the length of arcs because of new shorter path opportunities, as can be observed in the \texttt{hand point prep} model in \Cref{fig:i2:cs-hand-point-prep-0.025}-\ref{fig:i2:cs-hand-point-prep-0.1}.
However, in general, increases in contour spacing increase the number of thin features, which tend to be longer than their regular counterparts. Therefore, we often observed an overall increase, albeit small, in average arc length as contour spacing increased, which can be seen in \Cref{fig:issue2:CS:Vis:all}. 
\uline{Buffer Size:}
Increasing buffer size forces Reeb graph arcs to follow paths farther from the surface and critical points, resulting in longer arcs, as shown in \Cref{fig:issue2:buffer_size:all}. 
This trend is observed in the \Cref{fig:i2:buffer-0}-\ref{fig:i2:buffer-0.05}, where arc length slightly increases with buffer size. However, the change in the visualization is subtle and difficult to observe in the image.

\paragraph{Summary and Limitations}
Our evaluation shows that GASP generally produces shorter arcs than GB, and increases in both contour spacing and buffer size generally result in small increases in arc length. 
One important note about these results is that GASP does particularly well on nearby critical points because it connects the critical points more directly. However, some longer arcs, particularly in non-convex areas of high curvature, pose challenges to GASP that may result in producing longer arcs because the model's boundary constrains it. 
\Cref{fig:i2:Limitation:TTK} shows an example where the GB arcs in \textcolor{red}{\textit{red}} are shorter than the GASP arcs in \Cref{fig:i2:Limitation:ours-boundary}.

\begin{figure}[!b]
    \centering
    \begin{minipage}[m]{0.1\linewidth}
        \centering
        \includegraphics[width=\linewidth]{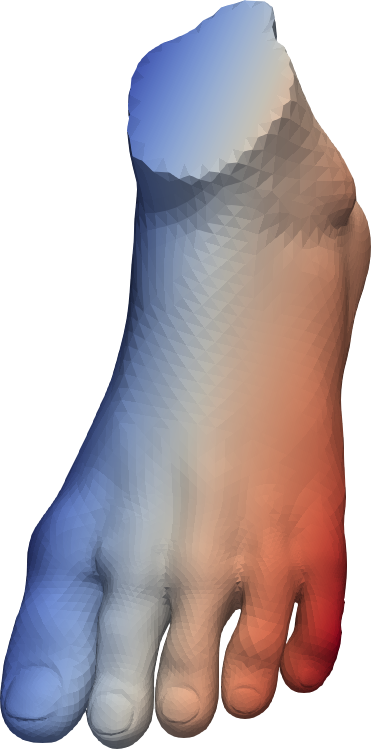} \\
        \tiny \texttt{foot1} \\
        $x$ \textsf{function}   
    \end{minipage}%
    \begin{minipage}[m]{0.75\linewidth}
        \hspace{5pt}
        \subfloat[GASP Boundary\label{fig:i3:BvC:ours:boundary}]{\hspace{4pt}\includegraphics[width=0.195\linewidth]{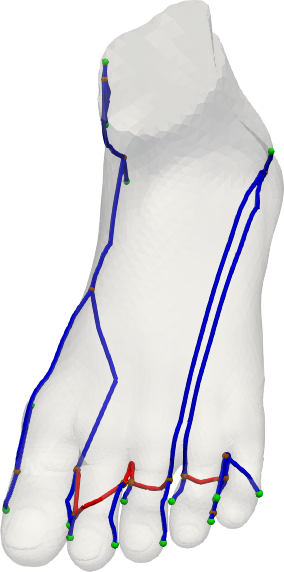}\hspace{4pt}}
        \hfill
        \subfloat[GASP Interior\label{fig:i3:BvC:ours:interior}]{\hspace{2pt}\includegraphics[width=0.195\linewidth]{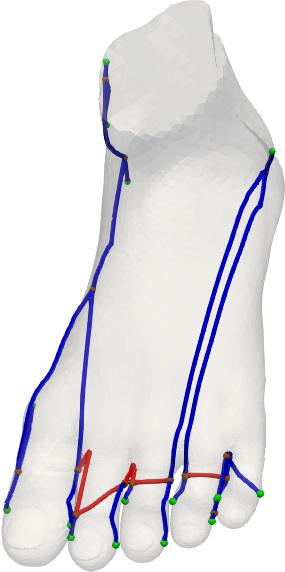}\hspace{2pt}}
        \hfill
        \subfloat[GB (TTK 5/15)\label{fig:i3:BvC:ttk:10}]{\includegraphics[width=0.195\linewidth]{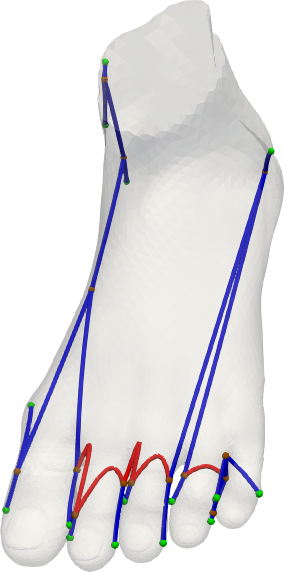}}
        \hfill
        \subfloat[GB (TTK 15/15)\label{fig:i3:BvC:ttk:15}]{\includegraphics[width=0.195\linewidth]{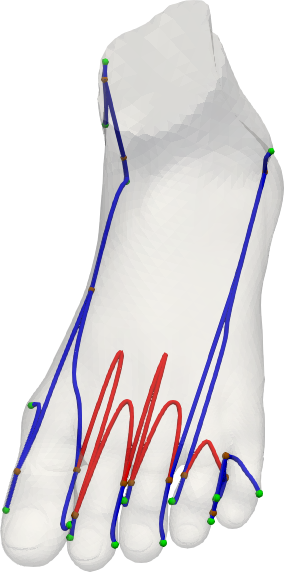}}
        \hspace{5pt}
    \end{minipage}
    \begin{minipage}[m]{1pt}
        \vspace{5pt}
        \hspace{-140pt} \rotatebox{90}{\scalebox{0.75}{\tiny $S\mathbin{=}0.05$}}

        \vspace{-14pt}
        \hspace{-93pt} \rotatebox{90}{\scalebox{0.75}{\tiny $S\mathbin{=}0.05$, $B\mathbin{=}0.05$}}
        \vspace{60pt}
    \end{minipage}   

    \vspace{-3pt}
    { \color{gray}\rule{0.8\linewidth}{0.4pt} }
    
    \vspace{5pt}
    \begin{minipage}[m]{0.125\linewidth}
        \centering
        \includegraphics[width=1\linewidth]{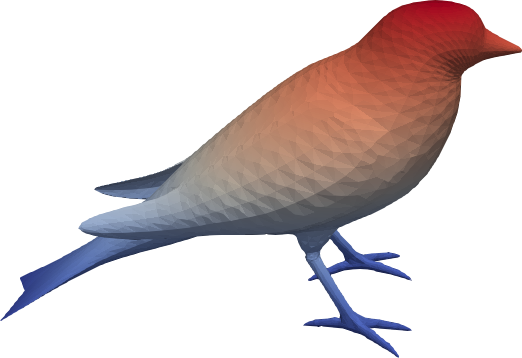} \\
        \tiny \texttt{bird} \\
        $y$ \textsf{function}   
    \end{minipage}%
    \hspace{5pt}
    \begin{minipage}[m]{0.7\linewidth}
        \subfloat[Spacing $S\mathbin{=}0.025$\label{fig:i3:cs-bird-0.025}]{\includegraphics[width=0.31\linewidth]{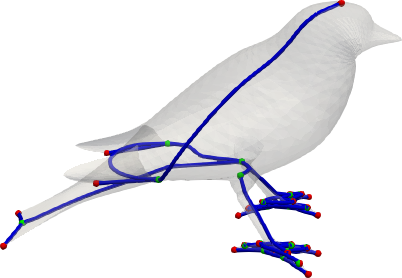}}
        \hfill
        \subfloat[Spacing $S\mathbin{=}0.05$\label{fig:i3:cs-bird-0.05}]{\includegraphics[width=0.31\linewidth]{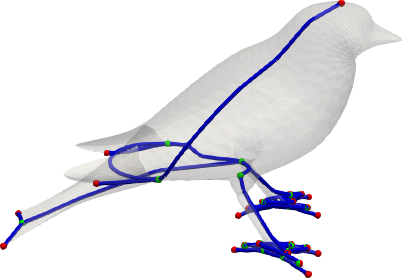}}
        \hfill
        \subfloat[Spacing $S\mathbin{=}0.1$\label{fig:i3:cs-bird-0.1}]{\includegraphics[width=0.31\linewidth]{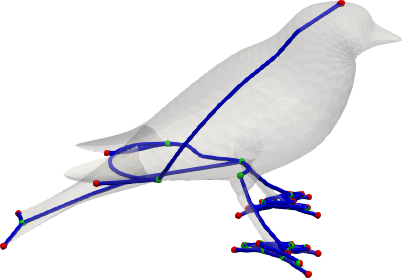}}
        \hfill
    \end{minipage}

    \begin{minipage}[m]{0.125\linewidth}
        \centering
        \includegraphics[width=0.8\linewidth]{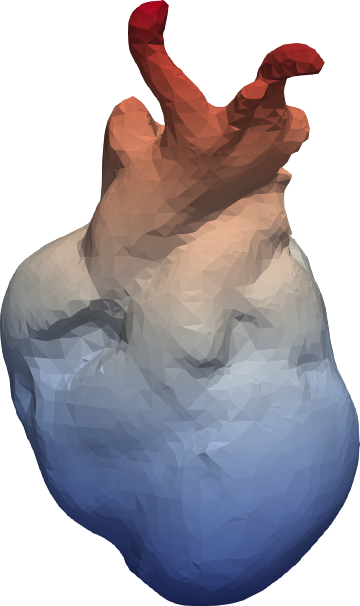} \\
        \tiny \texttt{heart} \\
        $y$ \textsf{function}   
    \end{minipage}
    \hspace{5pt}
    \begin{minipage}[m]{0.7\linewidth}
        \subfloat[Buffer $B\mathbin{=}0.0$\label{fig:i3:buffer-0}]{\hspace{5pt}\includegraphics[width=0.2\linewidth]{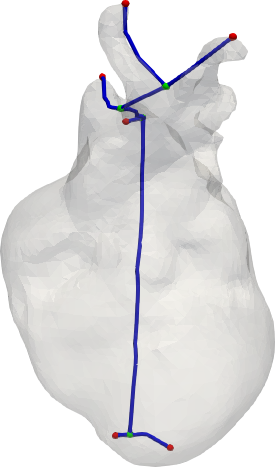}\hspace{5pt}}
        \hfill
        \subfloat[Buffer $B\mathbin{=}0.025$\label{fig:i3:buffer-0.025}]{\hspace{7pt}\includegraphics[width=0.2\linewidth]{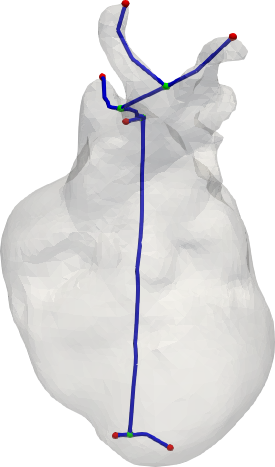}\hspace{7pt}}
        \hfill
        \subfloat[Buffer $B\mathbin{=}0.05$\label{fig:i3:buffer-0.05}]{\hspace{5pt}\includegraphics[width=0.2\linewidth]{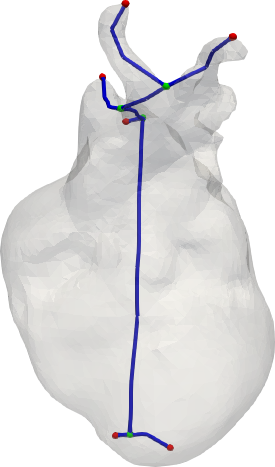}\hspace{5pt}}
        \hspace{15pt}
    \end{minipage}
    \begin{minipage}[m]{1pt}
        \hspace{-25pt} \scalebox{0.75}{\tiny $S\mathbin{=}0.1$}
        \vspace{-35pt}
    \end{minipage}

    \caption{Examples for Property 3, following function gradient.
    \textit{Top}:~Examples of GASP and GB show that the highlighted (c-d)~GB arcs do not do a good job traveling in the direction of the function gradient, while the (a-b)~GASP arcs do.
    \textit{Bottom}:
    (e-g)~Changes in GASP contour spacing have little impact on the alignment of the arcs with the gradient.
    (h-j)~Similarly, changes in buffer size have little impact on the alignment of the arcs with the gradient.
    Complete Reeb graphs include \textit{\textcolor{red}{red}} and \textit{\textcolor{blue}{blue}} arcs; \textit{\textcolor{red}{red}} arcs highlight comparisons. For additional examples, see supplement.
    }
    \label{fig:issue3:BvC:foot1}
\end{figure}

\subsection{Property 3: Following the Function Gradient}
\label{eval:issue3}
Reeb graph arcs represent the monotonic change of the function value between critical points. Therefore, the arc should ideally follow the gradient of the function. However, in practice, it is not always possible to draw a Reeb graph that completely follows the function's gradient direction because of the complex shape of the model and data.

\begin{figure}[!t]
    \centering

    \begin{minipage}[m]{0.025\linewidth}
        \rotatebox{90}{\hspace{15pt}\tiny \textsf{Mean Deviation from Gradient}}
    \end{minipage}
    \begin{minipage}[m]{0.5\linewidth}
        \subfloat[Comparison of GASP and GB (\textit{x} function)\label{fig:issue3:BvC}]{\includegraphics[width=\linewidth]{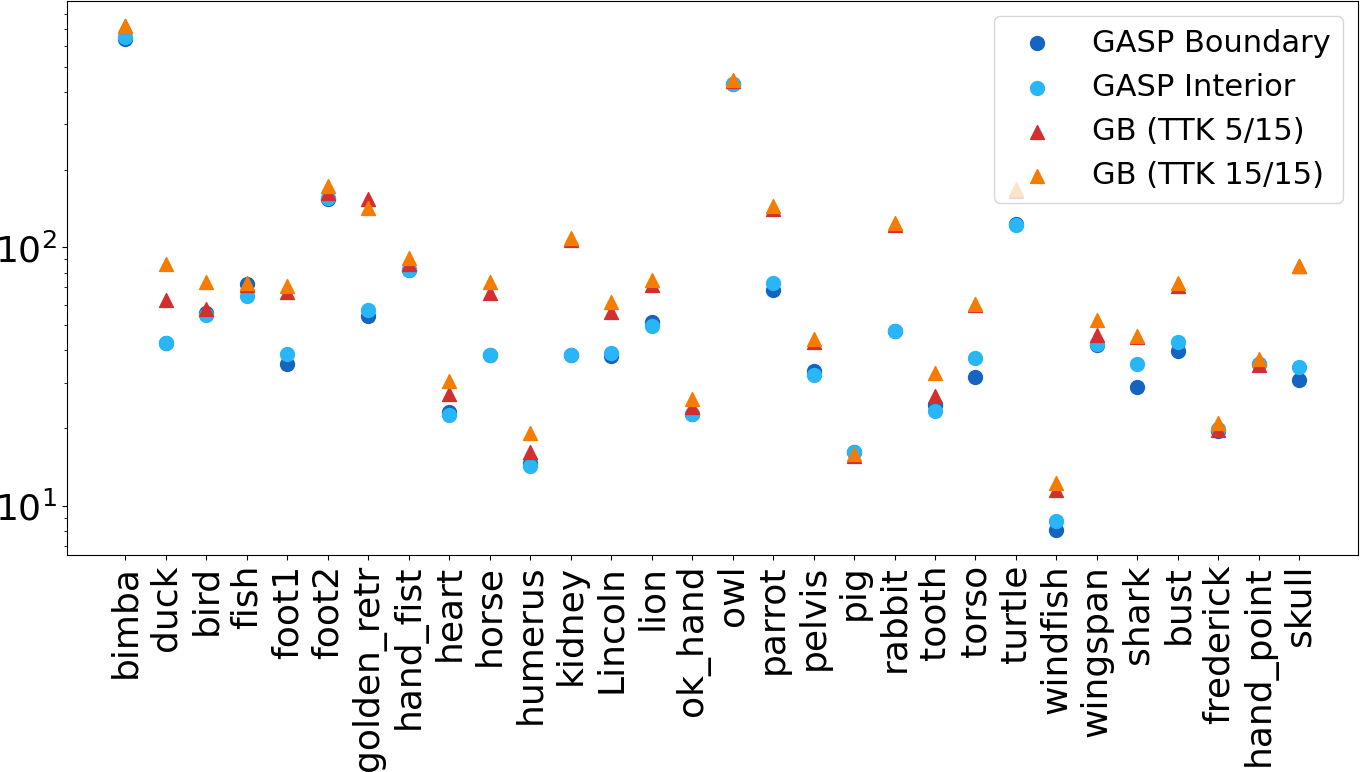}}        
    \end{minipage}
    \hfill
    \begin{minipage}[m]{0.29\linewidth}
        \subfloat[Contour Spacing (\textit{y} func)\label{fig:issue3:CS:all}]{\includegraphics[width=\linewidth]{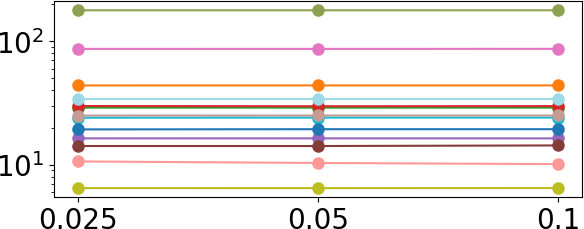}}        
        
        \subfloat[Buffer Size (\textit{y} func)\label{fig:issue3:buffer_size}]{\includegraphics[width=\linewidth]{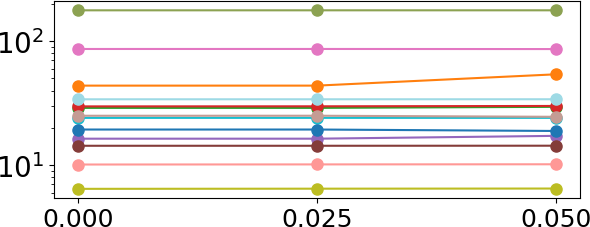}}
    \end{minipage}
    \hfill
    \begin{minipage}[m]{0.125\linewidth}
        \includegraphics[width=\linewidth]{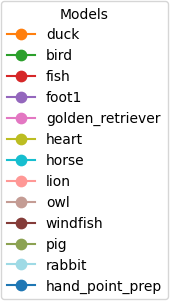}
        \vspace{10pt}
    \end{minipage}

    \caption{Quantitative analysis for Property 3, following function gradient, using the mean ratio of arc length in non-principal directions to the principal direction (log scale, lower is better). 
    (a)~Comparison of GASP and GB across all models shows that for the majority of configurations, GASP approaches yield better alignment with the gradient compared to both GB variations.
    (b)~Charts illustrating the effect of contour spacing on GASP  for the subset of models used in the paper. The near $0$ slope indicates a negligible impact on the alignment arcs with the gradient.
    (c)~Charts illustrating the effect of buffer size on GASP  for the same subset of models. The near $0$ slope indicates a negligible impact on the alignment arcs with the gradient.
    Additional examples in the supplement.
    }

\end{figure}

\paragraph{Comparison with GB}
GASP creates arcs that move contour to contour, meaning that its Reeb graphs move with the gradient, such as in \cref{fig:i3:BvC:ours:boundary}-\ref{fig:i3:BvC:ours:interior}. 
On the other hand, the sampling and smoothing of GB (TTK) cause its Reeb graph arcs (highlighted in \textcolor{red}{\textit{red}}) to move up and down (\cref{fig:i3:BvC:ttk:10}-\ref{fig:i3:BvC:ttk:15}), nearly orthogonal to the function gradient (from left to right).
To measure how well the direction of Reeb graph arcs reflects the flow direction of the function, we measure the total length in the principal direction of the function (i.e., $x$, $y$, or $z$) and compare it to the total length in non-principal directions (i.e., $y+z$, $x+z$, and $x+y$, respectively)\footnote{For ease of calculating the principal and non-principal direction, we perform this analysis only on height functions.}.  
Values are calculated per arc using the ratio of $(Distance_{NP_1} + Distance_{NP_2}) / Length_{PD}$, where $PD$ is the principal direction and $NP_1$ and $NP_2$ are the non-principal directions (lower is better). %
The results in \cref{fig:issue3:BvC} show that for most models, GASP produced arcs better aligned with the gradient than GB.

\paragraph{Contour Spacing and Buffer Size}
\uline{Contour Spacing:}
Our observation is that at lower contour spacing, arcs tend to be less smooth, while higher contour spacing produces longer, smoother arcs. However, in both cases, the arcs travel from contour to contour in the gradient direction, making their differences appear minimal (see \cref{fig:i3:cs-bird-0.025}-\ref{fig:i3:cs-bird-0.1}). This small effect is further confirmed by the negligible slope across all models in \cref{fig:issue3:CS:all}.
\uline{Buffer Size:}
Similarly, buffer size appears to have little to no effect on the Reeb graph’s alignment with the function gradient (see \cref{fig:i3:buffer-0}-\ref{fig:i3:buffer-0.05}). \Cref{fig:issue3:buffer_size} further confirms this, showing minimal slope across all models, indicating that buffer size has a negligible influence on gradient alignment.

\paragraph{Summary}
By focusing on contours for arc generation, GASP arcs align more with the gradient than GB, better representing the direction of the function of the scalar field. Further, contour spacings and buffer size show a negligible impact on the quality of GASP Reeb graphs. 

\subsection{Property 4: Optional Arc Smoothness}
\label{eval:aesthetic}

\begin{figure}[!b]
    \centering

    \begin{minipage}[m]{0.025\linewidth}
        \rotatebox{90}{\hspace{30pt}{\tiny \textsf{Mean Smoothness}}}
    \end{minipage}
    \begin{minipage}[m]{0.55\linewidth}

        \subfloat[Comparison of GASP and GB (\textit{y} function)\label{fig:Smoothness}]{\includegraphics[width=0.95\linewidth]{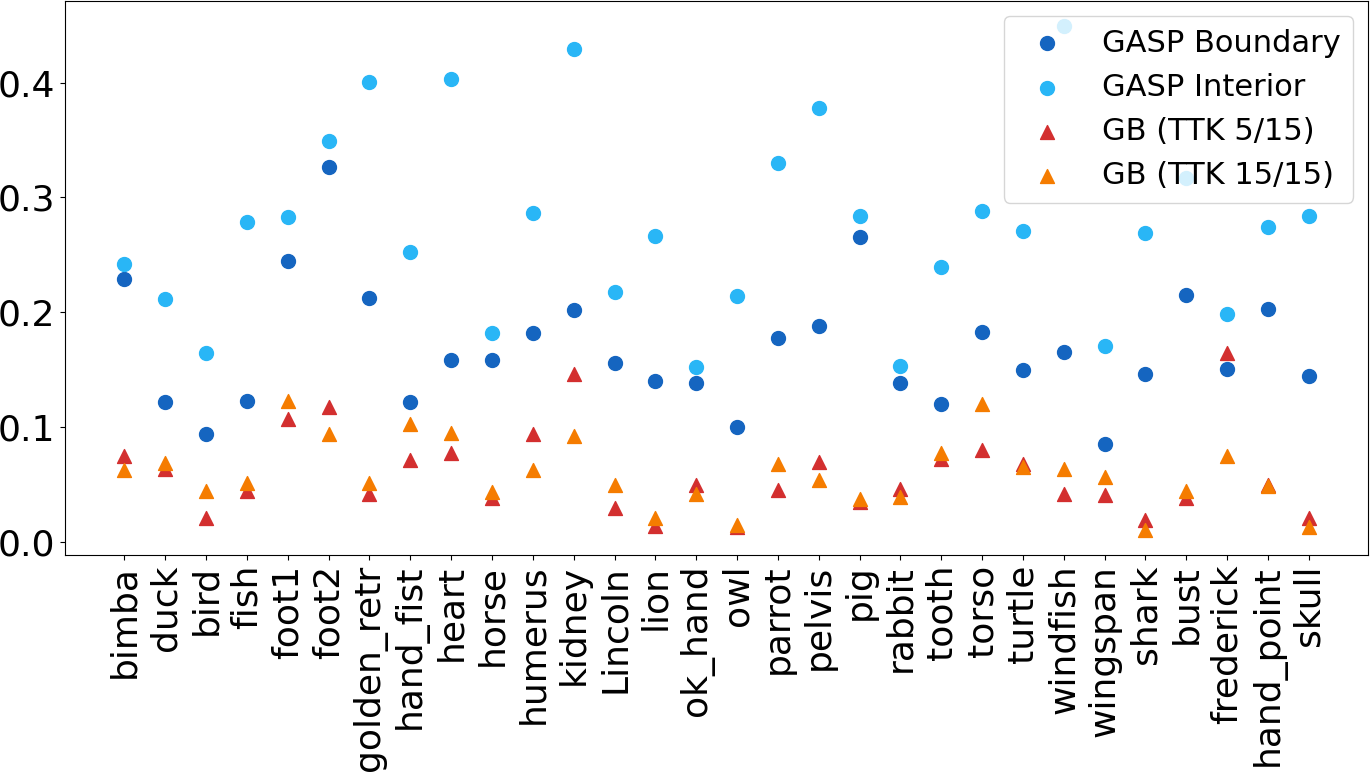}}  
    \end{minipage}
    \hfill
    \begin{minipage}[m]{0.375\linewidth}
        \centering
        
        \includegraphics[width=0.25\linewidth]{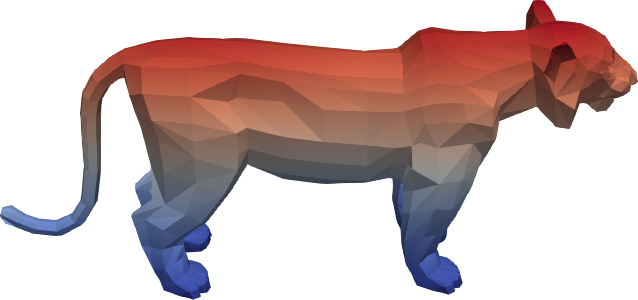}

        \vspace{-13pt}
        \hspace{40pt} \tiny \texttt{lion}
        
        \vspace{0pt}
        \hspace{51pt} $y$ \textsf{function}           
        
        \vspace{5pt}
        \subfloat[GASP Boundary\label{fig:smoothness:ours:boundary-regular}]{\includegraphics[width=0.48\linewidth]{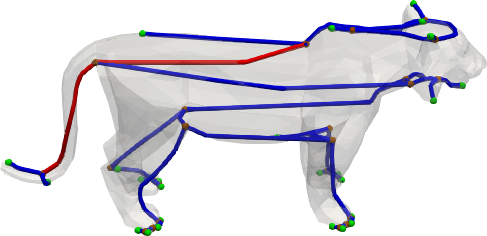}}
        \hfill
        \hfill
        \subfloat[GASP Interior\label{fig:smoothness:ours:interior-regular}]{\includegraphics[width=0.48\linewidth]{figs/Evaluation-Examples/Smoothness/lion_y_boundary_0.05_None_183.png}}
        
        \subfloat[GB (TTK 5/15)\label{fig:smoothness:ttk5}]{\includegraphics[width=0.48\linewidth]{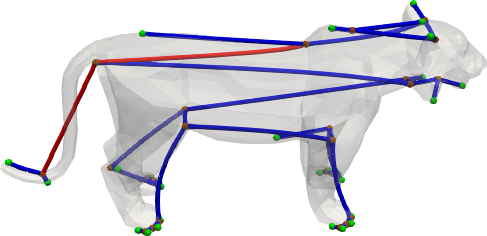}}
        \hfill
        \subfloat[GB (TTK 15/15)\label{fig:smoothness:ttk15}]{\includegraphics[width=0.48\linewidth]{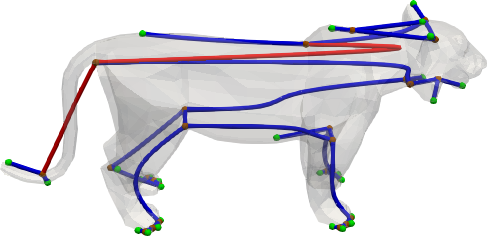}}
    \end{minipage}

    \caption{Quantitative analysis and examples of smoothness for the models used in the experiment using the smoothness aesthetic measurement (lower is better). 
    (a)~Comparison of GASP and GB shows that GB has better aesthetics than GASP. Further, the GASP boundary method has better aesthetics than the interior method. For additional functions, see the supplement.
    (b-e)~Examples of (b-c)~GASP and (d-e)~GB that show differences in the smoothness of arcs.
    Complete Reeb graphs include \textit{\textcolor{red}{red}} and \textit{\textcolor{blue}{blue}} arcs; \textit{\textcolor{red}{red}} arcs highlight comparisons. 
    }
    \label{fig:smoothness}
\end{figure}

The arc smoothness of the Reeb graph is an aesthetic quality that can provide the auxiliary benefit of a clearer representation~\cite{miura2014aesthetic, dev2016perceptual}. However, ideally, improvements in smoothness should be achieved without sacrificing the three faithfulness properties. Our primary goal was to improve the faithfulness of the Reeb graph visualization, a choice that necessitates a compromise in this aesthetic aspect. This trade-off is observed as lower \textit{smoothness} in some GASP Reeb graphs. To quantify this, we calculated smoothness, $M$, as the mean of the tangents of a Reeb graph arc, with a corresponding value $\alpha=\frac{1}{1+M}$ in the range $[0,1]$ (lower is smoother). \Cref{fig:Smoothness} compares the smoothness of GASP and GB for the models. The GASP boundary approach achieves better smoothness than the interior approach, as interior arcs must deviate from the boundary and critical points, leading to more abrupt path changes (see \cref{fig:smoothness:ours:boundary-regular}-\ref{fig:smoothness:ours:interior-regular}), whereas  GB produces smoother Reeb graphs for nearly all models (see \cref{fig:smoothness:ttk5}-\ref{fig:smoothness:ttk15}).

\subsection{Computational Performance}
\label{eval:comp}

We briefly discuss the computational complexity of GASP and demonstrate the scaling effects. 

\paragraph{Complexity Analysis}
We consider a mesh of $T$ triangles and $t$ triangles per cut. 
In the average case, each triangle appears in a small number of cylinders, so $T=t\times|\Cspace|$ (i.e., the number of topological cylinders or the number of Reeb graph edges).
Decomposing the model through cut operations iterates through all triangles, such that it takes on average $\Theta(T)$. 
However, in the worst-case situation, each cut operation can end up with all $t=T$ triangles, making the worst-case complexity $O(T\times|\Cspace|)$.
Identifying each topological cylinder uses a combination of disjoint sets, which are $O(t \alpha(t))$ ($\alpha$ is the inverse Ackermann function), and priority queues, which are $O(t \log(t))$, for a worst case complexity of approximately $O(T\times|\Cspace|\log(t))$ or average case of $\Theta(T\log(t))$. By using binning approach from \cref{sec:imp:decomp}, isocontouring takes approximately $O(T)$, while building path graphs and taking the shortest path is $O(|E|\log(|V|))$ per arc ($|E|$ edges and $|V|$ vertices). Unfortunately, $|E|$ and $|V|$ are dependent on the contour spacing. All of this is to say that the average case for GASP is mostly dependent upon the number of triangles and topological cylinders ($T \times |\Cspace|$), with some variation based on the data and parameters.

\begin{figure}[!ht]
    \centering

    \begin{minipage}[m]{0.02\linewidth}
        \rotatebox{90}{\hspace{20pt}{\tiny \textsf{Execution Time (s)}}}
    \end{minipage}
    \begin{minipage}[m]{0.31\linewidth}
        \centering
        \includegraphics[width=\linewidth]{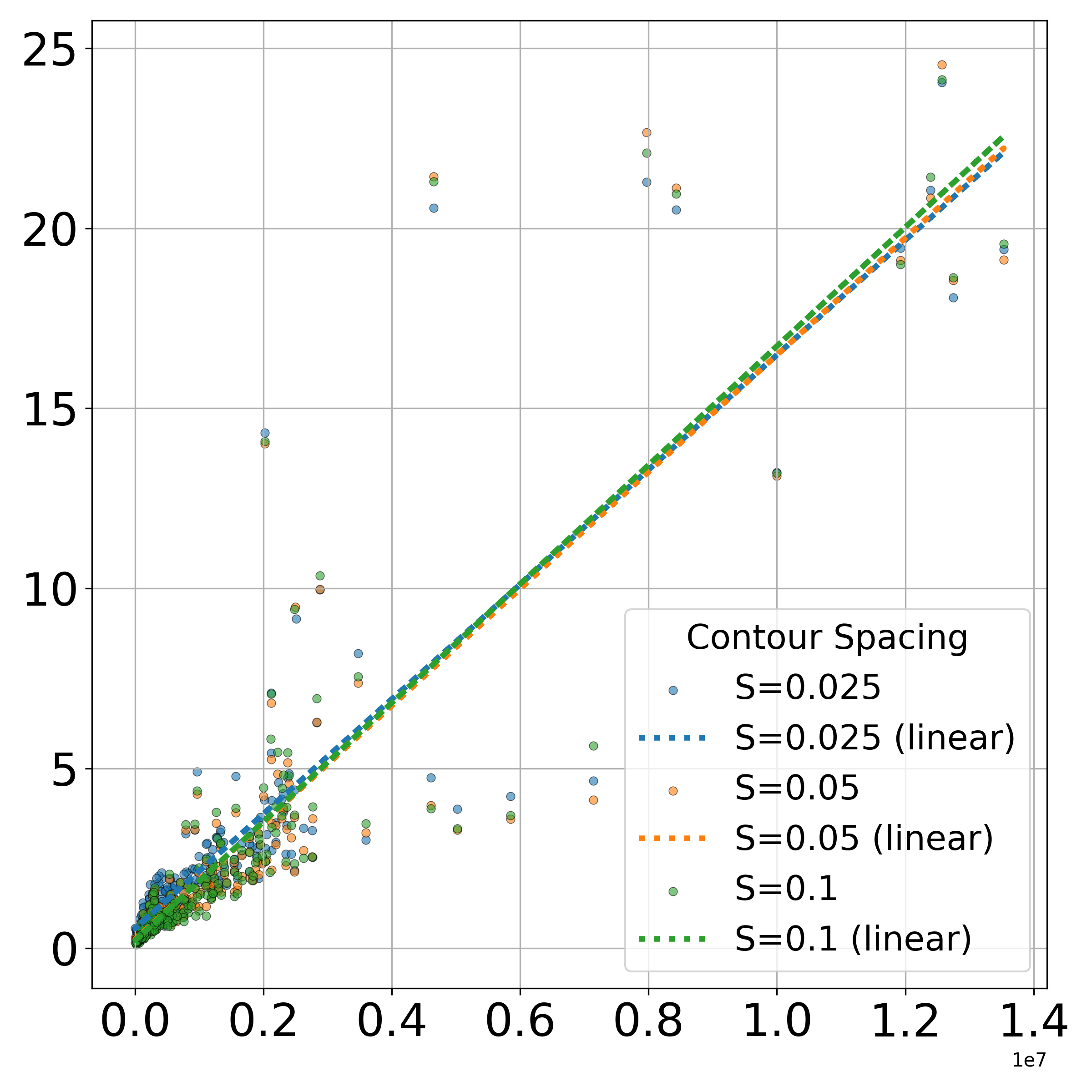}
        
        \vspace{-4pt}
        \tiny Triangles $\times$ Reeb graph arcs
    \end{minipage}    
    \begin{minipage}[m]{0.31\linewidth}
        \centering
        \includegraphics[width=\linewidth]{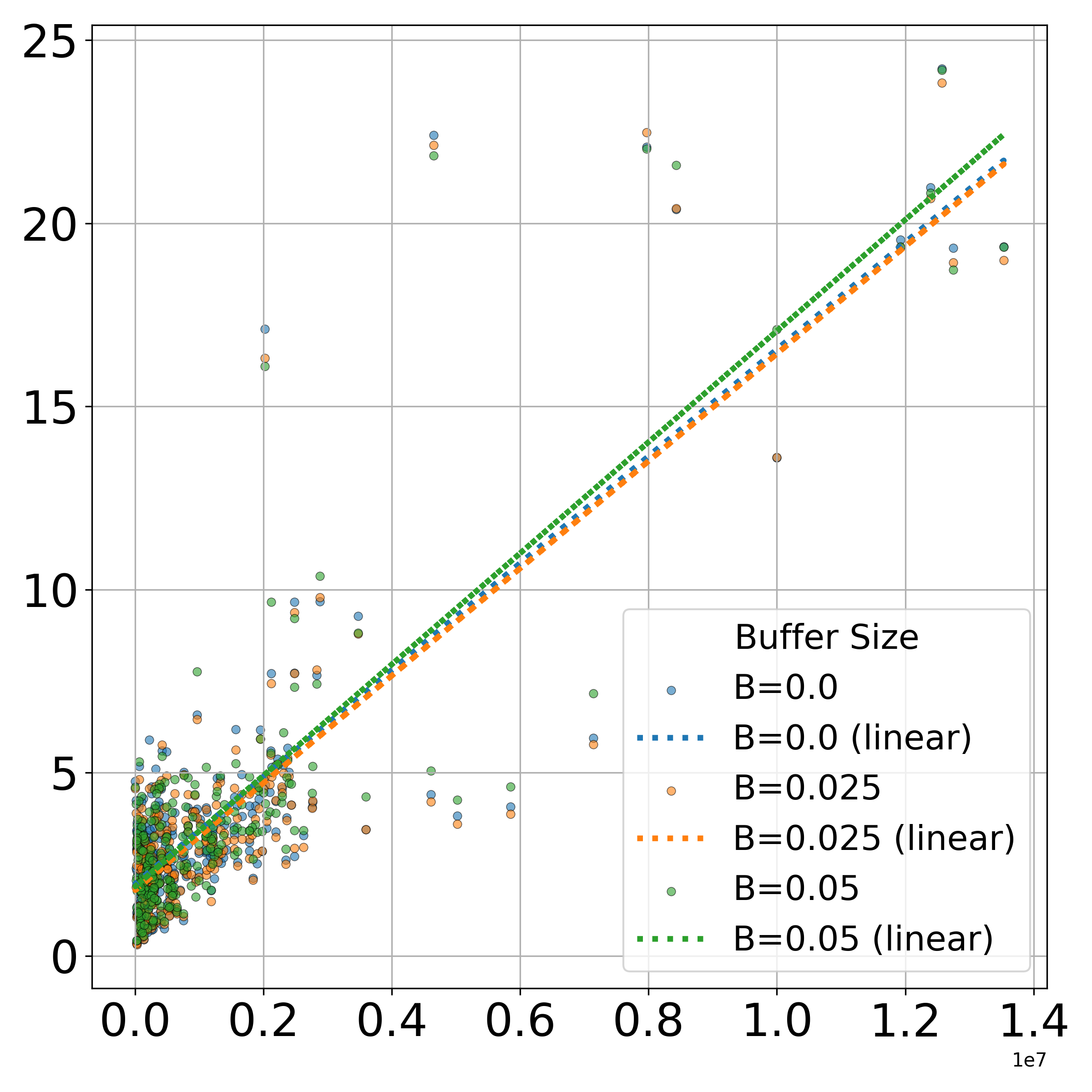}
        
        \vspace{-4pt}
        \tiny Triangles $\times$ Reeb graph arcs
    \end{minipage}    
    \begin{minipage}[m]{0.31\linewidth}
        \centering
        \includegraphics[width=\linewidth]{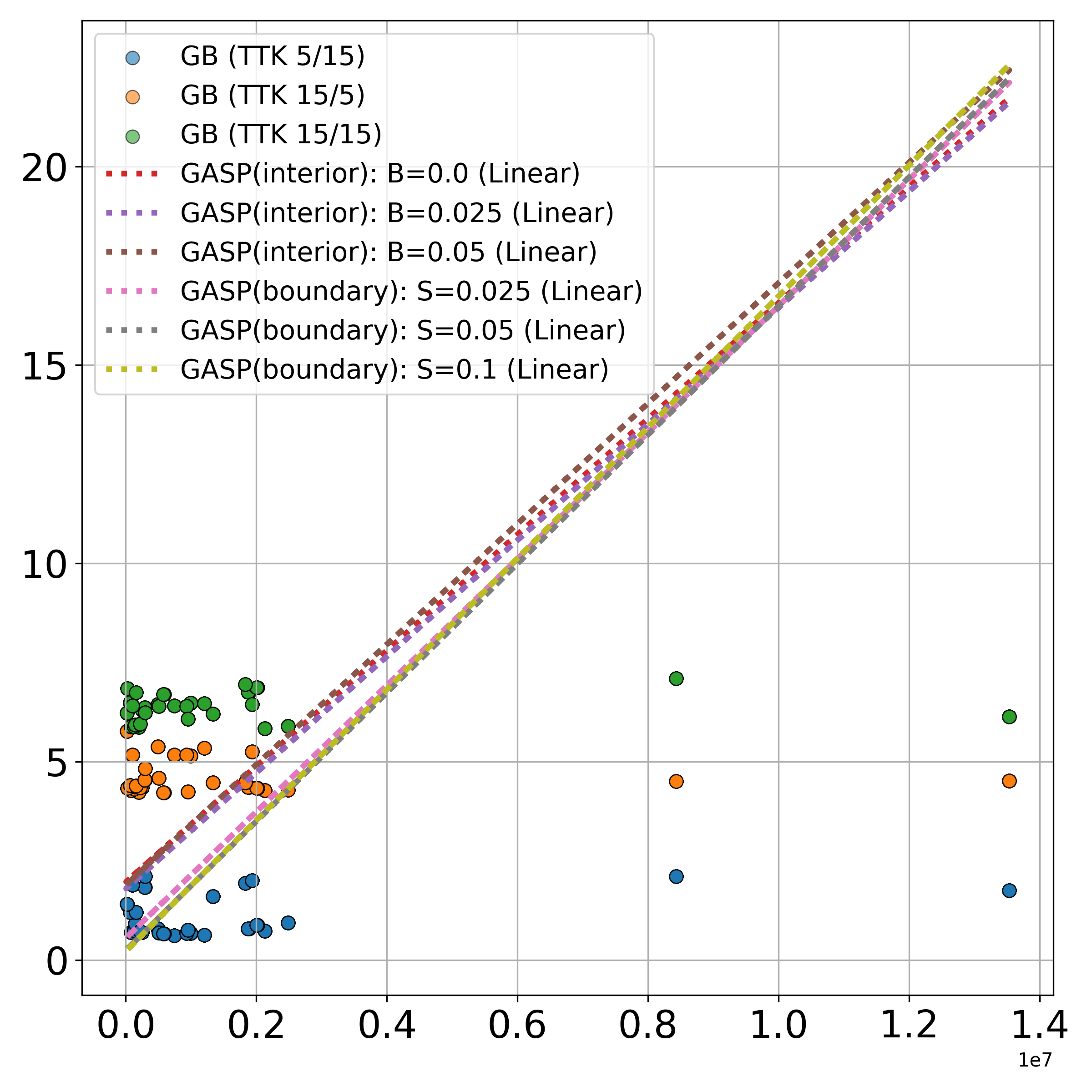}
        
        \vspace{-4pt}
        \tiny Triangles $\times$ Reeb graph arcs
    \end{minipage}    

    \vspace{-2pt}
    \hspace{0.02\linewidth}
    \subfloat[GASP Boundary\label{fig:time:boundary}]{\hspace{0.31\linewidth}}
    \hfill
    \subfloat[GASP Interior\label{fig:time:interior}]{\hspace{0.31\linewidth}}
    \hfill
    \subfloat[GB (TTK)\label{fig:time:ttk}]{\hspace{0.31\linewidth}}

    \caption{Computational time of (a-b) GASP for all models and functions for the (a)~boundary method (varied $S$) and (b)~interior method (fixed $S$, varied $B$). GASP showed highly linear scaling (goodness of fit, $R^2\approx0.8$ for boundary and $R^2\approx0.7$ for interior) with respect to the number of triangles $\times$ the number of Reeb graph arcs.  (c)~GB (TTK, implemented in C++) performance for all models for the $x$ function (other functions showed similar performance) along with the regression lines of GASP variations (implemented in Python). Unlike GASP, GB's performance is mostly related to parameter selection.}
    \label{fig:exec-time}
\end{figure}

\paragraph{Computational Performance}
To experimentally confirm this analysis, we recorded the execution time of all models and function directions on a M1 Mac Mini for the boundary approach (see \cref{fig:time:boundary}) and interior approach (see \cref{fig:time:interior}) across different contour spacing and buffer size. Reeb graph arcs were calculated using a thread pool of eight simultaneous threads. The resulting performance was linear with respect to the total number of triangles in the mesh times the number of Reeb graph edges (goodness of fit, $R^2\approx0.8$ for all boundary configurations, and $R^2\approx0.7$ for all interior configurations). The performance was slightly lower as contour spacing was reduced (i.e., more contours) and as the buffer size for the interior approach was reduced (i.e., more points inside the contours).

\paragraph{Comparison With GB}
Comparing the computational performance of GASP to the TTK implementation of GB is an apples-to-oranges comparison, as GASP is implemented as a standalone in Python, and
TTK utilizes C++ implementation of GB as part of Paraview.
Nevertheless, for context, \cref{fig:time:ttk} compares the computational performance of GB with regression lines for several variations of GASP's boundary and interior approaches.
GB's performance was, unsurprisingly, better than GASP. We also observed that GB's performance was largely parameter dependent, not data dependent.

\section{Conclusion and Future Work}

Efficient calculation of the Reeb graphs has been a well-studied problem. However, the actual visualization of Reeb graphs is understudied. When abstracting and representing data, accurately representing it is of the utmost importance~\cite{TA:2024:topoSensitivity}. 
The geometric barycenter algorithm used in the state-of-the-practice tool TTK has some drawbacks, including producing arcs outside of the model that are longer than necessary and do not optimally follow the function gradient.
GASP demonstrates significant qualitative and quantitative improvements over the geometric barycenter algorithm by constraining arcs to the model's surface and utilizing a gradient-aware shortest paths.

Despite the improvement provided by GASP, the approach still presents some limitations. 
First, the interior method assumes that the contour is approximately planar and the gradient of the function is well defined. 
For the height function, the contours are planar and the gradient is well-defined. However, for non-planar contours, determining interior points in the embedding space is not straightforward and the gradient is ill-defined. As the contour becomes less planar, the likelihood increases that the interior approach will produce a low-quality Reeb graph arc due to inaccurate approximation of interior points. Second, although GASP can technically work on any Morse function, we evaluated it only for the height and geodesic functions. Additional examples using a function defined as the sum of 3D Gaussians centered at random vertices are provided in the appendix, where our method produces Reeb graphs of similar or better quality compared to TTK.
In the future, we will investigate the extension of GASP methods to 3D meshes, with challenges associated with Reeb graph representation arising from tetrahedra and isosurface evolution.

Source code and data for the project are available at \url{https://github.com/shape-vis/gasp}.

\section*{Acknowledgments}
This project is supported in part by U.S. Department of Energy (DOE) RAPIDS-2 SciDAC project under contract number DE-AC0500OR22725 and the National Science Foundation (IIS-2316496).

\bibliographystyle{abbrv-doi}
\bibliography{main}

\begin{thebibliography}{10}

\bibitem{agarwal_et_al:LIPIcs.FSTTCS.2017.8}
P.~K. Agarwal, K.~Fox, and A.~Nath.
\newblock {Maintaining {R}eeb Graphs of Triangulated 2-Manifolds}.
\newblock In S.~Lokam and R.~Ramanujam, eds., {\em IARCS Annual Conference on
  Foundations of Software Technology and Theoretical Computer Science
  (FSTTCS)}, vol.~93 of {\em Leibniz International Proceedings in Informatics
  (LIPIcs)}, pp. 8:1--8:14. Schloss Dagstuhl -- Leibniz-Zentrum f{\"u}r
  Informatik, Dagstuhl, Germany, 2018. doi: {{%
10\hspace{.1pt}\discretionary{.}{%
}{.}\hspace{.4pt}4230\discretionary{/}{%
}{/}LIPIcs\hspace{.1pt}\discretionary{.}{%
}{.}\hspace{.4pt}FSTTCS\hspace{.1pt}\discretionary{.}{%
}{.}\hspace{.4pt}2017\hspace{.1pt}\discretionary{.}{%
}{.}\hspace{.4pt}8}}


\bibitem{TA:2005:paraview}
J.~Ahrens, B.~Geveci, and C.~Law.
\newblock {\em ParaView: An End-User Tool for Large Data Visualization},
  chap.~36, pp. 717--731.
\newblock Elsevier, 2005. doi: {{%
10\hspace{.1pt}\discretionary{.}{%
}{.}\hspace{.4pt}1016\discretionary{/}{%
}{/}B978\discretionary{%
}{-}{-}012387582\discretionary{%
}{-}{-}2\discretionary{/}{%
}{/}50038\discretionary{%
}{-}{-}1}}


\bibitem{TA:2024:topoSensitivity}
T.~M. Athawale, B.~Triana, T.~Kotha, D.~Pugmire, and P.~Rosen.
\newblock A comparative study of the perceptual sensitivity of topological
  visualizations to feature variations.
\newblock {\em IEEE Transactions on Visualization and Computer Graphics},
  30(1):1074--1084, 2024. doi: {{%
10\hspace{.1pt}\discretionary{.}{%
}{.}\hspace{.4pt}1109\discretionary{/}{%
}{/}TVCG\hspace{.1pt}\discretionary{.}{%
}{.}\hspace{.4pt}2023\hspace{.1pt}\discretionary{.}{%
}{.}\hspace{.4pt}3326592}}


\bibitem{Auber2004:tulip}
D.~Auber.
\newblock {\em Tulip --- A Huge Graph Visualization Framework}, pp. 105--126.
\newblock Springer Berlin Heidelberg, Berlin, Heidelberg, 2004. doi: {{%
10\hspace{.1pt}\discretionary{.}{%
}{.}\hspace{.4pt}1007\discretionary{/}{%
}{/}978\discretionary{%
}{-}{-}3\discretionary{%
}{-}{-}642\discretionary{%
}{-}{-}18638\discretionary{%
}{-}{-}7\_5}}


\bibitem{TA:2014:Bauer:reebGraphDistance}
U.~Bauer, X.~Ge, and Y.~Wang.
\newblock Measuring distance between {R}eeb graphs.
\newblock In {\em Symposium on Computational Geometry (SoCG)}, pp. 464--473.
  ACM, 2014. doi: {{%
10\hspace{.1pt}\discretionary{.}{%
}{.}\hspace{.4pt}1145\discretionary{/}{%
}{/}2582112\hspace{.1pt}\discretionary{.}{%
}{.}\hspace{.4pt}2582169}}


\bibitem{extendedSurfaceGraphs:SurfaceRendering}
S.~Biasotti, B.~Falcidieno, and M.~Spagnuolo.
\newblock Extended {R}eeb graphs for surface understanding and description.
\newblock In G.~Borgefors, I.~Nystr{\"o}m, and G.~S. di~Baja, eds., {\em
  Discrete Geometry for Computer Imagery}, pp. 185--197. Springer Berlin
  Heidelberg, Berlin, Heidelberg, 2000. doi: {{%
10\hspace{.1pt}\discretionary{.}{%
}{.}\hspace{.4pt}1007\discretionary{/}{%
}{/}3\discretionary{%
}{-}{-}540\discretionary{%
}{-}{-}44438\discretionary{%
}{-}{-}6\_16}}


\bibitem{BIASOTTI20085:shapeanalysis}
S.~Biasotti, D.~Giorgi, M.~Spagnuolo, and B.~Falcidieno.
\newblock {R}eeb graphs for shape analysis and applications.
\newblock {\em Theoretical Computer Science}, 392(1):5--22, 2008.
\newblock Computational Algebraic Geometry and Applications. doi: {{%
10\hspace{.1pt}\discretionary{.}{%
}{.}\hspace{.4pt}1016\discretionary{/}{%
}{/}j\hspace{.1pt}\discretionary{.}{%
}{.}\hspace{.4pt}tcs\hspace{.1pt}\discretionary{.}{%
}{.}\hspace{.4pt}2007\hspace{.1pt}\discretionary{.}{%
}{.}\hspace{.4pt}10\hspace{.1pt}\discretionary{.}{%
}{.}\hspace{.4pt}018}}


\bibitem{TA:2023:Bollen:stableDistance}
B.~Bollen, P.~Tennakoon, and J.~A. Levine.
\newblock Computing a stable distance on merge trees.
\newblock {\em IEEE Transactions on Visualization and Computer Graphics},
  29(1):1168--1177, 2023. doi: {{%
10\hspace{.1pt}\discretionary{.}{%
}{.}\hspace{.4pt}1109\discretionary{/}{%
}{/}TVCG\hspace{.1pt}\discretionary{.}{%
}{.}\hspace{.4pt}2022\hspace{.1pt}\discretionary{.}{%
}{.}\hspace{.4pt}3209395}}


\bibitem{Bremer09tvcg}
P.-T. Bremer, G.~Weber, V.~Pascucci, M.~Day, and J.~Bell.
\newblock Analyzing and tracking burning structures in lean premixed hydrogen
  flames.
\newblock {\em IEEE Transactions on Visualization and Computer Graphics},
  16(2):248--260, 2010. doi: {{%
10\hspace{.1pt}\discretionary{.}{%
}{.}\hspace{.4pt}1109\discretionary{/}{%
}{/}TVCG\hspace{.1pt}\discretionary{.}{%
}{.}\hspace{.4pt}2009\hspace{.1pt}\discretionary{.}{%
}{.}\hspace{.4pt}69}}


\bibitem{carr2003computing}
H.~Carr, J.~Snoeyink, and U.~Axen.
\newblock Computing contour trees in all dimensions.
\newblock {\em Computational Geometry}, 24(2):75--94, 2003. doi: {{%
10\hspace{.1pt}\discretionary{.}{%
}{.}\hspace{.4pt}1016\discretionary{/}{%
}{/}S0925\discretionary{%
}{-}{-}7721\discretionary{%
}{(}{(}02\discretionary{)}{%
}{)}00093\discretionary{%
}{-}{-}7}}


\bibitem{TA:Chen:2013:fluidParticleTracking}
F.~Chen, H.~Obermaier, H.~Hagen, B.~Hamann, J.~Tierny, and V.~Pascucci.
\newblock Topology analysis of time-dependent multi-fluid data using the {R}eeb
  graph.
\newblock {\em Computer Aided Geometric Design}, 30(6):557--566, 2013.
\newblock Foundations of Topological Analysis. doi: {{%
10\hspace{.1pt}\discretionary{.}{%
}{.}\hspace{.4pt}1016\discretionary{/}{%
}{/}j\hspace{.1pt}\discretionary{.}{%
}{.}\hspace{.4pt}cagd\hspace{.1pt}\discretionary{.}{%
}{.}\hspace{.4pt}2012\hspace{.1pt}\discretionary{.}{%
}{.}\hspace{.4pt}03\hspace{.1pt}\discretionary{.}{%
}{.}\hspace{.4pt}019}}


\bibitem{persplot_stability}
D.~Cohen-Steiner, H.~Edelsbrunner, and J.~Harer.
\newblock Stability of persistence diagrams.
\newblock In {\em Symposium on Computational Geometry (SoCG)}, pp. 263--271.
  ACM, 2005. doi: {{%
10\hspace{.1pt}\discretionary{.}{%
}{.}\hspace{.4pt}1145\discretionary{/}{%
}{/}1064092\hspace{.1pt}\discretionary{.}{%
}{.}\hspace{.4pt}1064133}}


\bibitem{TA:Kree:2004}
K.~Cole-McLaughlin, H.~Edelsbrunner, J.~Harer, V.~Natarajan, and V.~Pascucci.
\newblock Loops in {R}eeb graphs of 2-manifolds.
\newblock {\em Discrete and Computational Geometry}, 32(2):231--244, 2004. doi:
  {{%
10\hspace{.1pt}\discretionary{.}{%
}{.}\hspace{.4pt}1145\discretionary{/}{%
}{/}777792\hspace{.1pt}\discretionary{.}{%
}{.}\hspace{.4pt}777844}}


\bibitem{cornea2024curve}
N.~D. Cornea, D.~Silver, and P.~Min.
\newblock Curve-skeleton properties, applications, and algorithms.
\newblock {\em IEEE Transactions on visualization and computer graphics},
  13(3):530--548, 2024.

\bibitem{dev2016perceptual}
K.~Dev, M.~Lau, and L.~Liu.
\newblock A perceptual aesthetics measure for 3d shapes.
\newblock {\em arXiv preprint}, 2016.

\bibitem{TA:Doraiswamy:2008}
H.~Doraiswamy and V.~Natarajan.
\newblock Efficient output-sensitive construction of {R}eeb graphs.
\newblock In S.-H. Hong, H.~Nagamochi, and T.~Fukunaga, eds., {\em Algorithms
  and Computation}, pp. 556--567. Springer Berlin Heidelberg, Berlin,
  Heidelberg, 2008. doi: {{%
10\hspace{.1pt}\discretionary{.}{%
}{.}\hspace{.4pt}1007\discretionary{/}{%
}{/}978\discretionary{%
}{-}{-}3\discretionary{%
}{-}{-}540\discretionary{%
}{-}{-}92182\discretionary{%
}{-}{-}0\_50}}


\bibitem{DORAISWAMY2009606}
H.~Doraiswamy and V.~Natarajan.
\newblock Efficient algorithms for computing {R}eeb graphs.
\newblock {\em Computational Geometry}, 42(6):606--616, 2009. doi: {{%
10\hspace{.1pt}\discretionary{.}{%
}{.}\hspace{.4pt}1016\discretionary{/}{%
}{/}j\hspace{.1pt}\discretionary{.}{%
}{.}\hspace{.4pt}comgeo\hspace{.1pt}\discretionary{.}{%
}{.}\hspace{.4pt}2008\hspace{.1pt}\discretionary{.}{%
}{.}\hspace{.4pt}12\hspace{.1pt}\discretionary{.}{%
}{.}\hspace{.4pt}003}}


\bibitem{ReebGraphSimplification}
H.~Doraiswamy and V.~Natarajan.
\newblock Output-sensitive construction of {Reeb} graphs.
\newblock {\em IEEE Transactions on Visualization and Computer Graphics},
  18(1):146--159, 2012. doi: {{%
10\hspace{.1pt}\discretionary{.}{%
}{.}\hspace{.4pt}1109\discretionary{/}{%
}{/}TVCG\hspace{.1pt}\discretionary{.}{%
}{.}\hspace{.4pt}2011\hspace{.1pt}\discretionary{.}{%
}{.}\hspace{.4pt}37}}


\bibitem{doraiswamy2012computing}
H.~Doraiswamy and V.~Natarajan.
\newblock Computing {R}eeb graphs as a union of contour trees.
\newblock {\em IEEE Transactions on Visualization and Computer Graphics},
  19(2):249--262, 2013. doi: {{%
10\hspace{.1pt}\discretionary{.}{%
}{.}\hspace{.4pt}1109\discretionary{/}{%
}{/}TVCG\hspace{.1pt}\discretionary{.}{%
}{.}\hspace{.4pt}2012\hspace{.1pt}\discretionary{.}{%
}{.}\hspace{.4pt}115}}


\bibitem{computational_topology_intro}
H.~Edelsbrunner and J.~Harer.
\newblock {\em Computational Topology - an Introduction.}
\newblock American Mathematical Society, 2010.

\bibitem{TA:Edelsbrunner:2003:MorseSmaleComplexes}
H.~Edelsbrunner, J.~Harer, and A.~Zomorodian.
\newblock Hierarchical {M}orse---{S}male complexes for piecewise linear
  2-manifolds.
\newblock {\em Discrete and Computational Geometry}, 30(1):87--107, 2003. doi:
  {{%
10\hspace{.1pt}\discretionary{.}{%
}{.}\hspace{.4pt}1007\discretionary{/}{%
}{/}s00454\discretionary{%
}{-}{-}003\discretionary{%
}{-}{-}2926\discretionary{%
}{-}{-}5}}


\bibitem{EdelsbrunnerLetscherZomorodian2002}
H.~Edelsbrunner, D.~Letscher, and A.~J. Zomorodian.
\newblock Topological persistence and simplification.
\newblock {\em Discrete and Computational Geometry}, 28:511--533, 2002. doi:
  {{%
10\hspace{.1pt}\discretionary{.}{%
}{.}\hspace{.4pt}1109\discretionary{/}{%
}{/}SFCS\hspace{.1pt}\discretionary{.}{%
}{.}\hspace{.4pt}2000\hspace{.1pt}\discretionary{.}{%
}{.}\hspace{.4pt}892133}}


\bibitem{TA:Forsberg:2009:3dvectorFieldVisUserStudy}
A.~Forsberg, J.~Chen, and D.~Laidlaw.
\newblock Comparing 3{D} vector field visualization methods: A user study.
\newblock {\em IEEE Transactions on Visualization and Computer Graphics},
  15(6):1219--1226, 2009. doi: {{%
10\hspace{.1pt}\discretionary{.}{%
}{.}\hspace{.4pt}1109\discretionary{/}{%
}{/}TVCG\hspace{.1pt}\discretionary{.}{%
}{.}\hspace{.4pt}2009\hspace{.1pt}\discretionary{.}{%
}{.}\hspace{.4pt}126}}


\bibitem{dot:graphdrawing}
E.~Gansner, E.~Koutsofios, S.~North, and K.-P. Vo.
\newblock A technique for drawing directed graphs.
\newblock {\em IEEE Transactions on Software Engineering}, 19(3):214--230,
  1993. doi: {{%
10\hspace{.1pt}\discretionary{.}{%
}{.}\hspace{.4pt}1109\discretionary{/}{%
}{/}32\hspace{.1pt}\discretionary{.}{%
}{.}\hspace{.4pt}221135}}


\bibitem{Gelbukh_2023}
I.~Gelbukh.
\newblock {R}eeb graphs of circle-valued functions: A survey and basic facts.
\newblock {\em Topological Methods in Nonlinear Analysis}, 61(1):59--81, Mar.
  2023. doi: {{%
10\hspace{.1pt}\discretionary{.}{%
}{.}\hspace{.4pt}12775\discretionary{/}{%
}{/}TMNA\hspace{.1pt}\discretionary{.}{%
}{.}\hspace{.4pt}2022\hspace{.1pt}\discretionary{.}{%
}{.}\hspace{.4pt}023}}


\bibitem{gueunet2019task}
C.~Gueunet, P.~Fortin, J.~Jomier, and J.~Tierny.
\newblock Task-based augmented {R}eeb graphs with dynamic {ST}-trees.
\newblock In {\em Eurographics Symposium on Parallel Graphics and
  Visualization}, 2019.

\bibitem{TA:Gyulassy:2005:scalarFieldSimplification}
A.~Gyulassy and V.~Natarajan.
\newblock Topology-based simplification for feature extraction from 3{D} scalar
  fields.
\newblock In {\em IEEE Visualization}, pp. 535--542, 2005. doi: {{%
10\hspace{.1pt}\discretionary{.}{%
}{.}\hspace{.4pt}1109\discretionary{/}{%
}{/}VISUAL\hspace{.1pt}\discretionary{.}{%
}{.}\hspace{.4pt}2005\hspace{.1pt}\discretionary{.}{%
}{.}\hspace{.4pt}1532839}}


\bibitem{parallel_reeb}
M.~Hajij and P.~Rosen.
\newblock An efficient data retrieval parallel {R}eeb graph algorithm.
\newblock {\em Algorithms}, 13(10):258, 2020. doi: {{%
10\hspace{.1pt}\discretionary{.}{%
}{.}\hspace{.4pt}3390\discretionary{/}{%
}{/}a13100258}}


\bibitem{harvey2010randomized}
W.~Harvey, Y.~Wang, and R.~Wenger.
\newblock A randomized {O}(m log m) time algorithm for computing {R}eeb graphs
  of arbitrary simplicial complexes.
\newblock In {\em Proceedings of the twenty-sixth annual symposium on
  Computational geometry}, pp. 267--276, 2010.

\bibitem{Heine:contourtreedrawing}
C.~Heine, D.~Schneider, H.~Carr, and G.~Scheuermann.
\newblock Drawing contour trees in the plane.
\newblock {\em IEEE Transactions on Visualization and Computer Graphics},
  17(11):1599--1611, 2011. doi: {{%
10\hspace{.1pt}\discretionary{.}{%
}{.}\hspace{.4pt}1109\discretionary{/}{%
}{/}TVCG\hspace{.1pt}\discretionary{.}{%
}{.}\hspace{.4pt}2010\hspace{.1pt}\discretionary{.}{%
}{.}\hspace{.4pt}270}}


\bibitem{hilton1996marching}
A.~Hilton, A.~J. Stoddart, J.~Illingworth, and T.~Windeatt.
\newblock Marching triangles: range image fusion for complex object modelling.
\newblock In {\em IEEE International Conference on Image Processing}, vol.~2,
  pp. 381--384, 1996. doi: {{%
10\hspace{.1pt}\discretionary{.}{%
}{.}\hspace{.4pt}1109\discretionary{/}{%
}{/}ICIP\hspace{.1pt}\discretionary{.}{%
}{.}\hspace{.4pt}1996\hspace{.1pt}\discretionary{.}{%
}{.}\hspace{.4pt}560840}}


\bibitem{9926485}
J.~Khatkar, L.~Clemon, R.~Fitch, and R.~Mettu.
\newblock A {R}eeb graph approach for faster 3{D} printing.
\newblock In {\em IEEE International Conference on Automation Science and
  Engineering (CASE)}, pp. 277--282, 2022. doi: {{%
10\hspace{.1pt}\discretionary{.}{%
}{.}\hspace{.4pt}1109\discretionary{/}{%
}{/}CASE49997\hspace{.1pt}\discretionary{.}{%
}{.}\hspace{.4pt}2022\hspace{.1pt}\discretionary{.}{%
}{.}\hspace{.4pt}9926485}}


\bibitem{kitazawa2023notes}
N.~Kitazawa.
\newblock Notes on {R}eeb graphs of real algebraic functions which may not be
  planar.
\newblock {\em arXiv preprint}, 2023.

\bibitem{TA:Laidlaw:2005:2dvectorFieldVisUserStudy}
D.~Laidlaw, R.~Kirby, C.~Jackson, J.~Davidson, T.~Miller, M.~da~Silva,
  W.~Warren, and M.~Tarr.
\newblock Comparing 2{D} vector field visualization methods: a user study.
\newblock {\em IEEE Transactions on Visualization and Computer Graphics},
  11(1):59--70, 2005. doi: {{%
10\hspace{.1pt}\discretionary{.}{%
}{.}\hspace{.4pt}1109\discretionary{/}{%
}{/}TVCG\hspace{.1pt}\discretionary{.}{%
}{.}\hspace{.4pt}2005\hspace{.1pt}\discretionary{.}{%
}{.}\hspace{.4pt}4}}


\bibitem{TA:2023:Lan:labledInterleavingDistanceReebGraph}
F.~Lan, S.~Parsa, and B.~Wang.
\newblock Labeled interleaving distance for {R}eeb graphs, 2023. doi: {{%
10\hspace{.1pt}\discretionary{.}{%
}{.}\hspace{.4pt}48550\discretionary{/}{%
}{/}arXiv\hspace{.1pt}\discretionary{.}{%
}{.}\hspace{.4pt}2306\hspace{.1pt}\discretionary{.}{%
}{.}\hspace{.4pt}01186}}


\bibitem{Lorensen:1987:MCA}
W.~E. Lorensen and H.~E. Cline.
\newblock Marching cubes: A high resolution {3D} surface construction
  algorithm.
\newblock {\em SIGGRAPH Computer Graphics}, 21(4):163--169, August 1987. doi:
  {{%
10\hspace{.1pt}\discretionary{.}{%
}{.}\hspace{.4pt}1145\discretionary{/}{%
}{/}37402\hspace{.1pt}\discretionary{.}{%
}{.}\hspace{.4pt}37422}}


\bibitem{miura2014aesthetic}
K.~T. Miura and G.~RU.
\newblock Aesthetic curves and surfaces in computer aided geometric design.
\newblock {\em International Journal of Automation Technology}, 8(3):304--316,
  2014. doi: {{%
10\hspace{.1pt}\discretionary{.}{%
}{.}\hspace{.4pt}20965\discretionary{/}{%
}{/}ijat\hspace{.1pt}\discretionary{.}{%
}{.}\hspace{.4pt}2014\hspace{.1pt}\discretionary{.}{%
}{.}\hspace{.4pt}p0304}}


\bibitem{TA:2013:Morozov:interleavingDistance}
D.~Morozov, K.~Beketayev, and G.~Weber.
\newblock Interleaving distance between merge trees.
\newblock In {\em TopoInVis: Topological Methods in Data Analysis and
  Visualization}, 2013.

\bibitem{TA:Morse:1930:criticalPoints}
M.~Morse.
\newblock The critical points of a function of n variables.
\newblock {\em National Academy of Sciences}, 16(11):777, 1930. doi: {{%
10\hspace{.1pt}\discretionary{.}{%
}{.}\hspace{.4pt}1073\discretionary{/}{%
}{/}pnas\hspace{.1pt}\discretionary{.}{%
}{.}\hspace{.4pt}16\hspace{.1pt}\discretionary{.}{%
}{.}\hspace{.4pt}11\hspace{.1pt}\discretionary{.}{%
}{.}\hspace{.4pt}777}}


\bibitem{NATALI2011151:pointCloud}
M.~Natali, S.~Biasotti, G.~Patan{\`e}, and B.~Falcidieno.
\newblock Graph-based representations of point clouds.
\newblock {\em Graphical Models}, 73(5):151--164, 2011. doi: {{%
10\hspace{.1pt}\discretionary{.}{%
}{.}\hspace{.4pt}1016\discretionary{/}{%
}{/}j\hspace{.1pt}\discretionary{.}{%
}{.}\hspace{.4pt}gmod\hspace{.1pt}\discretionary{.}{%
}{.}\hspace{.4pt}2011\hspace{.1pt}\discretionary{.}{%
}{.}\hspace{.4pt}03\hspace{.1pt}\discretionary{.}{%
}{.}\hspace{.4pt}002}}


\bibitem{NatarajanMolecularDynamics}
V.~Natarajan, P.~Koehl, Y.~Wang, and B.~Hamann.
\newblock Visual analysis of biomolecular surfaces.
\newblock In L.~Linsen, H.~Hagen, and B.~Hamann, eds., {\em Visualization in
  Medicine and Life Sciences}, pp. 237--255. Springer Berlin Heidelberg,
  Berlin, Heidelberg, 2008. doi: {{%
10\hspace{.1pt}\discretionary{.}{%
}{.}\hspace{.4pt}1007\discretionary{/}{%
}{/}978\discretionary{%
}{-}{-}3\discretionary{%
}{-}{-}540\discretionary{%
}{-}{-}72630\discretionary{%
}{-}{-}2\_14}}


\bibitem{TA:2022:Nauleau:turbulentFlowTopoAnalysis}
F.~Nauleau, F.~Vivodtzev, T.~Bridel-Bertomeu, H.~Beaugendre, and J.~Tierny.
\newblock Topological analysis of ensembles of hydrodynamic turbulent flows --
  an experimental study.
\newblock In {\em Symposium on Large Data Analysis and Visualization (LDAV)},
  2022. doi: {{%
10\hspace{.1pt}\discretionary{.}{%
}{.}\hspace{.4pt}1109\discretionary{/}{%
}{/}LDAV57265\hspace{.1pt}\discretionary{.}{%
}{.}\hspace{.4pt}2022\hspace{.1pt}\discretionary{.}{%
}{.}\hspace{.4pt}9966403}}


\bibitem{landscape:ReebGraph}
K.~Ohmori and T.~L. Kunii.
\newblock Three dimensional sketch for a landscape using {M}orse theory and
  {R}eeb graphs.
\newblock In {\em International Conference on Cyberworlds}, pp. 178--183, 2012.
  doi: {{%
10\hspace{.1pt}\discretionary{.}{%
}{.}\hspace{.4pt}1109\discretionary{/}{%
}{/}CW\hspace{.1pt}\discretionary{.}{%
}{.}\hspace{.4pt}2012\hspace{.1pt}\discretionary{.}{%
}{.}\hspace{.4pt}32}}


\bibitem{parsa2012deterministic}
S.~Parsa.
\newblock A deterministic {O}(m log m) time algorithm for the {R}eeb graph.
\newblock In {\em Proceedings of the twenty-eighth annual symposium on
  Computational geometry}, pp. 269--276, 2012.

\bibitem{Pascucci2009}
V.~Pascucci, K.~Cole-McLaughlin, and G.~Scorzelli.
\newblock {\em The TOPORRERY: computation and presentation of multi-resolution
  topology}, pp. 19--40.
\newblock Springer Berlin Heidelberg, Berlin, Heidelberg, 2009. doi: {{%
10\hspace{.1pt}\discretionary{.}{%
}{.}\hspace{.4pt}1007\discretionary{/}{%
}{/}b106657\_2}}


\bibitem{pascucci2007robust}
V.~Pascucci, G.~Scorzelli, P.-T. Bremer, and A.~Mascarenhas.
\newblock Robust on-line computation of {R}eeb graphs: simplicity and speed.
\newblock In {\em ACM SIGGRAPH}, pp. 58--es, 2007. doi: {{%
10\hspace{.1pt}\discretionary{.}{%
}{.}\hspace{.4pt}1145\discretionary{/}{%
}{/}1275808\hspace{.1pt}\discretionary{.}{%
}{.}\hspace{.4pt}1276449}}


\bibitem{minimalContouringReebGraphs}
G.~Patane, M.~Spagnuolo, and B.~Falcidieno.
\newblock A minimal contouring approach to the computation of the {R}eeb graph.
\newblock {\em IEEE Transactions on Visualization and Computer Graphics},
  15(4):583--595, 2009. doi: {{%
10\hspace{.1pt}\discretionary{.}{%
}{.}\hspace{.4pt}1109\discretionary{/}{%
}{/}TVCG\hspace{.1pt}\discretionary{.}{%
}{.}\hspace{.4pt}2009\hspace{.1pt}\discretionary{.}{%
}{.}\hspace{.4pt}22}}


\bibitem{rehal2023generation}
A.~Rehal and D.~Sen.
\newblock Generation of homotopy classes for unconstrained 3d wire routing from
  characteristic loops.
\newblock {\em Computer-Aided Design}, 164:103607, 2023.

\bibitem{rosen2021using}
P.~Rosen, A.~Seth, E.~Mills, A.~Ginsburg, J.~Kamenetzky, J.~Kern, C.~R.
  Johnson, and B.~Wang.
\newblock Using contour trees in the analysis and visualization of radio
  astronomy data cubes.
\newblock In I.~Hotz, T.~Bin~Masood, F.~Sadlo, and J.~Tierny, eds., {\em
  Topological Methods in Data Analysis and Visualization VI -- Theory,
  Applications, and Software}, pp. 87--108. Springer, Cham, 2021. doi: {{%
10\hspace{.1pt}\discretionary{.}{%
}{.}\hspace{.4pt}1007\discretionary{/}{%
}{/}978\discretionary{%
}{-}{-}3\discretionary{%
}{-}{-}030\discretionary{%
}{-}{-}83500\discretionary{%
}{-}{-}2\_6}}


\bibitem{Shailja2022.03.11.482601}
S.~Shailja, S.~T. Grafton, and B.~S. Manjunath.
\newblock A robust {R}eeb graph model of white matter fibers with application
  to alzheimer{\textquoteright}s disease progression*.
\newblock {\em bioRxiv preprint}, 2022. doi: {{%
10\hspace{.1pt}\discretionary{.}{%
}{.}\hspace{.4pt}1101\discretionary{/}{%
}{/}2022\hspace{.1pt}\discretionary{.}{%
}{.}\hspace{.4pt}03\hspace{.1pt}\discretionary{.}{%
}{.}\hspace{.4pt}11\hspace{.1pt}\discretionary{.}{%
}{.}\hspace{.4pt}482601}}


\bibitem{TA:Shinagawa:1991}
Y.~Shinagawa and T.~Kunii.
\newblock Constructing a {R}eeb graph automatically from cross sections.
\newblock {\em IEEE Computer Graphics and Applications}, 11(6):44--51, 1991.
  doi: {{%
10\hspace{.1pt}\discretionary{.}{%
}{.}\hspace{.4pt}1109\discretionary{/}{%
}{/}38\hspace{.1pt}\discretionary{.}{%
}{.}\hspace{.4pt}103393}}


\bibitem{Shinagawa:Reebgraphdrawing}
Y.~Shinagawa, T.~Kunii, and Y.~Kergosien.
\newblock Surface coding based on {M}orse theory.
\newblock {\em IEEE Computer Graphics and Applications}, 11(5):66--78, 1991.
  doi: {{%
10\hspace{.1pt}\discretionary{.}{%
}{.}\hspace{.4pt}1109\discretionary{/}{%
}{/}38\hspace{.1pt}\discretionary{.}{%
}{.}\hspace{.4pt}90568}}


\bibitem{TA:2020:Sridharamurthy:editDistance}
R.~Sridharamurthy, T.~B. Masood, A.~Kamakshidasan, and V.~Natarajan.
\newblock Edit distance between merge trees.
\newblock {\em IEEE Transactions on Visualization and Computer Graphics},
  26(3):1518--1531, 2020. doi: {{%
10\hspace{.1pt}\discretionary{.}{%
}{.}\hspace{.4pt}1109\discretionary{/}{%
}{/}TVCG\hspace{.1pt}\discretionary{.}{%
}{.}\hspace{.4pt}2018\hspace{.1pt}\discretionary{.}{%
}{.}\hspace{.4pt}2873612}}


\bibitem{STRODTHOFF2015186:layeredReebGraph}
B.~Strodthoff and B.~J\"{u}ttler.
\newblock Layered {R}eeb graphs for three-dimensional manifolds in boundary
  representation.
\newblock {\em Computers and Graphics}, 46:186--197, 2015.
\newblock Shape Modeling International 2014. doi: {{%
10\hspace{.1pt}\discretionary{.}{%
}{.}\hspace{.4pt}1016\discretionary{/}{%
}{/}j\hspace{.1pt}\discretionary{.}{%
}{.}\hspace{.4pt}cag\hspace{.1pt}\discretionary{.}{%
}{.}\hspace{.4pt}2014\hspace{.1pt}\discretionary{.}{%
}{.}\hspace{.4pt}09\hspace{.1pt}\discretionary{.}{%
}{.}\hspace{.4pt}026}}


\bibitem{julien2024}
J.~Tierny.
\newblock Personal Communication, 2024.

\bibitem{ttk}
J.~Tierny, G.~Favelier, J.~A. Levine, C.~Gueunet, and M.~Michaux.
\newblock The topology toolkit.
\newblock {\em IEEE Transactions on Visualization and Computer Graphics},
  24(1):832--842, 2018. doi: {{%
10\hspace{.1pt}\discretionary{.}{%
}{.}\hspace{.4pt}1109\discretionary{/}{%
}{/}TVCG\hspace{.1pt}\discretionary{.}{%
}{.}\hspace{.4pt}2017\hspace{.1pt}\discretionary{.}{%
}{.}\hspace{.4pt}2743938}}


\bibitem{TA:Tierny:2009:LoopSurgery}
J.~Tierny, A.~Gyulassy, E.~Simon, and V.~Pascucci.
\newblock Loop surgery for volumetric meshes: {R}eeb graphs reduced to contour
  trees.
\newblock {\em IEEE Transactions on Visualization and Computer Graphics},
  15(6):1177--1184, 2009. doi: {{%
10\hspace{.1pt}\discretionary{.}{%
}{.}\hspace{.4pt}1109\discretionary{/}{%
}{/}TVCG\hspace{.1pt}\discretionary{.}{%
}{.}\hspace{.4pt}2009\hspace{.1pt}\discretionary{.}{%
}{.}\hspace{.4pt}163}}


\bibitem{9677901}
S.~Wang, W.~Wang, and H.~Zhao.
\newblock Using foliation leaves to extract {R}eeb graphs on surfaces.
\newblock {\em IEEE Transactions on Visualization and Computer Graphics},
  29(4):2117--2131, 2023. doi: {{%
10\hspace{.1pt}\discretionary{.}{%
}{.}\hspace{.4pt}1109\discretionary{/}{%
}{/}TVCG\hspace{.1pt}\discretionary{.}{%
}{.}\hspace{.4pt}2022\hspace{.1pt}\discretionary{.}{%
}{.}\hspace{.4pt}3141764}}


\bibitem{Weber2011}
G.~Weber, P.-T. Bremer, M.~Day, J.~Bell, and V.~Pascucci.
\newblock {\em Feature Tracking Using {R}eeb Graphs}, pp. 241--253.
\newblock Springer Berlin Heidelberg, Berlin, Heidelberg, 2011. doi: {{%
10\hspace{.1pt}\discretionary{.}{%
}{.}\hspace{.4pt}1007\discretionary{/}{%
}{/}978\discretionary{%
}{-}{-}3\discretionary{%
}{-}{-}642\discretionary{%
}{-}{-}15014\discretionary{%
}{-}{-}2\_20}}


\bibitem{TA:Yan:2021:ScalarFieldComparisonTopoDescriptors}
L.~Yan, T.~B. Masood, R.~Sridharamurthy, F.~Rasheed, V.~Natarajan, I.~Hotz, and
  B.~Wang.
\newblock Scalar field comparison with topological descriptors: Properties and
  applications for scientific visualization.
\newblock {\em Computer Graphics Forum}, 40(3):599--633, 2021. doi: {{%
10\hspace{.1pt}\discretionary{.}{%
}{.}\hspace{.4pt}1111\discretionary{/}{%
}{/}cgf\hspace{.1pt}\discretionary{.}{%
}{.}\hspace{.4pt}14331}}


\end{thebibliography}

\end{document}